\documentclass[aoas]{imsart}

\RequirePackage{amsthm,amsmath,amsfonts,amssymb,centernot,float,import,makeidx,subfiles, mathtools, epstopdf, hyperref, lscape, subcaption, accents, tabularx, threeparttable, varioref, longtable, natbib, graphicx}
\usepackage[nokeyprefix]{refstyle}
\startlocaldefs
\theoremstyle{plain}

\newtheorem{proposition}{Proposition}
\newcommand{\matr}[1]{\mathbf{#1}} % undergraduate algebra version
 
\theoremstyle{remark}
\newtheorem{remark}{remark}
\endlocaldefs

\makeatletter
\newcommand*{\addFileDependency}[1]{
  \typeout{(#1)}
  \@addtofilelist{#1}
  \IfFileExists{#1}{}{\typeout{No file #1.}}
}
\makeatother

\newcommand*{\myexternaldocument}[1]{
    \externaldocument{#1}
    \addFileDependency{#1.tex}
    \addFileDependency{#1.aux}
}
\begin{document}

\begin{frontmatter}
\title{Balancing weights for region-level analysis: the effect of Medicaid Expansion on the uninsurance rate among states that did not expand Medicaid}
%\title{A sample article title with some additional note\thanksref{T1}}
\runtitle{Medicaid Expansion}
%\thankstext{T1}{A sample of additional note to the title.}

\begin{aug}
\author[A]{\fnms{Max} \snm{Rubinstein}\ead[label = e1]{mrubinst@andrew.cmu.edu}},
\author[A]{\fnms{Amelia} \snm{Haviland}\ead[label = e2,mark]{amelia@andrew.cmu.edu}}, \and
\author[A]{\fnms{David} \snm{Choi}\ead[label = e3,mark]{davidch@andrew.cmu.edu}}
\address[A]{Carnegie Mellon University, Heinz College and Department of Statistics \& Data Science \\ \printead{e1,e2,e3}}
\end{aug}

\begin{flushleft}
We predict the average effect of Medicaid expansion on the non-elderly adult uninsurance rate among states that did not expand Medicaid in 2014 as if they had expanded their Medicaid eligibility requirements. Using American Community Survey data aggregated to the region level, we estimate this effect by finding weights that approximately reweight the expansion regions to match the covariate distribution of the non-expansion regions. Existing methods to estimate balancing weights often assume that the covariates are measured without error and do not account for dependencies in the outcome model. Our covariates have random noise that is uncorrelated with the outcome errors and our assumed outcome model has state-level random effects inducing dependence between regions. To correct for the bias induced by the measurement error, we propose generating our weights on a linear approximation to the true covariates, using an idea from measurement error literature known as ``regression-calibration'' (see, e.g., \cite{carroll2006measurement}). This requires auxiliary data to estimate the variability of the measurement error. We also modify the Stable Balancing Weights objective proposed by \cite{zubizarreta2015stable}) to reduce the variance of our estimator when the model errors are homoskedastic and equicorrelated within states. We show that these approaches outperform existing methods when attempting to predict observed outcomes during the pre-treatment period. Using this method we estimate that Medicaid expansion would have caused a -2.33 (-3.54, -1.11) percentage point change in the adult uninsurance rate among states that did not expand Medicaid.
\end{flushleft}


\end{frontmatter}
%%%%%%%%%%%%%%%%%%%%%%%%%%%%%%%%%%%%%%%%%%%%%%
%% Please use \tableofcontents for articles %%
%% with 50 pages and more                   %%
%%%%%%%%%%%%%%%%%%%%%%%%%%%%%%%%%%%%%%%%%%%%%%
%\tableofcontents

%%%%%%%%%%%%%%%%%%%%%%%%%%%%%%%%%%%%%%%%%%%%%%
%%%% Main text entry area:

\section{Introduction}

We study the effect of 2014 Medicaid expansion on the non-elderly adult uninsurance rate among states that did not expand Medicaid in 2014 as if they had expanded their Medicaid eligibility requirements. We use public-use survey microdata from annual American Community Survey (ACS) aggregated to the consistent public use microdata area (CPUMA) level, a geographic region that falls within states. We calculate weights that reweight expansion-state CPUMAs to approximately match the covariate distribution of CPUMAs in states that did not expand Medicaid in 2014. We then estimate our causal effect as the difference in means between the reweighted treated CPUMAs and the observed mean of the non-expansion CPUMAs. A key challenge is that our data consists of estimated covariates. The sampling variability in these estimates is a form of measurement error that may bias effect estimates calculated on the observed data. Additionally, CPUMAs fall within states and share a common policy-making environment. The data-generating process for the outcomes therefore may contain state-level random effects that can increase the variance of standard estimation procedures. Our study contributes to the literature on balancing weights by proposing approaches to address both of these problems. We also contribute to the literature on Medicaid expansion by estimating the foregone coverage gains of Medicaid among states that did not expand Medicaid in 2014, which to our knowledge has not yet been directly estimated.

Approximate balancing weights are an estimation method in causal inference that grew out of the propensity score weighting literature. Rather than iteratively modeling the propensity score until the inverse probability weights achieve a desired level of balance, recent papers propose using optimization methods to generate weights that enforce covariate balance between the treated and control units (see, e.g., \cite{hainmueller2012entropy}, \cite{imai2014covariate}, \cite{zubizarreta2015stable}).\footnote{These methods also use ideas from the survey literature, which had proposed similar approaches to adjust sample weights to enforce equality between sample totals and known population totals (see, e.g., \cite{haberman1984adjustment}, \cite{deville1992calibration}, \cite{deville1993generalized}, \cite{sarndal2005estimation}).} From an applied perspective, there are at least four benefits of this approach: first, it does not require iterating propensity score models to generate satisfactory weights. Second, these methods (and propensity score methods generally) do not use the outcomes to determine the weights, mitigating the risk of cherry-picking an outcome model specification to obtain a desired result. Third, these methods can constrain the weights to prevent extrapolation from the data, thereby reducing model dependence (\cite{zubizarreta2015stable}). Finally, the estimates are more interpretable: by making the comparison group explicit, it is easy to communicate exactly which units contributed to the counterfactual estimate.

Most proposed methods in this literature assume that the covariates are measured without error. For our application we assume that our covariates are measured with mean-zero additive error. This error could potentially bias standard estimation procedures. As a first contribution, we therefore propose generating our weights as a function of a linear approximation to the true covariate values, using an idea from the measurement error literature known as ``regression-calibration'' (see, e.g., \cite{gleser1992importance}, \cite{carroll2006measurement}). This method requires access to an estimate of the measurement error covariance matrix, which we estimate using the ACS microdata. The theoretic consistency of these estimates requires several assumptions, including that the covariate measurement errors are uncorrelated with any errors in the outcome model, the outcome model is linear in the covariates, and the covariates are i.i.d. gaussian. The first assumption is reasonable for our application since the covariates are measured on a different cross-sectional survey than our outcomes. The second is strong but somewhat relaxed because we prevent our weights from extrapolating beyond the support of the data. The third can be fully relaxed to obtain consistent estimates using ordinary least squares (OLS), but unfortunately not without additional modeling beyond our proposed method. Despite appearing costly, we show in Section~\ref{ssec:methodsmsrment} that this tradeoff is likely worth it in our application.

As a second contribution, we propose modifying the Stable Balancing Weights (SBW) objective (\cite{zubizarreta2015stable}) to account for possible state-level dependencies in our outcome model. We assume that the errors are homoskedastic with constant positive equicorrelation, though our general approach can accommodate other assumed correlation structures. In a setting without measurement error, we show that this modification can reduce the variance of the resulting estimates. We also connect these weights to the implied regression weights from Generalized Least Squares (GLS). Our overall approach provides a general framework that can be used by other applied researchers who wish to use balancing weights to estimate causal effects when their data are measured with error and/or the model errors are dependent.\footnote{Our approach also relates to the ``synthetic controls'' literature (see, e.g., \cite{abadie2010synthetic}). Synthetic controls are a popular balancing weights approach frequently used in the applied economics literature to estimate treatment effects on the treated (ETT) for region-level policy changes when using time series cross sectional data. Our application uses a similar data structure; however, we instead consider the problem of estimating the ETC. In contrast to much of the synthetic controls literature, which assumes that the counterfactual outcomes follow a linear factor model, we instead assume that the counterfactual outcomes are linear in the observed covariates (including pre-treatment outcomes).}

Section 2 begins with a more detailed overview of the policy problem, and then defines the study period, covariates, outcome, and treatment. Section 3 discusses our methods, including outlining our identification, estimation, and inferential procedures. Section 4 presents our results. Section 5 contains a discussion of our findings, and Section 6 contains a brief summary. The Appendices contain proofs, summary statistics, and additional results.

\section{Policy Problem and Data}

\subsection{Policy Problem Statement}

Under the Affordable Care Act (ACA), states were required to expand their Medicaid eligibility requirements by 2014 to offer coverage to all adults with incomes at or below 138 percent of the federal poverty level (FPL). The United States Supreme Court ruled this requirement unconstitutional in 2012, allowing states to decide whether to expand Medicaid coverage. In 2014, twenty-six states and the District of Columbia expanded their Medicaid programs. From 2015 through 2021, an additional twelve states elected to expand their Medicaid programs. Medicaid expansion has remained a debate among the remaining twelve states that have not expanded Medicaid as of 2021.\footnote{https://www.nbcnews.com/politics/politics-news/changed-hearts-minds-biden-s-funding-offer-shifts-medicaid-expansion-n1262229} The effects of Medicaid expansion on various outcomes, including uninsurance rates, mortality rates, and emergency department use, have been widely studied, primarily by using the initial expansions in 2014 and 2015 to define expansion states as ``treated'' states and non-expansion states as ``control'' states (see, e.g., \cite{courtemanche2017early}, \cite{miller2021medicaid}, \cite{ladhania2021effect}).

Medicaid enrollment is not automatic, and take-up rates have historically varied across states. This variation is partly a function of state discretion in administering programs: for example, program outreach, citizenship verification policies, and application processes differ across states (\cite{courtemanche2017early}). Estimating how Medicaid eligibility expansion actually affects the number of uninsured individuals is therefore not obvious. This is also important because many effects are mediated largely through reducing the number of uninsured individuals. Existing studies have estimated that Medicaid expansion reduced the uninsurance rate between three and six percentage points on average among states that expanded Medicaid. These estimates differed depending on the data used, specific target population, study design, and level of analysis (see, e.g., \cite{kaestner2017effects}, \cite{courtemanche2017early}, \cite{frean2017premium}). However, none of these studies have directly estimated the average treatment effect on the controls (ETC). 

We believe that the ETC may differ from the ETT. Every state had different coverage policies prior to 2014, and non-expansion states tended to have less generous policies than expansion states. ``Medicaid expansion'' therefore represents a set of treatments of varying intensities that are distributed unevenly across expansion and non-expansion states. Averaged over the non-expansion states, which tended to have less generous policies and higher uninsurance rates prior to Medicaid expansion, we might expect the average effect to be larger in absolute magnitude than among the expansion states, where ``Medicaid expansion'' on average reflected smaller policy changes.\footnote{As a part of our analysis strategy, we limit our pool of expansion states to those where the policy changes were comparable to the non-expansion states. We also control for pre-treatment uninsurance rates (see Section~\ref{sssec:txassign}).} Even limited to states with equivalent coverage policies prior to 2014 we still might expect the ETT to differ from the ETC. For example, all states that were entirely controlled by the Democratic Party at the executive and legislative levels expanded their Medicaid programs, while only states where the Republican Party controlled at least part of the state government failed to expand their programs. Prior to the 2014 Medicaid expansion, \cite{sommers2012understanding} found that conservative governance was associated with lower Medicaid take-up rates. This might reflect differences in program implementation, which could serve as effect modifiers for comparable policy changes.\footnote{Interestingly, \cite{sommers2012understanding} also find that the association between conservative governance and lower take-up rates prior to 2014 existed even after controlling for a variety of factors pertaining to state-level policy administration decisions. They conjecture that this may reflect cultural conservatism: people in conservative states are more likely to view enrollment in social welfare programs negatively, and therefore be less likely to enroll.} These factors may then attenuate the effects of Medicaid expansion averaged over non-expansion states relative to expansion states.

As a more general causal quantity, the ETC is also interesting in its own right: to the extent that the goal of studying Medicaid expansion is to understand the foregone benefits - or potential harms - of Medicaid expansion among non-expansion states, the ETC is the relevant quantity of interest. Authors have previously made claims about the ETC without directly estimating it. For example, \cite{miller2021medicaid} use their estimates of the ETT to predict that had non-expansion states expanded Medicaid, they would have seen 15,600 fewer deaths from 2014 through 2017. However, as these authors note, this estimate assumes that the ETT provides a comparable estimate to the ETC. From a policy analysis perspective, we recommend that researchers estimate the ETC directly when it answers a substantive question of interest. We therefore contribute to the literature on Medicaid expansion by directly estimating this quantity.

\subsection{Data Source and Study Period}\label{ssec:data}

Our primary data source is the annual household and person public use microdata files from the ACS from 2011 through 2014. The ACS is an annual cross-sectional survey of approximately three million individuals across the United States. The public use microdata files include information on individuals in geographic areas greater than 65,000 people. The smallest geographic unit contained in these data are public-use microdata areas (PUMAs), arbitrary boundaries that nest within states but not within counties or other more commonly used geographic units. One limitation of these data is a 2012 change in the PUMA boundaries, which do not overlap well with the previous boundaries. As a result, the smallest possible geographic areas that nest both PUMA coding systems are known as consistent PUMAs (CPUMAs). The United States contains 1,075 total CPUMAs, with states ranging from having one CPUMA (South Dakota, Montana, and Idaho) to 123 CPUMAs (New York). Our primary dataset contains 925 CPUMAs among 45 states (see also Section~\ref{sssec:txassign}). The average total number of sampled individuals per CPUMA across the four years is 1,001; the minimum number of people sampled was 334 and the maximum is 23,990. We aggregate the microdata to the CPUMA level using the survey weights.  

This aggregation naturally raises concerns about measurement error and hierarchy. Any CPUMA-level variable is an estimate, leading to concerns about measurement error. The hierarchical nature of the dataset -- CPUMAs within states -- raises concerns about geographic dependence.

Our study period begins in 2011, following \cite{courtemanche2017early}, who note that several other aspects of the ACA were implemented in 2010 -- including the provision allowing for dependent coverage until age 26 and the elimination of co-payments for preventative care -- and likely induced differential shocks across states. We also restrict our post-treatment period to 2014. We therefore avoid additional assumptions required for identification given that several states expanded Medicaid in 2015, including Indiana, Pennsylvania, and Alaska.

\subsection{Treatment assignment} \label{sssec:txassign}

Reducing the concept of ``Medicaid expansion'' to a binary treatment simplifies a more complex reality. There are at least three reasons to be cautious about this simplification. First, states differed substantially in their Medicaid coverage policies prior to 2014. Given perfect data we might ideally consider Medicaid expansion as a continuous treatment with values proportional to the number of newly eligible individuals. The challenge is correctly identifying newly eligible individuals in the data (though see \cite{frean2017premium} and \cite{miller2021medicaid}, who attempt to address this). Second, \cite{frean2017premium} note that five states (California, Connecticut, Minnesota, New Jersey, and Washington) and the District of Columbia adopted partial limited Medicaid expansions prior to 2014. The ``2014 expansion'' therefore actually occurred in part prior to 2014 for several states.\footnote{\cite{kaestner2017effects} and \cite{courtemanche2017early} also consider Arizona, Colorado, Hawaii, Illinois, Iowa, Maryland, and Oregon to have had early expansions.} Finally, timing is an issue: among the states that expanded Medicaid in 2014, Michigan's expansion did not go into effect until April 2014, while New Hampshire's expansion did not occur until September 2014.

Our primary analysis excludes New York, Vermont, Massachusetts, Delaware, and the District of Columbia from our pool of expansion states. These states had comparable Medicaid coverage policies prior to 2014 and therefore reflect invalid comparisons (\cite{kaestner2017effects}). We also exclude New Hampshire because it did not expand Medicaid until September 2014. While Michigan expanded Medicaid in April 2014, we leave this state in our pool of ``treated'' states. We consider the remaining expansion states, including those with early expansions, as ``treated'' and the non-expansion states, including those that later expanded Medicaid, as ``control'' states. We later consider the sensitivity of our results to these classifications by removing the early expansion states indicated by \cite{frean2017premium}. Our final dataset contains data for 925 CPUMAs, with 414 CPUMAs among 24 non-expansion states and 511 CPUMAs among 21 expansion states.\footnote{We additionally include the 4 CPUMAs from New Hampshire in the covariate adjustment procedure described in Section~\ref{sec:methods}.} When we exclude the early expansion states, we are left with 292 CPUMAs across 16 expansion states. We provide a complete list of states by Medicaid expansion classification in Appendix~\ref{app:sumstats}.

\subsection{Outcome}

Our outcome is the non-elderly (individuals aged 19-64) adult uninsurance rate in 2014. While take-up among the Medicaid-eligible population is a more natural outcome, we choose the non-elderly adult uninsurance rate for two reasons, one theoretic and one practical. First, Medicaid eligibility post-expansion is likely endogenous: Medicaid expansion may affect an individual's income and poverty levels, which often define Medicaid eligibility. Second, we can better compare our results with the existing literature, including \cite{courtemanche2017early}, who also use this outcome. One drawback is that the simultaneous adoption of other ACA provisions by all states in 2014 also affects this outcome. As a result, we only attempt to estimate the effect of Medicaid expansion in 2014 in the context of this changing policy environment. We discuss this further in Sections~\ref{ssec:estimand} and ~\ref{ssec:identification}. 

\subsection{Covariates}

We choose our covariates to approximately align with those considered in \cite{courtemanche2017early} and that are likely to be potential confounders. Specifically, using the ACS public use microdata, we calculate the unemployment and uninsurance rates for each CPUMA from 2011 through 2013. We also estimate a variety of demographic characteristics averaged over this same time period, including percent female, white, married, Hispanic ethnicity, foreign-born, disabled, students, and citizens. We estimate the percent in discrete age categories, education attainment categories, income-to-poverty ratio categories, and categories of number of children. Finally, we calculate the average population growth and number of households to adults. We provide a more extensive description of our calculation of these variables in Appendix~\ref{app:adjustmentdetails}.

In addition to the ACS microdata we use 2010 Census data to estimate the percentage living in an urban area for each CPUMA. Lastly, we include three state-level covariates reflecting the partisan composition of each state's government in 2013. These include an indicator for states with a Republican governor, Republican control over the lower legislative chamber, and Republican control over both legislative chambers and the governorship.\footnote{Nebraska is the only state with a unicameral legislature and the legislature is technically non-partisan. We nevertheless classified them as having Republican control of the legislature for this analysis.} 

\section{Methods}\label{sec:methods}

In this section we present our causal estimand, identifying assumptions, estimation strategy, and inferential procedure. Our primary methodological contributions are contained in the subsection on estimation. We begin by outlining notation.

\subsection{Notation}
We let $s$ index states, and $c$ index CPUMAs within states. Let $m$ denote the number of states, $p_s$  the number of CPUMAs in state $s$, and $n = \sum_{s=1}^m p_s$ the total number of CPUMAs. For each state $s$, let $A_s$ denote its treatment assignment according to the discussion given in Section \ref{sssec:txassign}, with $A_s = 1$ indicating treatment and $A_s=0$ indicating control. For each CPUMA $c$ in state $s$, let $Y_{sc}$ denote its uninsurance rate in 2014; let $X_{sc}$ denote a q-dimensional covariate vector; and let $A_{sc} = A_{s}$ denote its treatment status. We assume potential outcomes (\cite{rubin2005causal}), defining a CPUMA's potential uninsurance under treatment by $Y^1_{sc}$, and under control by $Y^0_{sc}$. Finally, let $n_1$ and $n_0$ denote the number of treated and control CPUMAs, and define $m_1$ and $m_0$ analagously for states.

Given a collection of objects indexed over CPUMAs, we denote the complete set by removing the subscript. For example, $X$ denotes $\{X_{sc}: sc \in \mathcal{C}\}$, where $\mathcal{C}$ is the set of all CPUMAs. Subscripting by the labels $A=1$ or $A=0$ denotes the subset corresponding to the treated or control units; for example, $X_{A=0}$ denotes $\{X_{sc}: A_{sc}=0\}$, the covariates of the control units. To denote averaging, we will use an overbar, while also abbreviating $A=0$ and $A=1$ respectively by $0$ and $1$. For example, $\bar{X}_0$ (which abbreviates $\bar{X}_{A=0}$) denotes the average covariate vector for the control units, and $\bar{Y}_0^1$ denotes the average potential outcome under treatment for the control units.

\subsection{Estimand} \label{ssec:estimand}

% Dave's note: I removed the equality in the estimand definition since it only holds under assumptions that we haven't introduced yet
Letting $X$, $Y^1$, and $Y^0$ be random, we define the causal estimand $\psi_0$

\begin{align} \label{eqn:psi}
    \psi_0&= \mathbb{E}[\bar{Y}_0^1 - \bar{Y}_0^0 \mid X_{A=0}] \\ 
    &= \psi_0^1 - \psi_0^0
\end{align}

where $\psi_0^a$ denotes the expectation $\mathbb{E}[\bar{Y}_0^a \mid X_{A=0}]$. The estimand $\psi_0$ represents the expected treatment effect on non-expansion states conditioning on $X_{A=0}$, the observed covariate values of the non-expansion states (see, e.g., \cite{imbens2004nonparametric}). The challenge is that we do not observe $Y^1_{A=0}$, the counterfactual outcomes for non-expansion CPUMAs had their states expanded their Medicaid programs, nor their average $\bar{Y}^1_0$. We therefore require causal assumptions to identify this counterfactual quantity using our observed data.\footnote{As noted previously, the 2014 Medicaid expansion occurred simultaneously with the implementation of several other major ACA provisions, including (but not limited to) the creation of the ACA-marketplace exchanges, the individual mandate, health insurance subsidies, and community-rating and guaranteed issue of insurance plans (\cite{courtemanche2017early}). Almost all states broadly implemented these reforms beginning January 2014. Conceptually we think of the other ACA components as a state-level treatment ($R$) separate from Medicaid expansion ($A$). Our total estimated effect may include interactions between these policy changes; however, we do not attempt to separately identify these effects. Without further assumptions, we therefore cannot generalize these results beyond 2014.} 

\subsection{Identification} \label{ssec:identification}

We appeal to the following causal assumptions to identify $\psi_0$ from our observed data: the stable unit treatment value assumption (SUTVA), no unmeasured confounding given the true covariates, and no anticipatory treatment effects. We also invoke parametric assumptions to model the measurement error and to express our estimand in terms of parameters from a linear model. We conclude by using ideas from the ``regression-calibration'' literature (see, e.g., \cite{gleser1992importance}) to ensure that identifying our target estimand is possible given auxiliary data on the measurement error covariance matrix.

We first assume the SUTVA at the CPUMA level. Assuming the SUTVA has two implications for our analysis: first, that there is only one version of treatment; second, that each unit's potential outcome only depends on its treatment assignment. We discussed potential violations of the first implication previously when considering how to reduce Medicaid expansion to a binary treatment. The second implication could be violated if one CPUMA's expansion decision affected uninsurance rates in another CPUMA (see, e.g., \cite{frean2017premium}). On the other hand, our assumption allows for interference among individuals living within CPUMAs and is therefore weaker than assuming no interference among any individuals. Further addressing this is beyond the scope of this paper.

Second, we assume no effects of treatment on the observed covariates. This includes assuming no anticipatory effects on pre-2014 uninsurance rates. This is violated in our study, as some treated states allowed specific counties to expand their Medicaid programs prior to 2014, thereby affecting their pre-2014 uninsurance rates. We later test the sensitivity of our results to the exclusion of these states.

Third, we assume no unmeasured confounding. Specifically, we posit that in 2014 the potential outcomes for each CPUMA are jointly independent of the state-level treatment assignment conditional on CPUMA and state-level covariates $X_{sc}$:

\begin{equation}\label{eqn:unconfoundedness}
(Y_{sc}^1, Y_{sc}^0) \perp A_s \mid X_{sc} 
\end{equation}

The covariate vector $X_{sc}$ includes both time-varying pre-treatment covariates, including pre-treatment outcomes, and covariates averaged across 2011-2013, such as demographic characteristics, and the state-level governance indicators discussed in Section~\ref{ssec:data}. We believe this assumption is reasonable given our rich covariate set. 

Fourth, we assume that the outcomes for each treatment group are linear in the true covariates:
\begin{equation}\label{eqn:linmod}
Y_{sc}^a = \alpha_a + X_{sc}^\top\beta_a + \epsilon_{sc} + \varepsilon_s \qquad a = 0, 1
\end{equation}
where $^\top$ denotes vector transpose, and the errors $\epsilon_{sc}$ and $\varepsilon_{s}$ are mean-zero; independent from the covariates, treatment assignment, and each other; and have finite variances $\sigma^2_{\epsilon}$ and $\sigma^2_{\varepsilon}$, respectively.\footnote{Because our covariates include pre-treatment outcomes, this assumption also implies that $\epsilon_{sc}$ and $\varepsilon_{sc}$ are uncorrelated with pre-treatment outcomes, including any error terms that might appear in their generative models.} This implies that the errors for each CPUMA within a given state have a constant within-state correlation $\sigma^2_{\varepsilon}/(\sigma^2_{\varepsilon} + \sigma^2_{\epsilon})$, which we denote as $\rho$. To fix ideas, $\epsilon_{sc}$ may capture time-specific idiosyncracies at the local level, possibly due to the local policy or economic conditions. By contrast $\varepsilon_s$ captures time-specific idiosyncracies at the state-level that are common across CPUMAs within a state due to the shared policy and economic environment.

Fifth, we assume that the covariates $X$ and outcomes $Y$ are not observed directly. Instead, survey sampled versions $W$ and $J$ are available, with additive gaussian measurement error arising due to sample variability.

\begin{align} \label{eqn:additivenoise}
	J_{sc} & = Y_{sc} + \xi_{sc} & \text{and} & & W_{sc} & = X_{sc} + \nu_{sc}
\end{align}
where $(\xi_{sc}, \nu_{sc})$ is independent of $(X, Y)$ and has distribution

\begin{equation} \label{eqn:gaussiannoise}
 (\xi_{sc}, \nu_{sc}) \stackrel{\text{indep}}{\sim} \operatorname{MVN}\left(0, \left[\begin{array}{cc} \sigma_{\xi,sc}^2 & 0 \\ 0 & \Sigma_{\nu, sc} \end{array}\right] \right)
\end{equation}

We believe equations (\ref{eqn:additivenoise}) and (\ref{eqn:gaussiannoise}) are reasonable because measurement error in our context is sampling variability. While \eqref{eqn:gaussiannoise} further implies the measurement errors in our covariates and outcomes are uncorrelated, this is also reasonable because our outcomes are measured on a different cross-sectional survey than our covariates.\footnote{Our covariates are almost all ratio estimates, which are in general biased. This bias, however, is $O(1/n)$ and therefore decreases quickly with the sample size; given that our sample sizes are all over 300, we treat these estimates as unbiased.} 

Sixth, we assume that the covariates for the treated units $X_{sc}$ are drawn i.i.d. multivariate normal conditional on treatment:

\begin{align} \label{eqn:Xgaussian}
    X_{sc}|A_{sc} = 1 & \stackrel{\text{iid}}{\sim} MVN(\upsilon_1, \Sigma_{X|1})%, \qquad \forall\, sc: A_{sc} = a,
\end{align}
Under equations (\ref{eqn:gaussiannoise})-(\ref{eqn:Xgaussian}), the conditional expectation of $X_{sc}$ given noisy observation $W$ among the treated units can be seen to equal 

\begin{equation} \label{eqn:regcal}
\mathbb{E}[X_{sc}| W, A] = \upsilon_1 + \Sigma_{X|1} \left(\Sigma_{X|1} + \Sigma_{\nu, sc}\right)^{-1}  (W_{sc} - \upsilon_1), \qquad \forall\, sc: A_{sc} = 1
\end{equation}
Equation (\ref{eqn:Xgaussian}) is a convenient simplification to motivate \eqref{eqn:regcal}. For example, $X_{sc}$ includes state-level covariates, so the covariates cannot be independent. More generally, many of the covariates are bounded, and therefore cannot be gaussian. In fact, assuming (\ref{eqn:Xgaussian}) is not strictly necessary for consistent estimation of $\psi_0^1$ (see, e.g., \cite{gleser1992importance}); however, it is required by the weighting approaches that we consider here. In our validation experiments described in Section~\ref{sec:validation}, we find that our approaches which assume (\ref{eqn:Xgaussian}) outperform those that do not, as the latter group evidently relies more heavily on the linearity assumption (\ref{eqn:linmod}).\footnote{In principle it may be possible to generalize equations (\ref{eqn:additivenoise})-(\ref{eqn:Xgaussian}) to settings where the conditional expectation $\mathbb{E}[X_{sc}|W,A]$ follows a different form than \eqref{eqn:regcal}, but is still accessible given auxiliary data. For example, to make the linearity assumption of equation (\ref{eqn:linmod}) more credible, $X_{sc}$ might include transformations or a basis expansion of the covariate, so that $X_{sc} = \phi(U_{sc})$ for some function $\phi$ of the untransformed covariates $U_{sc}$. Under assumptions analogous to (\ref{eqn:additivenoise})-(\ref{eqn:Xgaussian}) for $U_{sc}$, we may still be able to estimate $\mathbb{E}[X_{sc} \mid W, A]$. We give some preliminary findings in Appendix \ref{app:AsecI}, Remark \ref{remark:basis expansion}. Developing this idea further would be an interesting area for future work.}
 
Regardless, to use \eqref{eqn:regcal} to estimate $\mathbb{E}[X_{sc}|W_{sc}, A_{sc}=1]$, we require $\upsilon_1$, $\Sigma_{\nu,sc}$, and $\Sigma_{X|1}$. $\bar{W}_1$ provides a consistent estimate of $\upsilon_1$. However, the data does not identify $\Sigma_{\nu,sc}$ and $\Sigma_{X|1}$. Our final assumption is that we can consistently estimate the covariance matrices $\Sigma_{\nu,sc}$ and $\Sigma_{X|1}$ using auxiliary data, so that we can use (\ref{eqn:regcal}) to estimate the conditional mean. The ACS microdata serves as our auxiliary data; further details are discussed in Section~\ref{ssec:methodsmsrment}.

Under these assumptions we can rewrite our causal estimand in terms of the model parameters. As $\epsilon_{sc}$ and $\varepsilon_s$ are zero-mean and independent of covariates,  we can rewrite $\psi_0^a = \mathbb{E}[\bar{Y}_0^a \mid X_{A=0}]$ by applying expectations to (\ref{eqn:linmod}), yielding

\begin{equation}\label{eqn:outcome}
\psi_0^a = \alpha_a + \bar{X}_0^\top\beta_a
\end{equation}
If we observed $(A,Y,X)$, we would have $Y_{A=a}$ and $X_{A=a}$ for $a=\{0,1\}$. Therefore, by (\ref{eqn:linmod}) the data would identify $(\alpha_a, \beta_a)$, which identifies $\psi_0^a$ since $\bar{X}_0$ is observed. However, we only observe the noisy measurements $J$ and $W$. As $\xi_{sc}$ is zero-mean, equation (\ref{eqn:additivenoise}) implies that $\bar{J}_0$ estimates $\bar{Y}_0^0$ and therefore $\psi_0^0$. Estimating $\psi_0^1$ remains challenging: as $Y_{sc}^1 \perp A_{sc} \mid X_{sc} \centernot\implies J_{sc}^1 \perp A_{sc} \mid W_{sc}$, it is well-known that noisy measurements will bias standard estimation procedures, such as linear regression, that naively use them without adjustment (see also Appendix~\ref{app:AsecI}).

Let $\tilde{X}_{sc} = \mathbb{E}[X_{sc} |W, A]$ abbreviate the conditional expectation of the covariates given the noisy observations, as given by \eqref{eqn:regcal}. Substituting $X_{sc} = \tilde{X}_{sc} + X_{sc} - \tilde{X}_{sc}$ into the outcome model given by (\ref{eqn:linmod}) and then (\ref{eqn:linmod}) into (\ref{eqn:additivenoise}) yields

\begin{equation} \label{eqn:JXtilde}
    J_{sc} = \alpha_1 + \tilde{X}_{sc}^\top\beta_1 + (X_{sc} - \tilde{X}_{sc})^\top\beta_1 + \xi_{sc} + \epsilon_{sc} + \varepsilon_s \qquad\forall\, sc: A_{sc} = 1
\end{equation}
As $X_{sc} - \tilde{X}_{sc}$ equals $X_{sc} - \mathbb{E}[X_{sc}|W,A]$, this quantity is zero-mean conditioned on $(W,A)$. The term $(X_{sc} - \tilde{X}_{sc})^\top\beta_1$ appearing in (\ref{eqn:JXtilde}) is therefore also conditionally zero-mean. Finally, the outcome noise $\xi_{sc}$ as well as the noise terms $\epsilon_{sc}$ and $\varepsilon_{sc}$ appearing in this equation are also zero-mean and independent of $(W,A)$. It follows that if we observed $(\tilde{X}, J, A)$, we would have $\tilde{X}_{A=1}$ and $J_{A=1}$; therefore, by (\ref{eqn:JXtilde}) the data would identify $(\alpha_1, \beta_1)$. In turn, \eqref{eqn:outcome} implies that $\alpha_1$, $\beta_1$ and $\bar{X}_{A=0}$ (which can be estimated without bias from $W_{A=0}$) identifies $\psi_0^1$. Since have assumed that $\tilde{X}$ follows \eqref{eqn:regcal}, and that we have auxiliary data available to estimate this equation, we therefore have sufficient data to estimate $\psi_0$ under our models and assumptions. We now discuss estimation.

\subsection{Estimation}\label{ssec:estimation}

We propose to use approximate balancing weights to estimate $\psi_0^1$. We first review approximate balancing weights and the SBW objective proposed by \cite{zubizarreta2015stable}. These methods typically assume that the covariates are measured without error. We will show in Proposition~\ref{cl1} that under the classical-errors-in-variables model, the SBW estimate of $\psi_0^1$ has the same bias as the OLS estimate.

We first attempt to remove this bias by estimating \eqref{eqn:regcal}, leveraging the ACS microdata replicate survey weights to estimate this model. We consider two adjustments: (a) a homogeneous adjustment that assumes the noise covariance $\Sigma_{\nu, sc}$ is constant across all CPUMAs; and (b) a heterogeneous adjustment that allows $\Sigma_{\nu,sc}$ to vary according to the sample sizes associated with each CPUMA. We next propose a modification to SBW that we call H-SBW, which accounts for the state-level random effects $\varepsilon_s$. Using SBW and H-SBW we generate weights that balance the adjusted data to the mean covariate values of the non-expansion states. To further reduce imbalances that remain after weighting, we consider bias-corrections using ridge-regression augmentation, following \cite{ben2021augmented}. 

\subsubsection{Stable balancing weights}\label{ssec:SBW}

\cite{zubizarreta2015stable} proposes the Stable Balancing Weights (SBW) algorithm to generate a set of weights $\gamma$ that reweight a set of covariates $Z = \{Z_{sc}\}$ to a target covariate vector $\upsilon$ within a tolerance parameter $\delta$ by solving the optimization problem:

\begin{equation}\label{eqn:SBWobjective}
 \min_{\gamma} \sum_{sc} \gamma_{sc}^2 \quad \text{such that} \quad \gamma \in \Gamma(Z, \upsilon, \delta)
\end{equation}
where the constraint set $\Gamma(Z, \upsilon, \delta)$ is given by

\[ \Gamma(Z, \upsilon, \delta) = \left\{\gamma: \left|\sum \gamma_{sc} Z_{sc}  - \upsilon\right| \leq \delta,\, \gamma \geq 0,\, \sum_{sc} \gamma_{sc} = 1\right\}\]
and where $\delta$ may be a $q$-dimensional vector if non-uniform tolerances are desired. To estimate $\psi_0$ given the true covariates $X$ and outcomes $Y$, one can use SBW to reweight the treated units to approximately equal the mean covariate value of the control units by finding $\hat{\gamma}$ solving \eqref{eqn:SBWobjective} with $\upsilon_0 = \bar{X}_0$ and $Z = X_{A=1}$ for some feasible $\delta$. One can then use $\hat{\gamma}$ to estimate $\psi_0^1$ and $\psi_0$:

\begin{align}\label{eqn:estimators}
\hat{\psi}_0^1 &= \sum_{A_{sc}=1} \hat{\gamma}_{sc} Y_{sc}, & \hat{\psi}_0^0 & = \bar{Y}_0^0, & \hat{\psi} = \hat{\psi}_0^1 - \hat{\psi}_0^0
\end{align}
In the case where the potential outcomes follow the linear model specified in ~\eqref{eqn:linmod}, the bias of $\hat{\psi}^1_0$ is less than or equal to $\lvert\beta_1\rvert^\top\delta$, and therefore equal to zero if $\delta = 0$ \citep{zubizarreta2015stable}. Moreover, $\hat{\psi}_0^1$ produces the minimum variance estimator within the constraint set - conditional on $X$ - assuming that the errors in the outcome model are independent and identically distributed.

\subsubsection{Measurement error}\label{ssec:methodsmsrment} 

In the presence of measurement error the estimation procedure described in Section \ref{ssec:SBW} will be biased. We show in Appendix~\ref{app:AsecI}, Proposition \ref{cl1}, that under the classical errors-in-variables model where $\Sigma_{\nu,sc} = \Sigma_{\nu}$ for all units, if $\hat{\gamma}$ is found by solving the SBW objective (\ref{eqn:SBWobjective}) with $Z$ equal to the noisy covariates $W_{A=1}$, $\upsilon$ equal to the estimated mean $\bar{W}_0$ of the control units, and $\delta=0$, the estimator $\hat{\psi}_0^1$ in \eqref{eqn:estimators} has bias
\begin{align*}
\mathbb{E}[\hat{\psi}_0^1] - \psi_0^1 = (\bar{X}_0 - \upsilon_1)^\top(\kappa_1 - I_q)\beta_1 
\end{align*}
where $\kappa_1 = (\Sigma_{X|1} + \Sigma_{\nu})^{-1}\Sigma_{X|1}$. This is equivalent to the bias for an OLS estimator of $\psi_0^1$, where $(\alpha_1, \beta_1)$ are estimated by regression of $Y_{A=1}$ on $W_{A=1}$.

We mitigate this bias by setting $Z = \hat{X}_{A=1}$, where $\hat{X}_{A=1}$ is an estimate of $\tilde{X}_{A=1}$ given by \eqref{eqn:regcal}. This requires estimating $\upsilon_1$, $\Sigma_{\nu, sc}$ and $\Sigma_{X|1}$. To estimate $\upsilon_1$ we simply use $\bar{W}_1$. To estimate $\Sigma_{X|1}$ and $\Sigma_{\nu,sc}$ we use the ACS microdata's set of 80 replicate survey weights to construct 80 additional CPUMA-level datasets. For each CPUMA among the treated states, we take the empirical covariance matrix of its covariates over the datasets to derive unpooled etimates $\hat{\Sigma}_{\nu,sc}^{\text{raw}}$, which we average over CPUMAs to create $\hat{\Sigma}_{\nu}$. We then estimate $\Sigma_{X|1}$ by subtracting $\hat{\Sigma}_{\nu}$ from the empirical covariance matrix of $W_{A=1}$,
\[ \hat{\Sigma}_{X|1} = \frac{1}{n_1} \sum_{sc:A_{sc}=1} (W_{sc} - \bar{W}_1)(W_{sc} - \bar{W}_1)^\top - \hat{\Sigma}_{\nu}\]
We consider two estimates of $\Sigma_{\nu, sc}$: first, where we let $\hat{\Sigma}_{\nu,sc} = \hat{\Sigma}_{\nu}$ for all units, which we call the homogeneous adjustment; second, where each $\hat{\Sigma}_{\nu, sc}$ equals $\hat{\Sigma}_{\nu}$ rescaled according to the sample size of the estimate $W_{sc}$, which we call the heterogeneous adjustment. We describe these adjustments fully in Appendix~\ref{app:adjustmentdetails}. Using $\hat{\Sigma}_{X|1}$ and $\hat{\Sigma}_{\nu, sc}$, we estimate $\tilde{X}_{A=1}$ using the empirical version of \eqref{eqn:regcal}, inducing estimates $\hat{X}_{A=1}$ given by
\begin{equation}\label{eqn:hatX}
\hat{X}_{sc} = \bar{W}_1 + \hat{\Sigma}_{X|1} (\hat{\Sigma}_{X|1} + \hat{\Sigma}_{\nu,sc})^{-1}  (W_{sc} - \bar{W}_1), \qquad \forall\, sc: A_{sc}=1
\end{equation}
We then compute debiased balancing weights $\hat{\gamma}$ by solving \eqref{eqn:SBWobjective} with $Z = \hat{X}_{A=1}$, $\upsilon = \bar{W}_0$, and tuning parameter $\delta$ chosen as described in Section \ref{ssec:delta}. Given $\hat{\gamma}$, we find $\hat{\psi}_0^1$ and $\hat{\psi}$ again using \eqref{eqn:estimators}.

The homogeneous adjustment approximately aligns with the adjustments suggested by \cite{carroll2006measurement} and \cite{gleser1992importance}. In Appendix~\ref{app:AsecI}, Propositions \ref{cl2}-\ref{cl3}, we show that this procedure returns consistent estimates of $\psi_0^1$ and $\psi_0$ under the identifying assumptions discussed and assuming $\delta = 0$ is feasible. This is the first application we are aware of to apply regression calibration in the context of balancing weights to address measurement error. However, this method requires access to knowledge about $\Sigma_{\nu}$. We use survey microdata to identify this parameter for our application. Alternatively, region-level datasets often contain region-level variance estimates. If a researcher is willing to assume $\Sigma_{\nu}$ is diagonal, she could leverage this information to use this approach. If no auxiliary data is available, she could also consider $\Sigma_{\nu}$ to be a sensitivity parameter and conduct estimates over a range of possible values (see, e.g., \cite{huque2014impact}, \cite{illenberger2020impact}). 

\subsubsection{H-SBW criterion}\label{sssec:hsbw}

Unlike the setting outlined in \cite{zubizarreta2015stable}, our application likely has state-level dependencies in the error terms which may increase the variance of the SBW estimator. We therefore add a tuning parameter $\rho \in [0, 1)$ to penalize the within-state cross product of the weights, as detailed in ~\eqref{eqn:hsbwobjective}, representing a constant within-state correlation of the errors.

\begin{equation}\label{eqn:hsbwobjective}
\min_{\gamma} \quad \sum_{s=1}^{m}(\sum_{c = 1}^{p_s}\gamma_{sc}^2 + \sum_{c \ne d}\rho \gamma_{sc}\gamma_{sd}) \quad \text{such that} \quad \gamma \in \Gamma(Z, \upsilon, \delta)\\
\end{equation}
To build intuition about this objective, for $\delta \to \infty$, the following solution is attained:

\begin{equation}\label{eqn:sbwsol}
\hat{\gamma}_{sc} \propto \frac{1}{(p_s - 1)\rho + 1}
\end{equation}
Setting $\rho = 0$ returns the SBW solution: $\hat{\gamma}_{sc} \propto 1$. When setting $\rho \approx 1$, we get $\hat{\gamma}_{sc} \propto \frac{1}{p_s}$. In other words, as we increase $\rho$, this objective downweights CPUMAs in states with large numbers of CPUMAs and upweights CPUMAs in states with small numbers of CPUMAs (assigning each CPUMA within a state equal weight). As we increase $\rho$, the objective will therefore more uniformly disperse weights across states. We show in Appendix~\ref{app:AsecIII} that solving the H-SBW produces the minimum conditional variance estimator of $\psi_0^1$ within the constraint set assuming homoskedasticity and equicorrelated errors. We also highlight the connection between the H-SBW solution and the implied regression weights from GLS.

An important caveat emerges in the context of measurement error. In settings where the covariates are dependent, the conditional expectation $\mathbb{E}[X_{sc}|W,A]$ is no longer given by \eqref{eqn:regcal}, and therefore $\hat{X}_{sc}$ given by \eqref{eqn:hatX} is no longer consistent for $\tilde{X}_{sc}$. As a result, finding weights solving H-SBW \eqref{eqn:hsbwobjective} with $Z = \hat{X}_{A=1}$ is not unbiased in these settings. A simulation study in Appendix~\ref{app:simstudy} also shows that this bias increases with $\rho$, and that the SBW solution remains approximately unbiased. To regain unbiasedness in general, $\hat{X}_{sc}$ must be modified from \eqref{eqn:hatX} to account for dependencies, requiring new modeling assumptions. We demonstrate this more formally in Appendix~\ref{app:AsecIII} and propose an adjustment to account for dependent gaussian covariates in Appendix~\ref{app:adjustmentdetails}.

\subsubsection{Hyperparameter selection} \label{ssec:delta}

Practical guidance in the literature is that $\delta$ should reduce the standardized mean differences to be less than 0.1 (see, e.g., \cite{zhang2019balance}). In our application, all of our covariates are measured on the same scale. Additionally, because some of these covariates have very small variances (for example, percent female), we instead target the percentage point differences. We can then estimate $\psi_0$ using ~\eqref{eqn:estimators}, substituting $J_{sc}$ for $Y_{sc}$ and using the weights $\hat{\gamma}$.

We choose the vector $\delta$ using domain knowledge about which covariates are most likely to be important predictors of the potential outcomes under treatment, again recalling that the bias of our estimate is bounded by $\sum_{d=1}^q \delta_d \lvert \beta_{1,d} \rvert$ (where we index the covariates from $d = 1, ..., q)$. Specifically, we know that pre-treatment outcomes are often strong predictors of post-treatment outcomes, so we constrain the imbalances to fall within 0.05 percentage points (out of 100) for pre-treatment outcomes.\footnote{While both our approach and the synthetic controls literature prioritizes balancing pre-treatment outcomes, the motivation is somewhat different. The synthetic controls literature frequently assumes that the potential outcomes absent treatment follow a linear factor model. Under this model a heuristic is that the effects of covariates must show up via the pre-treatment outcomes, so that balancing pre-treatment outcomes is sufficient to balance any other relevant covariates (see \cite{botosaru2019role}, who formalize this idea). However, both the theory and common practice recommends balancing a large vector of pre-treatment outcomes. We instead have very limited pre-treatment data. We therefore do not assume a factor model, and instead treat the pre-treatment outcomes simply as covariates. We then justify prioritizing balancing these pre-treatment outcomes assuming they have large coefficients relative to other covariates in the assumed model for the (post-treatment) potential outcomes under treatment.} Because health insurance is often tied to employment, we also prioritize balancing pre-treatment uninsurance rates, seeking to reduce imbalances below 0.15 percentage points. On the opposite side of the spectrum, we constrain the Republican governance indicators to fall within 25 percentage points. While we believe that these covariates are important to balance, given the data we are unable to reduce the constraints further without generating extreme weights. We detail the remaining constraints in Appendix~\ref{app:weightdiagnostics}.\footnote{Our key estimand - $\psi_0^1$ - differs from the traditional target of the synthetic controls literature - $\psi^0_1$. If there exist covariates that are not relevant for producing $Y^0$ but are relevant for $Y^1$, then balancing pre-treatment outcomes alone will not in general balance these covariates (\cite{botosaru2019role}). Therefore, even with access to a large vector of pre-treatment outcomes and assuming that $Y^1$ follows a factor model, explicitly balancing such covariates in addition to pre-treatment outcomes may be necessary to estimate the ETC well.}

We consider $\rho \in \{0, 1/6\}$. The first choice is equivalent to the SBW objective, while the second assumes constant equicorrelation of $1/6$. We choose $\rho$ to be small to limit additional bias induced by H-SBW in the context of dependent data and measurement error.

Data-driven approaches to select these parameters could also be used. For example, absent measurement error if pre-treatment outcomes and covariates were available one could use the residuals from GLS to estimate $\rho$. Data-driven procedures for $\delta$ are also possible. \cite{wang2020minimal} propose a data-driven approach that only uses the covariate information. When data exists for a long pre-treatment period, \cite{abadie2015comparative} propose tuning their weights with respect to covariate balance using a ``training'' and ``validation'' period, an idea that could be adapted to choose $\delta$. Expanding these ideas to this setting would be an interesting area for future work.

\subsection{Sensitivity to covariate imbalance}

Our initial set of weights allow for some large covariate imbalances. We follow the proposal of \cite{ben2021augmented} and use ridge-regression augmentation to reduce these imbalances. While these weights achieve better covariate balance, this comes at the cost of extrapolating beyond the support of the data. Letting $\hat{\gamma}^{H-SBW}$ be our H-SBW weights, and $\hat{X}_1$ denote the matrix whose columns are the members of $\hat{X}_{A=1}$, we define these weights as:

\begin{equation}
\hat{\gamma}^{BC-HSBW} = \hat{\gamma}^{H-SBW} + (\hat{X}_1\hat{\gamma}^{H-SBW} - \bar{W}_0)^\top(\hat{X}_1\Omega^{-1}\hat{X}_1^\top + \lambda I_q)^{-1}\hat{X}_1\Omega^{-1}
\end{equation}
where $\Omega$ is a block diagonal matrix with diagonal entries equal to one and the within-group off diagonals equal to $\rho$. We choose $\lambda$ so that all imbalances fall within 0.5 percentage points. We refer to \cite{ben2021augmented} for more details about this procedure. For our results we consider estimators using SBW weights ($\rho = 0$), H-SBW weights ($\rho = 1/6$), and their ridge-augmented versions that we respectively call BC-SBW and BC-HSBW.

\subsection{Model validation}

To check model validity, we rerun our procedures on pre-treatment data to compare the performance of our models for a fixed $\delta$. In particular, we train our model on 2009-2011 data to predict 2012 outcomes, and 2010-2012 data to predict 2013 outcomes. We limit to one-year prediction error since our estimand is only one-year forward. We examine the performance of SBW against H-SBW and their bias-corrected versions BC-SBW and BC-HSBW, using the covariate adjustment methods described in Section \ref{ssec:methodsmsrment} to account for measurement error. In Appendix~\ref{app:allresults}, we additionally compare to ``Oaxaca-Blinder'' OLS and GLS weights (see, e.g, \cite{kline2011oaxaca}). The OLS weights do not require the gaussian assumption of (\ref{eqn:Xgaussian}) for consistency, but rely more heavily on the linear model (\ref{eqn:linmod}).

\subsection{Inference}

We use the leave-one-state-out jackknife to estimate the variance of $\hat{\psi}_0^1$ (see, e.g., \cite{cameron2015practitioner}).\footnote{The jackknife approximates the bootstrap, which is sometimes used to estimate the variance of the OLS-based estimates using regression-calibration in the standard setting with i.i.d. data (\cite{carroll2006measurement}).} Specifically, we take the pool of treated states and generate a list of datasets that exclude each state. For each dataset in this list we calculate the weights and the leave-one-state-out estimate of $\psi_0^1$. Throughout all iterations we hold our targeted mean fixed at $\bar{W}_0$.\footnote{That is, we treat $\bar{W}_0$ as identical to $\bar{X}_0$ and ignore the variability of the estimate. This variability is of smaller order than the variability in $\hat{\psi}_0^1$: the former does not depend on the number of states but instead decreases with the number of CPUMAs among the control states and the sample sizes used to estimate each CPUMA-level covariate. Recall that $\bar{X}_0$ is fixed because our estimand is conditional on the observed covariate values of the non-expansion states.} When generating these estimates, if our preferred initial choice of $\delta$ does not converge, we gradually reduce the constraints (increase $\delta$) until we can obtain a solution. For each dataset we also re-estimate $\hat{X}_{A=1}$ before estimating the weights to account for the variability in the covariate adjustment procedure. We then estimate the variance:

\begin{equation}\label{eqn:jackknife}
 \hat{Var}(\hat{\psi}_0^1) = \frac{m_1 - 1}{m_1} \sum_{s:A_s = 1} \left( S_{(s)} - S_{(\cdot)} \right)^2
\end{equation}
where $S_{(s)}$ is the estimator of $\psi_0^1$ with treated state $s$ removed, and $S_{(\cdot)} = \frac{1}{m_1} \sum_{s:A_s=1} S_{(s)}$. In other settings the jackknife has been shown to be a conservative approximation of the bootstrap, such as in \cite{efron1981jackknife}, which we apply in Appendix \ref{app:AsecI}, Proposition \ref{prop:jackknife} to give a partial result for our application. In a simulation study mirroring our setting with $m_1 = 25$ units (available in Appendix~\ref{app:simstudy}), we obtain close to nominal coverage rates using these variance estimates.\footnote{When more substantial undercoverage occurs it is likely due to bias.}

To estimate the variance of $\hat{\psi}_0^0$ we run an auxiliary regression model on the non-expansion states and estimate the variance of the linear combination $\bar{W}_0^\top\hat{\beta}_0$ using cluster-robust standard errors. We do not need to adjust the non-expansion state data to estimate this quantity: a linear regression line always contains the point $(\bar{W}_0, \bar{J}_0)$, which are unbiased estimates of $(\bar{X}_0, \psi_0^0)$. Therefore, $\mathbb{E}\{\bar{W}_0^\top\hat{\beta}_0 \mid X\} = \psi_0^0$. Our final variance estimate $\hat{Var}(\hat{\psi})$ is the sum of $\hat{Var}(\hat{\psi}_0^1)$ and $\hat{Var}(\hat{\psi}_0^0)$.\footnote{The latter is much smaller than the former -- specifically, we estimate $\hat{Var}(\hat{\psi}_0^0) = 0.017$.} We use the t-distribution with $m_1 - 1$ degrees of freedom to generate 95 percent confidence intervals.

\section{Results}\label{sec:results}

We first present summary statistics regarding the variability of the pre-treatment outcomes on our adjusted and unadjusted datasets. The second sub-section contains covariate balance diagnostics. The third sub-section contains a validation study, and the final sub-section contains our ETC estimates.

\subsection{Covariate adjustment}

Table~\ref{tab:adjust1} displays the sample variance of our pre-treatment outcomes among the expansion states using each covariate adjustment strategy (see Section~\ref{ssec:methodsmsrment} for definitions of these adjustments). Although we most heavily prioritize balancing these covariates, they are also among the least precisely estimated, as most of our other covariates average over multiple years of data. Both the homogeneous and heterogeneous adjustments reduce the variability in the data by comparable amounts. Intuitively, these adjustment reduce the likelihood that our balancing weights will fit to noise in the covariate measurements. These results are consistent across most of our other covariates. Tables containing distributional information for each covariate are available in Appendix~\ref{app:sumstats}. 

\begin{table}[ht]
\caption{Pre-treatment outcome sample variance by adjustment strategy (primary dataset)}\label{tab:adjust1}
\begin{tabular}{lrrr}
  \hline
Variable & Unadjusted & Heterogeneous & Homogeneous \\ 
  \hline
Uninsured Pct 2011 & 8.35 & 8.04 & 8.05 \\ 
  Uninsured Pct 2012 & 8.20 & 7.89 & 7.90 \\ 
  Uninsured Pct 2013 & 8.09 & 7.78 & 7.79 \\ 
   \hline
\end{tabular}
\end{table}

\subsection{Covariate balance}

\begin{figure}[H]
\begin{center}
    \caption{Balance plot, homogeneous adjustment (primary dataset)}\label{fig:loveplotc1a}
    \includegraphics[scale=0.45]{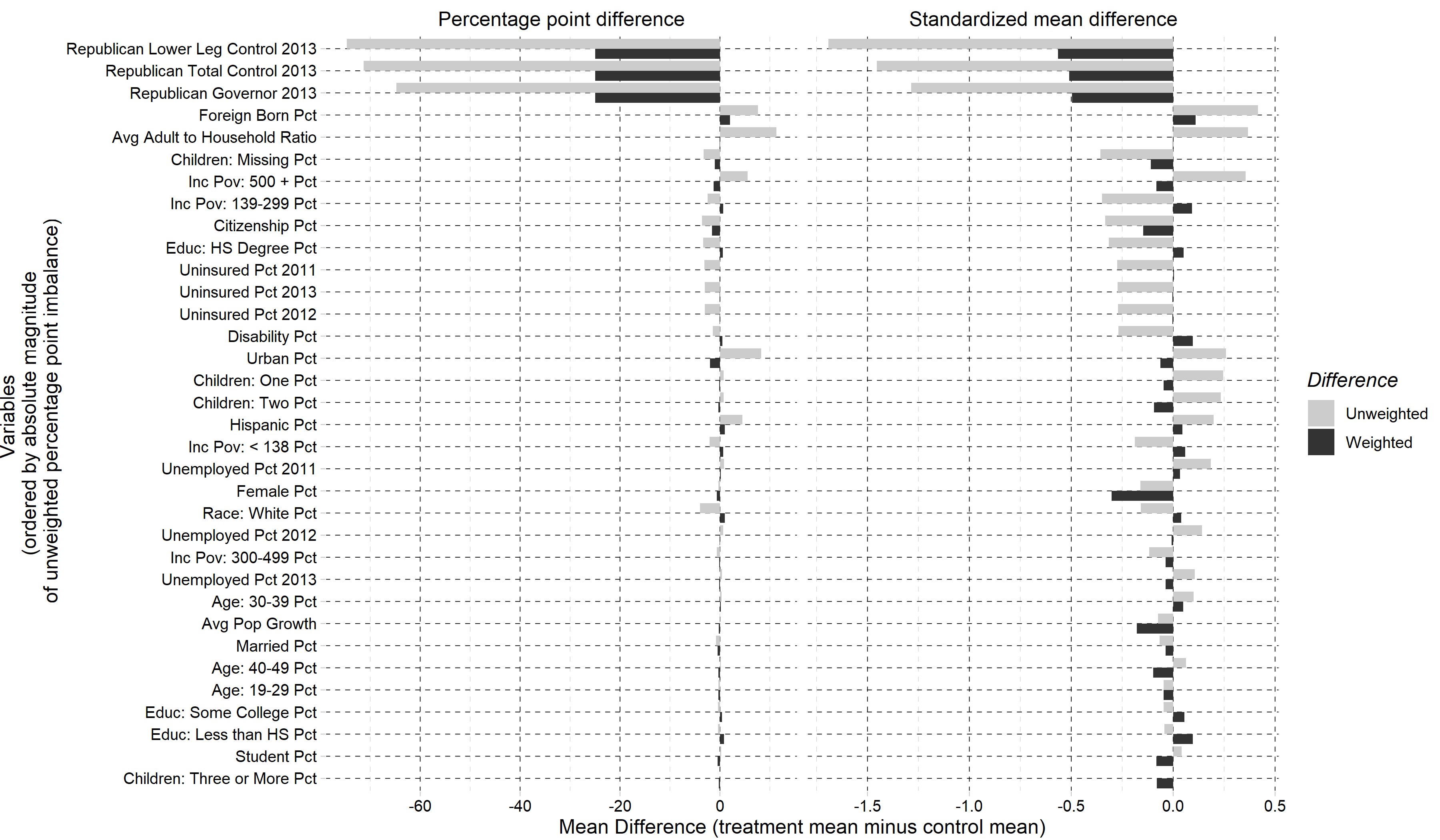}
\end{center}
\end{figure}

Figure~\ref{fig:loveplotc1a} displays the weighted and unweighted imbalances in our adjusted covariate set (using the homogeneous adjustment) using the H-SBW weights. Our unweighted data shows substantial imbalances in the Republican governance indicators as well as pre-treatment uninsurance rates. H-SBW reduces these differences; however, some remain, particularly among the Republican governance indicators. All other imbalances are relatively small, both on the absolute difference and standardized mean difference scale. In fact, despite not targeting the SMD, all but five remaining covariate imbalances fall within 0.1 SMD, compared to 23 covariates prior to reweighting. The largest remaining imbalance on the SMD scale is among percent female (-0.3), though, as noted previously, this variable has low variance in our dataset. A complete balance table is in Appendix~\ref{app:weightdiagnostics}. 

\begin{figure}[H]
\begin{center}
    \caption{Weight comparison, weights summed by state (primary dataset)}
    \label{fig:sbwvhsbw1}
    \includegraphics[scale=0.55]{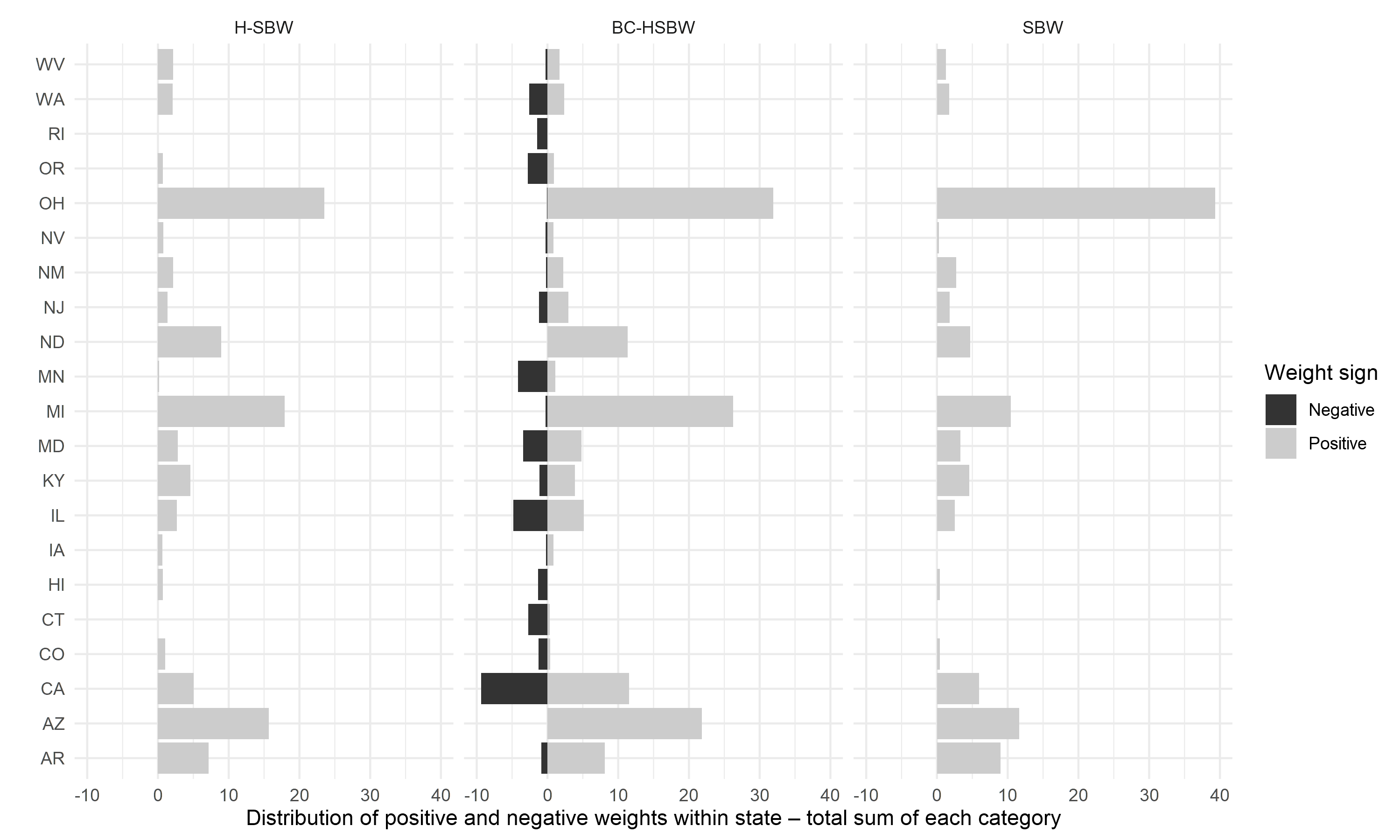}
\end{center}
\end{figure}

To further reduce these imbalances we augment these weights using ridge-regression. Figure~\ref{fig:sbwvhsbw1} shows the total weights summed across states (and standardized to sum to 100) for three estimators: H-SBW, BC-HSBW, and SBW. For BC-HSBW we also display the sum of the negative weights separately from the sum of the positive weights to highlight the extrapolation. This figure illustrates two key points: first, that H-SBW more evenly disperses the weights across states relative to SBW; second, that BC-HSBW extrapolates somewhat heavily in order to achieve balance, particularly from CPUMAs in California. This is likely in part because California has the most CPUMAs of any state in the dataset.

\subsection{Model validation}\label{sec:validation}

We compare the performance of our models by repeating the covariate adjustments and calculating our procedure on 2009-2011 ACS data to predict 2012 outcomes, and similarly for 2010-2012 data to predict 2013 outcomes for the non-expansion states. We run this procedure using the primary dataset and excluding the early expansion states. Table~\ref{tab:pretxpred} displays the mean error and RMSE of these results, with the rows ordered by RMSE for each dataset. We see that the estimators trained on the covariate adjusted data perform substantially better than the unadjusted data and the estimators trained on the homogeneous adjustment outperform their counterparts on the heterogeneous adjustment. We therefore prioritize the results using the homogeneous adjustment. We also observe that H-SBW has comparable performance to SBW throughout.

Interestingly, the bias-corrected estimators perform relatively poorly on the primary dataset, but are the best performing estimators when excluding early expansion states -- with RMSEs lower than any other estimator on the primary dataset. As the early expansion states include California, and Figure~\ref{fig:sbwvhsbw1} shows that the bias-corrected estimators extrapolate heavily from this state, the differences in these results suggest that preventing this extrapolation may improve the model's performance. While these results do not imply that the bias-corrected models will perform poorly when predicting $\psi^1_0$ in the post-treatment period on the primary dataset (or that they will perform especially well when excluding the early expansion states), it does highlight the dangers of extrapolation: linearity may approximately hold on the support of the data where we have sufficient covariate overlap, but beyond this region this may be a more costly assumption.\footnote{In Table~\ref{tab:pretxpredfull} in Appendix~\ref{app:allresults}, we also compare the performance against the implied regression weights from OLS and GLS. These weights exactly balance the observed covariates, but are almost always the worst performing estimators in the validation study on either dataset. This also illustrates the benefits of the regularization inherent in the bias-corrected weights.}

\begin{table}\caption{Estimator pre-treatment outcome mean prediction error and RMSE (in \% pts)}\label{tab:pretxpred}
\centering
\begin{tabular}{llrrllrr}
\hline
\multicolumn{4}{c}{Primary dataset} & \multicolumn{4}{c}{Early expansion 
 excluded} \\ 
 \hline
% \cmidrule(lr){1-4} \cmidrule(lr){5-8}
Sigma estimate & Estimator & Mean Error & RMSE & Sigma estimate & Estimator & Mean Error & RMSE \\ 
\hline
Homogeneous & SBW & -0.20 & 0.20 & Homogeneous & BC-HSBW & -0.02 & 0.07 \\ 
Homogeneous & H-SBW & -0.23 & 0.23 & Homogeneous & BC-SBW & -0.03 & 0.12 \\ 
Heterogeneous & SBW & -0.27 & 0.27 & Heterogeneous & BC-HSBW & -0.08 & 0.14 \\ 
Heterogeneous & H-SBW & -0.35 & 0.36 & Heterogeneous & BC-SBW & -0.07 & 0.15 \\ 
Homogeneous & BC-SBW & -0.39 & 0.39 & Homogeneous & H-SBW & 0.01 & 0.25 \\ 
Heterogeneous & BC-SBW & -0.42 & 0.42 & Homogeneous & SBW & 0.07 & 0.26 \\ 
Unadjusted & SBW & -0.56 & 0.56 & Heterogeneous & SBW & 0.04 & 0.28 \\ 
Unadjusted & H-SBW & -0.57 & 0.57 & Heterogeneous & H-SBW & -0.04 & 0.29 \\ 
Homogeneous & BC-HSBW & -0.58 & 0.58 & Unadjusted & SBW & -0.37 & 0.42 \\ 
Heterogeneous & BC-HSBW & -0.63 & 0.63 & Unadjusted & H-SBW & -0.43 & 0.46 \\ 
Unadjusted & BC-SBW & -0.88 & 0.88 & Unadjusted & BC-HSBW & -0.60 & 0.60 \\ 
Unadjusted & BC-HSBW & -0.96 & 0.96 & Unadjusted & BC-SBW & -0.70 & 0.71 \\ 
 \hline
\end{tabular}
\end{table}

Finally, we observe that the mean errors for the estimators generated on the unadjusted dataset are all negative, reflecting that these estimates under-predict the true uninsurance rate (see also Table~\ref{tab:pretxpredfull} in Appendix~\ref{app:allresults}, which shows similar results for each year individually). These results likely reflect a form of regression-to-the-mean caused by overfitting our weights to noisy covariate measurements. More formally, we can think of the uninsurance rates in time period $t$ in expansion and non-expansion regions as being drawn from separate distributions with means $(\upsilon_1, \upsilon_0)$, respectively, where $\upsilon_1 < \upsilon_0$. For simplicity assume that the only covariate in the outcome model at time $t$ is $Y_{sct-1}$. Under \eqref{eqn:linmod}, we obtain $Y_{sc} = Y_{sct} = \alpha_1 + \beta_1Y_{sct-1} + \epsilon_{sct} + \varepsilon_{st}$. The pre-treatment outcomes are likely positively correlated with the post-treatment outcomes, implying that $\beta_1 > 0$. Because $\upsilon_1 < \upsilon_0$, when reweighting the vector of noisy pre-treatment outcomes $J_{A=1}$ to $\upsilon_0$ the expected value of the weighted measurement error $\mathbb{E}[\sum_{A_{sc} = 1}\hat{\gamma}_{sc}\nu_{sct-1}]$ should be positive. In other words, our weights are likely to favor units with covariate measurements that are greater than their true covariate values. The expected value of the weighted pre-treament outcome will then be less than the target $\upsilon_0$. This implies that our estimates will be negatively biased, since $\mathbb{E}[\beta_1(\sum_{A_{sc} = 1}\hat{\gamma}_{sc}Y_{sct-1} - \upsilon_0)] \le 0$.\footnote{This phenomenon has also been discussed in the difference-in-differences and synthetic controls literature (see, e.g., \cite{daw2018matching}).} While our covariate adjustments are meant to eliminate this bias, in practice they appear more likely only to reduce it. Assuming these errors reflect a (slight) negative bias that will also hold for our estimates of $\psi^1_0$, these results suggest that the true treatment effect may be closer to zero than our estimates. 

\subsection{Primary Results}

Table~\ref{tab:mainresults} presents all of our estimates of the ETC. Using H-SBW we estimate an effect of -2.33 (-3.54, -1.11) percentage points on our primary dataset. The SBW results are almost identical with -2.35 (-3.76, -0.95) percentage points. Compared to the unadjusted data we find very similar point estimates at -2.34 (-2.88, -1.79) percentage points for H-SBW and -2.39 (-2.99, -1.79) percentage points for SBW. We see that H-SBW reduces the confidence interval length relative to SBW on our primary dataset, though the lengths are nearly identical when excluding early expansion states. This suggests that H-SBW had at best only modest variance improvements relative to SBW in this setting. Using the adjusted covariate set also increases the length of the confidence intervals relative to the unadjusted data. This increase in the estimated variance is expected in part because the adjustment procedure generally reduces the variability in the data, as we saw in Table~\ref{tab:adjust1}, requiring that the variance of the weights to increase to achieve approximate balance. More generally this increase also reflects the additional uncertainty due to the measurement error.

Adding the bias-correction decreases the absolute magnitude of the estimates: we estimate effects of -2.05 (-3.30, -0.80) percentage points for BC-HSBW and -2.07 (-3.14, -1.00) percentage points for BC-SBW. This contrasts to our validation study, where the bias-corrected estimators tended to predict lower uninsurance rates than the other estimators (implying we might see larger absolute magnitude effect estimates).

\begin{table}[ht]\caption{ETC estimates by weight type and adjustment strategy}\label{tab:mainresults}
\begin{tabular}{lllrlr} 
\hline
 &  & \multicolumn{2}{c}{Primary dataset} & \multicolumn{2}{c}{Early expansion 
 excluded} \\
  \hline
Weights & Adjustment & Estimate  & Difference & Estimate & Difference\\ 
 &  & (95\% CI) &  & (95\% CI) & \\
  \hline
H-SBW & Homogeneous & -2.33 (-3.54, -1.11) & 0.01 & -2.09 (-3.24, -0.94) & 0.19 \\ 
  H-SBW & Unadjusted & -2.34 (-2.88, -1.79) & - & -2.28 (-2.87, -1.70) & - \\ 
  BC-HSBW & Homogeneous & -2.05 (-3.30, -0.80) & 0.17 & -1.94 (-3.27, -0.61) & 0.28 \\ 
  BC-HSBW & Unadjusted & -2.22 (-2.91, -1.52) & - & -2.22 (-3.14, -1.31) & - \\ 
  SBW & Homogeneous & -2.35 (-3.76, -0.95) & 0.04 & -2.05 (-3.19, -0.91) & 0.16 \\ 
  SBW & Unadjusted & -2.39 (-2.99, -1.79) & - & -2.21 (-2.75, -1.68) & - \\ 
  BC-SBW & Homogeneous & -2.07 (-3.14, -1.00) & 0.13 & -1.99 (-3.33, -0.66) & 0.23 \\ 
  BC-SBW & Unadjusted & -2.19 (-2.94, -1.45) & - & -2.23 (-3.12, -1.33) & - \\    
  \hline
\end{tabular}
\subcaption{``Difference'' column reflects difference between adjusted and unadjusted estimators}
\end{table}

All adjusted estimators were closer to zero than the corresponding unadjusted estimators. This includes estimates using the heterogeneous adjustment (see Appendix~\ref{app:allresults}). However, the difference between the adjusted SBW and H-SBW and the unadjusted versions is close to zero. This contrasts with our validation study where the unadjusted SBW and H-SBW estimators ranged from about 0.3 to 0.4 percentage points lower than the adjusted estimators.\footnote{When excluding the early expansion states, the difference between the estimates on the adjusted and unadjusted data persist but are also smaller in magnitude.} We interpret this difference as due to chance: while theory and our validation study suggests that our unadjusted estimators are biased, bias only exists in expectation. Our primary dataset is a random draw where our unadjusted and adjusted estimators give similar estimates.

We next consider the sensitivity of our analysis with respect to the no anticipatory treatment effects assumption by excluding the early expansion states (California, Connecticut, Minnesota, New Jersey, and Washington) and re-running our analyses. The columns under ``Early expansion excluded'' in Table~\ref{tab:mainresults} reflects these results. The overall patterns of the results are consistent with our primary estimates; however, our point estimates almost all move somewhat closer to zero. This may indicate that either the primary estimates have a slight negative bias, or that these estimates have a slight positive bias. Given our analysis of the validation study we view the first case as more likely. This would imply that our primary estimators reflect a lower bound on the true treatment effect. Regardless, the differences are small relative to our uncertainty estimates. Overall we view our primary results as relatively robust to the exclusion of these states. Additional results are available in Appendix~\ref{app:allresults}.

\section{Discussion}

We divide our discussion into two sections: methodological considerations and policy considerations. 

\subsection{Methodological considerations and limitations}

We make multiple contributions to the literature on balancing weights. First, our estimation procedure accounts for mean-zero random noise in our covariates that is uncorrelated with the outcome errors. We modify the SBW constraint set to balance on a linear approximation to the true covariate values, applying the idea of regression calibration from the measurement error literature to the context of balancing weights. Our results illustrate the benefits of this procedure: using observed pre-treatment outcomes generated by an unknown data generating mechanism, Table~\ref{tab:pretxpred} demonstrates that our proposed estimators have better predictive performance when balancing on the adjusted covariates. This finding is consistent with concerns about overfitting to noisy covariate measurements and subsequent regression-to-the-mean.

This approach has several limitations: first, it requires access to auxiliary data with which to estimate the measurement error covariance matrix $\Sigma_{\nu}$. Many applications may not have access to such information. Even without such data, $\Sigma_{\nu}$ could also be considered a sensitivity parameter to evaluate the robustness of results to measurement error (see, e.g., \cite{huque2014impact}, \cite{illenberger2020impact}). Second, from a theoretic perspective, we require strong distributional assumptions on the covariates to consistently estimate $\psi_0^1$ using convex balancing weights. This contrasts to \cite{gleser1992importance}, who shows that the OLS estimates are consistent with only very weak distributional assumptions on the data (see also Propositions~\ref{cl8} and ~\ref{cl9} in Appendix~\ref{app:AsecI}). This relates to a third limitation: we require strong outcome modeling assumptions. Yet by preventing extrapolation, SBW and H-SBW estimates may be less sensitive than OLS estimates to these assumptions. Our validation results support this: the standard regression calibration adjustment using OLS and GLS weights performs the worst out of any methods we consider (see Table~\ref{tab:pretxpredfull} in Appendix~\ref{app:allresults}). In contrast, using regression-calibration with balancing weights -- even when allowing for limited extrapolation using ridge-augmentation -- performs better.\footnote{We also provide a suggestion of how to adapt our procedure to accommodate a basis expansion of gaussian covariates in Appendix~\ref{app:AsecI} Remark~\ref{remark:basis expansion}} Developing methods to relax these assumptions further would be a valuable area for future work.

A final concern is that this procedure may be sub-optimal with respect to the mean-square error of our estimator. In particular, the bias induced by the measurement error decreases with the sample size used to calculate each CPUMA's covariate values, the minimum of which were over three hundred. Yet the variance of our counterfactual estimate decreases with the number of treated states. From a theoretic perspective, the variance is of a larger order than the bias, so perhaps the bias from measurement error should not be a primary concern. Our final results support this: the changes in our results on the adjusted versus unadjusted data are of smaller magnitude than the associated uncertainty estimates. Other studies have proposed tests of whether the measurement error corrections are ``worth it,'' though we do not do this here (see, e.g., \cite{gleser1992importance}). However, as an informal observation, our simulation study in Appendix~\ref{app:simstudy} shows that the MSE of the SBW estimator that naively balances on the noisy covariates $W$ is comparable to the MSE of the SBW estimator that balances on the adjusted covariates $\hat{X}$ when the ratio of the variance of $X$ to the variance of $W$ is 0.95. However, even in this setting we find that confidence interval coverage can fall below nominal rates when balancing on $W$, and that the measurement error correction can improve our ability to construct valid confidence intervals.

Our second contribution is to introduce the H-SBW objective. This objective can improve upon the SBW objective assuming that the errors in the outcome model are homoskedastic with constant positive equicorrelation $\rho$ within known groups of units. Assuming no measurement error and that $\rho$ is known, we show that H-SBW returns the minimum variance estimator within the constraint set by more evenly dispersing weights across the groups. We also demonstrate the connection between these weights and the implied weights from GLS (see Propositions~\ref{cl4} and \ref{cl56} in Appendix~\ref{app:AsecII}). While studies have considered balancing weights in settings with hierarchical data (see, e.g., \cite{keele2020hospital}, \cite{ben2021multilevel}), we are the first to our knowledge to propose changing the criterion to account for correlated outcomes.

This estimation procedure has at least three potential drawbacks. First, we make a very specific assumption on the covariance structure of the error terms that is useful for our application. For applications where a different structure $\Omega$ is more appropriate one can still follow our approach and minimize the more general criterion $f(\gamma) = \gamma^\top\Omega\gamma$. Second, we require specifying the parameters $(\rho, \delta)$ in advance. Choosing $\delta$ is a challenging problem shared with SBW and we do not offer any new suggestions (though see \cite{wang2020minimal}, who offer an interesting approach to this problem.) Choosing $\rho$ is a new problem of this estimation procedure. Encouragingly, our simulation study shows the H-SBW estimator for any chosen $\rho$ almost always has lower variance than SBW in the presence of state-level random effects (see Appendix~\ref{app:simstudy}).\footnote{We caution that this finding solely reflects the simulation space we examined.} Even so, identifying a principled approach to choosing this parameter would be a useful future contribution. Third, in the presence of both measurement error and dependent data, using H-SBW in combination with the standard regression-calibration adjustment may be biased. This bias arises because the standard adjustment assumes independent data. Our simulations show that this bias can increase with $\rho$ and that SBW remains approximately unbiased if the covariates are gaussian - though even in this setting H-SBW may still yield modest MSE improvements. We also show a theoretical modification to the adjustment where H-SBW would return unbiased estimates in this setting (see Propositions~\ref{cl7} and ~\ref{cl7hsbw}).

\subsection{Policy considerations and limitations}

We estimate that had states that did not expand Medicaid in 2014 instead expanded their programs, they would have seen a -2.33 (-3.54, -1.11) percentage point change in the average adult uninsurance rate. Our validation study and robustness checks indicate that this estimate may be biased downwards (away from zero), in which case we can interpret this estimate as a lower bound on the ETC. Existing estimates place the ETT between -3 and -6 percentage points. These estimates vary depending on the targeted sub-population of interest, the data used, the level of modeling (individuals or regions), and the modeling approach (see, e.g., \cite{courtemanche2017early}, \cite{kaestner2017effects}, \cite{frean2017premium}). Our estimate of the ETC are closer to zero than these estimates. When we attempt to estimate the ETT using our proposed method, the resulting estimates have high uncertainty due to limited covariate overlap.\footnote{In particular, there were no states entirely controlled by Democrats that did not expand Medicaid. Even allowing for large imbalances, our standard error estimates were approximately three percentage points and our confidence intervals all contained zero. When estimating a simple difference-in-differences model on the unadjusted dataset we estimate that the ETT is -2.05 (-3.23, -0.87), where the standard errors account for clustering at the state level.\label{footnote_did}} The differences may reflect different modeling strategies and data, or it may suggest that the ETC is smaller in absolute magnitude than the ETT.

We are ultimately unable to draw any statistical conclusions about these differences; nevertheless, we continue to emphasize caution about assuming these estimands are equal. Due in part to the different covariate distributions of expansion versus non-expansion states, we may still suspect that the ETC differs from the ETT with respect to the uninsurance rates. Moreover, because almost every outcome of interest is mediated through increasing the number of insured individuals, such a difference may be important. For example, \cite{miller2021medicaid} study the effect of Medicaid expansion on mortality. Using their estimate of the ETT with respect to mortality they project that had all states expanded Medicaid, 15,600 deaths would have been avoided from 2014 through 2017. If we believe that this number increases monotonically with the number of uninsured individuals, this estimate may be an overestimate if the ETC with respect to the uninsurance rate is less than the ETT, or an underestimate if the ETC is greater than the ETT. Obtaining more precise inferences on the ETC with respect to uninsurance rates, if possible, would therefore be valuable future work.

Our analytic approach is not without limitations. Specifically, we require strong assumptions, including SUTVA, no anticipatory treatment effects, no unmeasured confounding conditional on the true covariates, and several parametric assumptions regarding the outcome and measurement error models. We address some concerns about possible violations of these assumptions. For example, our results were qualitatively similar whether we excluded possible ``early expansion states,'' or used different weighting strategies (including relaxing the positivity restrictions and changing the tuning parameter $\rho$). However, we do not attempt to address concerns about the impact of spillovers across regions. And while we believe that no unmeasured confounding is reasonable for this problem, we did not conduct a sensitivity analysis (see, e.g., \cite{bonvini2021sensitivity}) with respect to this assumption. 

Medicaid expansion remains an ongoing policy debate in the United States, where as of 2022 twelve states have not expanded their eligibility requirements. Our study estimates the effect of Medicaid expansion on adult uninsurance rates; however, the primary reason this effect interesting is that Medicaid enrollment is not automatic for eligible individuals. If the goal of Medicaid expansion is to increase insurance access for low-income adults, state policy-makers also may wish to make it easier or even automatic to enroll in Medicaid. 

\section{Conclusion}

We predict the average change in the non-elderly adult uninsurance rate in 2014 among states that did not expand their Medicaid eligibility thresholds as if they had. We use survey data aggregated to the CPUMA-level to estimate this quantity. The resulting dataset has both measurement error in the covariates that may bias standard estimation procedures, and a hierarchical structure that may increase the variance of these same approaches. We therefore propose an estimation procedure that uses balancing weights that accounts for these problems. We demonstrate that our bias-reduction approach improves on existing methods when predicting observed outcomes from an unknown data generating mechanism. Applying this method to our problem, we estimate that states that did not expand Medicaid in 2014 would have seen a -2.33 (-3.54, -1.11) percentage point change in their adult uninsurance rates had they done so. This is the first study we are aware of that directly estimates the treatment effect on the controls with respect to Medicaid expansion. From a methodological perspective, we demonstrate the value of our proposed method relative to existing methods. From a policy-analysis perspective, we emphasize the importance of directly estimating the relevant causal quantity of interest. More generally if the goal of Medicaid expansion is to improve access to insurance, state and federal policy-makers should consider policies that make Medicaid enrollment easier if not automatic.

\section*{Acknowledgements}

We gratefully acknowledge invaluable advice and comments from Zachary Branson, Riccardo Fogliato, Edward Kennedy, Brian Kovak, Akshaya Jha, Lowell Taylor, Jose Zubizaretta. 

\begin{supplement}
Analyses conducted using R Version 4.0.2 (\cite{r}), and the optweight (\cite{optweight}) and tidyverse (\cite{tidyverse}) packages. Programs and supporting materials are available at \url{github.com /mrubinst757/medicaid-expansion}. Proofs and additional results are available in the Appendix. 
\end{supplement}

\bibliographystyle{imsart-nameyear} % Style BST file
\bibliography{research.bib}       % Bibliography file (usually '*.bib')

\clearpage

\appendix

\section{Proofs}\label{ssec:proof}

We divide our proofs into three sections: the first two consist of propositions and the third contains the proofs of the propositions. In the first section our propositions pertain to the performance of SBW under the classical measurement error model. Our key results are that the bias of the SBW estimator is equivalent to the bias of the OLS estimator and that regression-calibration techniques can be used in this setting to obtain consistent estimators. However, these results assume that the data are gaussian. We also show that if the data are not gaussian, the OLS estimator using regression-calibration remains consistent, while the SBW estimator may be biased; however, this can be corrected if other distributional assumptions are made in place of gaussianity. In our second section we consider the properties of the H-SBW objective when the true covariates $X$ are observed. We show that if our assumed covariance structure for the outcome errors is correct, H-SBW produces the minimum conditional-on-X variance estimator within the constraint set. We also show how a generalized form of H-SBW weights relate to the implied regression weights from Generalized Least Squares (GLS). We conclude by showing that H-SBW may yield biased estimates if we do not correctly model the dependence structure of the data. Section~\ref{app:AsecIII} contains all of the proofs.

\subsection{SBW and classical measurement error}\label{app:AsecI}

We begin by showing several results regarding the bias of the OLS and SBW estimators under the classical errors-in-variables model. First, we show that without adjustment for errors-in-covariates, the bias of the SBW estimator that sets $\delta = 0$ (i.e. reweights the treated units to exactly balance the control units) is equal to the bias of the OLS estimator. Second, we show that if the observed covariate values for the treated data can be replaced by their conditional expectations $\tilde{X}$ given the noisy observations, then the SBW estimator will be unbiased and consistent. Third, we consider the case where $\tilde{X}$ must be estimated, and show that the SBW estimator is consistent if we replace $\tilde{X}$ by a consistent estimate $\hat{X}$. Finally, we remove the assumption that $X$ is gaussian, and show that while the OLS estimator remains unbiased under weaker assumptions, the SBW estimator does not, and we show a general expression for the asymptotic bias. We take the perspective throughout that $X$ is random among the treated units but fixed for the control units.

We assume that equations (\ref{eqn:unconfoundedness}) - (\ref{eqn:Xgaussian}) hold. For simplicity, we additionally assume that
\begin{equation}\label{eqn:simplifications}
\epsilon_{sc} = 0, \quad \varepsilon_s = 0,\quad \xi_{sc} = 0,\quad  \Sigma_{\nu,sc} = \Sigma_\nu \qquad \forall s,c
\end{equation}
noting that $\xi_{sc}=0$ implies $J_{sc} = Y_{sc}$. These assumptions imply that the data from the treated units are i.i.d., though for consistency of notation we continue to index the data by $s$ and $c$ and assume that the state-membership for each CPUMA is known. The covariate observations of the treated units can then be seen to have covariance matrix
\[ \Sigma_{W|1} = \Sigma_{X|1} + \Sigma_\nu\]
and the conditional expectation of $X_{sc}$ given $W_{sc}$ for the treated units can be seen to equal
\[ \tilde{X}_{sc} = v_1 + \kappa^\top (W_{sc} - v_1), \qquad \forall sc: A_{sc}=1\]
where
\[ \kappa = (\Sigma_{X|1} + \Sigma_{\nu})^{-1} \Sigma_{X|1}\]
To ease notation, we abbreviate $\Sigma_X = \Sigma_{X \mid 1}$ and similarly $ \Sigma_W = \Sigma_{W \mid 1}$. 

In Propositions \ref{cl8}, \ref{cl9}, and part of Proposition \ref{cl1}, we will remove the Gaussian covariate assumption given by \eqref{eqn:Xgaussian}. In its place, we will instead consider the weaker assumption that the empirical covariance of $X$ has a limit $S_X$,

\begin{equation}\label{eqn:limitX}
 \frac{1}{n_1} \sum_{A_{sc}=1} (X_{sc} - \bar{X}_1)(X_{sc} - \bar{X}_1)^\top \rightarrow^p S_X
\end{equation}
which implies a similar limit $S_W$ for the noisy observations $W$,

\begin{equation}\label{eqn:limitW}
 \frac{1}{n_1} \sum_{A_{sc}=1} (W_{sc} - \bar{W}_1)(W_{sc} - \bar{W}_1)^\top \rightarrow^p S_W = S_X + \Sigma_{\nu}
\end{equation}
where we have used the independence of the noise terms $\nu_{sc}$, and similarly that 
\begin{equation}\label{eqn:limitWY}
 \frac{1}{n_1} \sum_{A_{sc}=1} (W_{sc} - \bar{W}_1)(Y_{sc} - \bar{Y}_1)^\top \rightarrow^p S_X \beta_1
\end{equation}
where we have additionally used the linear model for $Y_{sc}$ given by \eqref{eqn:linmod}.

We first consider estimation without adjustment for errors in covariates. 
Proposition \ref{cl1} states that the unadjusted OLS and SBW estimators have equal bias, with the bias of the OLS estimator remaining unchanged if the gaussian assumption of \eqref{eqn:gaussiannoise} is removed.

\begin{proposition}\label{cl1}
Let (\ref{eqn:unconfoundedness}) - (\ref{eqn:Xgaussian}) and (\ref{eqn:simplifications}) hold.
Let $(\hat{\alpha}, \hat{\beta})$ denote the unadjusted OLS estimator of $(\alpha_1, \beta_1)$, 
\begin{equation}\label{eqn:prop1.beta}
(\hat{\alpha}, \hat{\beta}) = \arg \min_{\alpha, \beta} \sum_{sc:A_{sc}=1} (Y_{sc} - \alpha -  W_{sc}^\top \beta)^2
\end{equation}
which induces the OLS estimator of $\psi_0^1$ given by

\begin{align*}
\hat{\psi}^{1,\textup{ols}}_0 = \bar{Y}_1 + (\bar{W}_0 - \bar{W}_1)^\top\hat{\beta}_1
\end{align*}
Let ${\gamma}$ denote the unadjusted SBW weights under exact balance, found by solving \eqref{eqn:SBWobjective} with constraint set $\Gamma( W_{A=1}, \bar{W}_0, 0)$, which induces the SBW estimator of $\psi_0^1$ given by

\begin{align*}
\hat{\psi}^{1,\textup{sbw}}_0 = \sum_{sc: A_{sc} = 1} {\gamma}_{sc} Y_{sc}
\end{align*}
Then the estimators $\hat{\psi}^{1, \textup{ols}}_0$ and $\hat{\psi}^{1, \textup{sbw}}_0$ have equal bias, satisfying

\begin{align*}
\mathbb{E}[\hat{\psi}_0^{1,\textup{ols}}] &= \mathbb{E}[\hat{\psi}^{1, \textup{sbw}}_0]  = \psi_0^1 + (\bar{X}_0 - \upsilon_1)^\top(\mathbf{\kappa} - I_q)\beta
\end{align*}
Additionally, the bias of $\hat{\psi}_0^{1,\textup{ols}}$ is asymptotically unchanged if the gaussian covariate assumption given by \eqref{eqn:Xgaussian} is replaced by \eqref{eqn:limitX}.
\end{proposition}

To study the SBW estimator with covariate adjustment, we first consider an idealized version where $\Sigma_X$ and $\Sigma_\nu$ are known, so that $\tilde{X}_{A=1}$ is also known. Proposition \ref{cl2} shows that the resulting estimate of $\psi_0^1$ is unbiased if $\delta = 0$.

\begin{proposition}\label{cl2}
Let (\ref{eqn:unconfoundedness}) - (\ref{eqn:Xgaussian}) and (\ref{eqn:simplifications}) hold. Let $\tilde{X}_{A=1}$ equal the conditional expectation of $X_{A=1}$ given $W$,

\[ \tilde{X}_{sc} = \upsilon_1 + \kappa^\top (W_{sc} - \upsilon_1), \qquad \forall sc: A_{sc} = 1\] let $\gamma^*$ be the solution to the SBW objective defined over the constraint set $\Gamma(\tilde{X}_{A=1}, \bar{X}_0, 0)$, and let $\hat{\psi}^{1, \textup{ideal}}_0$ be the SBW estimator $\sum_{sc: A_{sc} = 1}\gamma^\star_{sc}Y_{sc}$. This estimator is unbiased for $\psi_0^1$.
\end{proposition}

Proposition \ref{prop:variance_rate} shows that the variance of this idealized SBW estimator goes to zero, implying consistency. 
\begin{proposition}\label{prop:variance_rate}
Let (\ref{eqn:unconfoundedness}) - (\ref{eqn:Xgaussian}) and (\ref{eqn:simplifications}) hold, and let $\gamma^*$ and $\hat{\psi}_0^{1, \textup{ideal}}$ be defined as in Proposition \ref{cl2}. Then the conditional variance of the estimation error is given by

\begin{align*}
\operatorname{Var}\left( \hat{\psi}_0^{1, \textup{ideal}} - \psi_0^1| W\right)  = \|\gamma^*\|^2 \cdot \beta_1^\top (\Sigma_{X} - \Sigma_{X}\Sigma_{W}^{-1}\Sigma_{X})\beta_1
\end{align*}
with $\operatorname{Var}\left( \hat{\psi}_0^{1, \textup{ideal}} - \psi_0^1| W\right)$ and $\operatorname{Var}(\hat{\psi}_0^{1,\textup{ideal}})$ both behaving as $O_P(n_1^{-1})$ as $n_1 \rightarrow \infty$.
\end{proposition}

In practice, the idealized SBW estimator considered in Propositions \ref{cl2} and \ref{prop:variance_rate} cannot be used, as $\Sigma_X$ and $\Sigma_{\nu}$ are not known, but instead must be estimated from auxiliary data. Proposition \ref{cl3} states that if these estimates are consistent, then the resulting adjusted SBW estimator for $\psi_0^1$ is also consistent if $\delta = 0$.

\begin{proposition}\label{cl3}
Let (\ref{eqn:unconfoundedness}) - (\ref{eqn:Xgaussian}) and (\ref{eqn:simplifications}) hold. Given estimates $\hat{\Sigma}_X$ and $\hat{\Sigma}_\nu$ that are consistent for $\Sigma_X$ and $\Sigma_\nu$, let $\hat{X}_{A=1}$ be given by 
\[ \hat{X}_{sc} = \bar{W}_1 + \hat{\kappa}^\top(W_{sc} - \bar{W}_1), \]
where $\hat{\kappa} = (\hat{\Sigma}_X + \hat{\Sigma}_{\nu})^{-1} \hat{\Sigma}_X$. Let $\hat{\gamma}$ be the weights that solve the SBW objective over the constraint set $\Gamma(\hat{X}_{A=1}, \bar{W}_0, 0)$, and let $\hat{\psi}^{1, \textup{adjusted}}_0 = \sum_{sc: A_{sc} = 1} \hat{\gamma}_{sc} Y_{sc}$ be the corresponding SBW estimator. This estimator is consistent for $\psi_0^1$ as $n_1 \to \infty$.
\end{proposition}

In (\ref{eqn:jackknife}) we propose a leave-one-state-out jackknife estimate of variance. Following \cite{efron1981jackknife}, this estimate can be decomposed a conservatively biased estimate of the variance of $\hat{\psi}_0^{1, \textup{adjusted}}$ given a sample size of $(m_1-1)$ treated states, plus a heuristic adjustment to go from sample size $(m_1-1)$ to sample size $m_1$, when treating the observations of the control states as fixed.

\begin{proposition}\label{prop:jackknife}
Let (\ref{eqn:unconfoundedness}) - (\ref{eqn:Xgaussian}) hold, and additionally assume that $p_s$, the number of CPUMAs, is i.i.d. in the treated states. Let $\hat{\operatorname{Var}}(\hat{\psi}_0^{1, \textup{adjusted}}) = \frac{m_1-1}{m_1} \cdot \tilde{\operatorname{Var}}(\hat{\psi}_0^{1, \textup{adjusted}})$, where

\begin{equation} \label{eqn:prop.jackknife}
\tilde{\operatorname{Var}}(\hat{\psi}_0^{1, \textup{adjusted}}) = \sum_{s:A_{s}=1} (S_{(s)} - S_{(\cdot)})^2
\end{equation}
with $S_{(s)}$ and $S_{(\cdot)}$ as defined for \eqref{eqn:jackknife}. Then $\tilde{\operatorname{Var}}$ is conservatively biased for the variance of the leave-one-state-out estimate,

\[ \mathbb{E}\left[ \tilde{\operatorname{Var}}(\hat{\psi}_0^{1, \textup{adjusted}})\right] \geq \operatorname{Var}(S_{(1)} | \bar{W}_0),\]
where $S_{(1)}$ can be seen to equal the estimator $\hat{\psi}_0^{1,\textup{adjusted}}$ under a sample size of $(m_1-1)$ treated states.

\end{proposition}

As the gaussian covariate assumption given by \eqref{eqn:Xgaussian} is strong, it would be desirable if the adjusted OLS or SBW estimators were consistent even for non-gaussian $X$. Proposition \ref{cl8} shows under mild assumptions that this is in fact true when running OLS on the adjusted covariates. 

\begin{proposition}\label{cl8}
Let (\ref{eqn:unconfoundedness}) - (\ref{eqn:gaussiannoise}), and (\ref{eqn:simplifications})- (\ref{eqn:limitX}) hold, with $S_X$ invertible. Let $(\check{\alpha}, \check{\beta})$ denote the adjusted OLS estimates of $(\alpha_1, \beta_1)$, solving

\[ \min_{\alpha,\beta} \sum_{A_{sc}=1} (Y_{sc} - \alpha - \check{X}_{sc}^\top \beta)^2 \]
where $\check{X}_{sc} = \bar{W}_1 + \check{\kappa}^\top(W_{sc} - \bar{W}_1)$ with $\check{\kappa} = (S_X + \Sigma_\nu)^{-1}S_X$. Then the adjusted OLS estimator of $\psi_0^1$ given by
\[ \bar{Y}_1 - (\bar{W}_0 - \bar{W}_1)^\top \check{\beta}\]
remains consistent if the gaussian assumption given by \eqref{eqn:gaussiannoise} is removed.
\end{proposition}

However, the same does not hold for the adjusted SBW estimator. Proposition \ref{cl9} gives an expression for its bias when the covariates are non-gaussian. 

\begin{proposition}\label{cl9}
Let the assumptions of Proposition \ref{cl8} hold. Let $\check{\gamma}$ solve the SBW objective over the constraint set $\Gamma(\check{X}_{A=1}, \bar{W}_0, 0)$ where $\check{X}_{A=1}$ and $\check{\kappa}$ are defined as in Proposition \ref{cl8}. Let $Q$ denote the set of indices where $\check{\gamma}$ is non-zero,

\[ Q = \{sc: \check{\gamma}_{sc} > 0\}\]
with cardinality $n_Q = |Q|$, and let $\bar{W}_Q$ and $S_{W_Q}$ denote the empirical mean and covariance of $\{W_{sc}:sc \in Q\}$,

\[ \bar{W}_Q = \frac{1}{n_Q}\sum_{sc \in Q} W_{sc},\qquad S_{W_Q} = \frac{1}{n_Q} \sum_{sc \in Q} (W_{sc} - \bar{W}_Q)(W_{sc} - \bar{W}_Q)^\top\]
with $\bar{X}_Q$ the analogous empirical mean of $\{X_{sc}:sc \in Q\}$ and $S_{XW_Q}$ the empirical cross covariance,
\[ S_{XW_Q} = \frac{1}{n_Q} \sum_{sc \in Q} (X_{sc} - \bar{X}_Q)(W_{sc} - \bar{W}_Q)^\top\]
Then if the gaussian assumption given by \eqref{eqn:gaussiannoise} is removed, the adjusted SBW estimator for $\psi_0^1$ given by 

\[\sum_{A_{sc}=1} Y_{sc} \check{\gamma}_{sc}\]
may be biased for $\psi_0^1$, with estimation error given by 

\begin{align} 
\nonumber \sum_{A_{sc}=1} Y_{sc} \check{\gamma}_{sc} - \psi_0^1 & = \beta_1^\top \Big[(S_{XW_Q}S_{W_Q}^{-1}S_WS_X^{-1} - I)\bar{X}_0  + (\bar{X}_Q - S_{XW_Q}S_{W_Q}^{-1} S_W S_X^{-1} \bar{X}_1) \\
& \hskip1cm {} - S_{XW_Q}S_{W_Q}^{-1}(\bar{X}_Q - \bar{X}_1)\Big](1 + o_P(1))  \label{eqn:cl9.error}
\end{align}
which need not converge to zero unless $\bar{X}_Q \to \bar{X}_1$, $S_{XW_Q} \to S_X$, and $S_{W_Q} \to S_W$.
\end{proposition}

Proposition \ref{cl7} shows that if the conditional expectations can be computed for the treated units (which may be computationally difficult or require strong modeling assumptions if the data is non-gaussian, or if dependencies exist between CPUMAs), then SBW yields unbiased estimates. 

\begin{proposition}\label{cl7}
    Let equations (\ref{eqn:unconfoundedness})-(\ref{eqn:linmod}) hold. Let $\tilde{X}^*$ denote the conditional expectation,
    \[\tilde{X}^*_{sc} = \mathbb{E}[X_{sc} | W, A_{sc}=1]\]
    let weights $\tilde{\gamma}^*$ solve the SBW objective (\ref{eqn:SBWobjective}) with constraint set $\Gamma(\tilde{X}^\star_{A=1}, \bar{X}_0, 0)$, and consider the estimator of $\psi_0^1$ given by $\sum_{A_{sc}=1} Y_{sc} \tilde{\gamma}^*_{sc}$. This estimator is unbiased for $\psi_0^1$.
\end{proposition}

\begin{remark}
    While we have assumed that $\epsilon_{sc}=0$ for simplicity in our propositions, removing this assumption simply leads to the additional term $\sum_{sc: A_{sc} = 1}\gamma_{sc}\epsilon_{sc}$ in the error of the SBW estimator of $\psi_0^1$. This again has expectation zero, because the weights remain independent of the error $\epsilon_{sc}$ in the outcomes. Allowing non-zero $\epsilon_{sc}$ also adds a term to the estimator variance (conditional on $W$) equal to $\sigma^2_{\epsilon}\cdot \|\gamma^*\|^2$,    which does not change the variance bound given by Proposition \ref{prop:variance_rate}.
\end{remark}

\begin{remark}
    For the adjusted OLS estimator, in which $\beta_1$ is estimated using the adjusted covariates $\tilde{X}_{A=1}$, in practice we must estimate $\tilde{X}$ with some estimator $\hat{X}$ that relies on an estimate $\hat{\kappa}$. As long as $\hat{\kappa}$ is consistent for $\kappa$ then the OLS estimator will also be consistent by the continuous mapping theorem.
\end{remark}

\begin{remark}
As Proposition \ref{prop:jackknife} implies that $\hat{\operatorname{Var}}$ is conservatively biased only up the heuristic $(m_1-1)/m_1$ scaling term, it may be preferable to remove this scaling term entirely, inflating the variance estimate slightly. While the proposition considers the marginal variance of the estimator $\hat{\psi}_0^{1,\textup{adjusted}}$, a confidence interval using the conditional variance $\operatorname{Var}(\hat{\psi}_0^{1 \textup{adjusted}}|X)$ (see, e.g., \cite{buonaccorsi2010measurement}, who discuss using a modification of the parametric bootstrap for parameters estimated via OLS in this setting) may be of interest, potentially leading to smaller intervals and more precise inference. 
\end{remark}

\begin{remark}
To see how Proposition \ref{cl9} implies that the adjusted SBW estimate may be biased in non-gaussian settings, we observe that as the set $Q$ in Proposition \ref{cl9} will depend on the values of the covariates $X$ and observation noise $\nu$, the values of $\{X_{sc}: sc \in Q\}$ and $\{\nu_{sc}: sc \in Q\}$ may differ systematically from their population, so that $\bar{X}_Q$, $S_{XW_Q}$ and $S_{W_Q}$ may not converge to their desired counterparts. While the expression for the estimation error given by (\ref{eqn:cl9.error}) is asymptotic, an exact formula is given in \eqref{eqn:cl9.proof3} which is very similar; the only asymptotic approximations are the convergence of $\bar{W}_1$ to $\bar{X}_1$ and $\bar{W}_0$ to $\bar{X}_0$. 
\end{remark}

\begin{remark}\label{remark:basis expansion}
We describe a possible direction for future work that utilizes Proposition \ref{cl7}. Suppose that in place of equations (\ref{eqn:additivenoise})-(\ref{eqn:Xgaussian}), we instead assume that $X_{sc}$ is a transformation of the covariate, so that $X_{sc} = \phi(U_{sc})$ for some transformation $\phi$, and that the untransformed $U_{sc}$ is observed with additive noise, so that $W_{sc} = U_{sc} + \nu_{sc}$. For example, to make the linear model (\ref{eqn:linmod}) more credible, $\phi(U_{sc})$ might denote a basis expansion applied to the survey sampled covariates for each unit. If, analogous to assumptions (\ref{eqn:gaussiannoise}) and (\ref{eqn:Xgaussian}), the original covariates $U_{sc}$ and measurement error $\nu_{sc}$ can be assumed to be i.i.d. gaussian, so that the treated units satisfy

\begin{align*}
    U_{sc} & \sim \mathcal{N}(v_1, \Sigma_{U|1}), & \nu_{sc} & \sim \mathcal{N}(0, \Sigma_{\nu}), \qquad \forall\ sc: A_{sc}=1
\end{align*}
then the posterior distribution of $U_{sc}$ given $W$ for the treated units will also be gaussian

\[ U_{sc}|W_{sc} \sim \mathcal{N}(\tilde{U}_{sc}, \Sigma_{\tilde{U}|1}), \qquad \forall\ sc:A_{sc}=1 \]
where $\tilde{U}_{sc}$ and $\Sigma_{\tilde{U}|1}$ are given for the treated units by

\begin{align*}
\tilde{U}_{sc} &= v_{1} + \Sigma_{U|1} (\Sigma_{U|1} + \Sigma_{\nu})^{-1}(W_{sc} - v_1), & \Sigma_{\tilde{U}|1} & = \Sigma_{U|1} - \Sigma_{U|1} (\Sigma_{U|1} + \Sigma_{\nu})^{-1} \Sigma_{U|1}
\end{align*}
with analogous expressions for the control units. This suggests that if auxiliary data can be used to find $\Sigma_{U|1}$, $\Sigma_{U|0}$, and $\Sigma_{\nu}$ as before, then  $\tilde{X}^*_{sc} = \mathbb{E}[\phi(U_{sc})|W,A]$ could be estimated by using monte carlo methods. Specifically, for each unit $sc$ we can generate random variates $\{u_{i}\}$ that are i.i.d. normal with mean $\tilde{U}_{sc}$ and covariance $\Sigma_{\tilde{U}|A_{sc}}$, and estimate $\tilde{X}_{sc}^*$ by the average of $\{\phi(u_{i})\}$. To estimate the SBW constraint set $\Gamma(\tilde{X}^*_{A=1}, \bar{X}_0, 0)$, then $\bar{X}_0$ could be estimated by averaging $\tilde{X}^*_{A=0}$. By Proposition \ref{cl7} the resulting SBW weights would yield unbiased estimates.
\end{remark}

\subsection{Properties of H-SBW}\label{app:AsecII}

Here we consider an H-SBW setting where $\nu_{sc}=0$ so that the true covariates are observed. By \eqref{eqn:linmod}, the outcomes have CPUMA level noise terms  $\epsilon_{sc}$, and also state-level noise terms $\varepsilon_s$ that correlate the outcomes of CPUMAs in the same state. Proposition \ref{cl4} states that if $\rho$ is the within-state correlation of these error terms, the H-SBW estimator produces the minimum conditional-on-X variance estimator of $\psi_0^1$ within the constraint set.

\begin{proposition}\label{cl4}
    Consider the outcome model in ~\eqref{eqn:linmod}. Assume the errors are homoskedastic and have finite variance $\sigma^2_{\epsilon}$ and $\sigma^2_{\varepsilon}$, and let $\rho$ be the within-state correlation of the error terms. Let $\hat{\gamma}^{\textup{hsbw}}$ be the weights that solve \eqref{eqn:hsbwobjective} for known parameter $\rho$ across the constraint set $\Gamma(X_{A=1}, \bar{X}_0, \delta)$ for any $\delta$. Then the H-SBW estimator of $\psi_0^1$,

    \[\sum_{s: A_s = 1}\sum_{c=1}^{p_s}\hat{\gamma}_{sc}^{\textup{hsbw}}Y_{sc}\] 
    is the minimum conditional-on-X variance estimator of $\psi_0^1$ within the constraint set $\Gamma(X_{A=1}, \bar{X}_0, \delta)$.
\end{proposition}

The SBW and H-SBW objective functions take the generic form $\gamma^\top\Omega\gamma$: SBW takes $\Omega = I_n$, while H-SBW specifies an $\Omega$ that allows for homoskedastic errors with positive within-state equicorrelation. Analogous versions hence exist for any assumed covariance structure $\Omega$. Proposition \ref{cl56} highlights connections between this generic form and GLS, showing that when exact balance is possible we can express the generic form of the problem as the implied regression weights from GLS estimated on a subset of the data. Similar results connecting balancing weights to regression weights can be found throughout the literature. For example, see \cite{kline2011oaxaca}, \cite{ben2021augmented}, \cite{chattopadhyay2021implied}, who connect balancing weights to regression weights for OLS, ridge-regression, and weighted least-squares and two-stage-least-squares; our result adds by connecting balancing weights to regression weights for GLS.

\begin{proposition}\label{cl56}
Let $\gamma^*$ solve the optimization problem

\begin{equation}\label{eqn:a1.1}
 \min_\gamma \gamma^\top \Omega \gamma \quad \text{ subject to } \quad  \sum_i \gamma_i Z_i = v,\ \sum_i \gamma_i = 1 \ \textup{ and } \gamma \geq 0%\gamma \in \Phi(Z, v, 0),
\end{equation}
with $\Omega$ positive definite. Let $Q = \{i: \gamma^*_i > 0\}$ denote the indices of its non-zero entries. Then $\gamma^*$ also solves the problem
  
  \begin{equation}\label{eqn:a1.2}
   \min_{\gamma}  \ \gamma^\top \Omega \gamma  \quad \textup{subject to }\quad \sum_{i \in Q} \gamma_i Z_i = v,\ \sum_{i \in Q} \gamma_i = 1,\ \textup{ and }   \gamma_i = 0\  \forall\ i \not\in Q
  \end{equation}
and hence has non-zero entries $\gamma^*_Q = \{\gamma_i^*: i \in Q\}$ satisfying
 
 \begin{equation}\label{eqn:a1.3}
 \gamma^*_{Q} = \Omega_{Q}^{-1} (Z_{Q} - \mu)^\top\left[ (Z_Q - \mu) \Omega_{Q}^{-1} (Z_Q - \mu)^\top\right]^{-1} (v - \mu) + \frac{\Omega^{-1}_Q {\bf 1} }{{\bf 1}^\top \Omega^{-1}_Q {\bf 1}}
 \end{equation}
where $Z_{Q}$ is the matrix whose columns are $\{Z_i: i \in Q\}$, $\Omega_Q$ is the submatrix of $\Omega$ whose rows and columns are in $Q$, ${\bf 1}$ is the column vector of ones, and $\mu$ is the vector $\frac{Z_{Q}\Omega_{Q}^{-1} {\bf 1}}{ {\bf 1}^\top \Omega^{-1}_Q {\bf 1}}$. These weights are equivalent to the implied regression weights when running GLS on the subset $Q$.
\end{proposition}

\begin{remark}
To lighten notation, we have used $Z_Q - \mu$ (a vector subtracted from a matrix) to mean $Z_Q - \mu{\bf 1}^\top$, so that each column of $Z_{Q}$ is centered by $\mu$. 
\end{remark}

\begin{remark}
Removing the positivity constraint from the generic form of the SBW objective implies that the resulting weights are equivalent to the implied regression weights from GLS. This follows by noting that removing the positivity constraint from the SBW objective removes the solution's dependence on the set $Q$. This result is a natural generalization of the connections between the implied regression weights of OLS and SBW also noted by, for example, \cite{chattopadhyay2021implied}.
\end{remark}

Proposition \ref{cl7hsbw} simply states that the conclusion of Proposition \ref{cl7} holds not only for SBW, but for H-SBW as well.

\begin{proposition}\label{cl7hsbw}
    Let the assumptions of Proposition \ref{cl7} hold. Let $\tilde{X}^*$ be defined as in Proposition \ref{cl7}, let weights $\tilde{\gamma}^{\textup{hsbw}*}$ solve the H-SBW objective \eqref{eqn:hsbwobjective} with constraint set $\Gamma(\tilde{X}^\star_{A=1}, \bar{X}_0, 0)$, and consider the estimator of $\psi_0^1$ given by $\sum_{A_{sc}=1} Y_{sc} \tilde{\gamma}^{\textup{hsbw}*}$. This estimator is unbiased for $\psi_0^1$.
\end{proposition}

\begin{remark}
    In Proposition \ref{cl4}, we assumed the outcomes followed \eqref{eqn:linmod} and the constraints balanced the means of the covariates; however, we can allow for any outcome model and our balance constraints can include any function of the covariate distribution and this result still holds conditional on $X$ (though of course the estimator may be badly biased). The key assumption is that the variability in the estimates comes from the outcome model errors, which are assumed to be homoskedastic and equicorrelated within state for known parameter $\rho$.
\end{remark}

\begin{remark}\label{remark:obgls}
    Assuming that $(X_{sc}, W_{sc}) \mid A_{sc} = 1$ are gaussian but dependent, Proposition \ref{cl7hsbw} implies that if we correctly model the correlations between the CPUMAs within states in our regression calibration step, we can use GLS or H-SBW without inducing asymptotic bias (assuming all of our models are correct). This is similar to the approach followed in \cite{huque2014impact}, who consider parameter estimation using GLS in the context of a one-dimensional spatially-correlated covariate measured with error. We also outline in Appendix~\ref{app:adjustmentdetails} a potential adjustment when we assume the units have a covariance structure similar to our assumptions for the outcome model errors. We evaluate this adjustment in simulations in Appendix~\ref{app:simstudy}. 
    
    To be clear if we do not model this dependence structure we cannot generally use the simple adjustment provided in \eqref{eqn:regcal} in combination with GLS to obtain asymptotically unbiased estimates. Intuitively this is because the implied weights from GLS depend on the dependence between CPUMAs within states, which \eqref{eqn:regcal} does not correctly account for. By contrast, Proposition \ref{cl8} shows that we safely can ignore such dependence when using regression-calibration with OLS (as long as a probability limit exists for the empirical covariance matrix).
\end{remark}

\begin{remark}\label{remark:sbwspeculation}
    In our simulation study in Appendix~\ref{app:simstudy} we obtain an approximately unbiased estimate when using SBW using the simple adjustment provided in \eqref{eqn:regcal} with dependent gaussian data. We conjecture that the set $Q$ may have some limiting boundary. If true, the characterization of SBW weights as regression weights in Proposition~\ref{cl56} would imply that the SBW weight $\hat{\gamma}_{sc}$ is fixed conditional on the input data point $W_{sc}$ asymptotically. The error of the estimator could then decompose as a function of $(X_{sc} - \tilde{X}_{sc})$, which is independent of $\gamma_{sc}^{sbw}$ given $W_{sc}$. This implies that it would suffice to balance on $\tilde{X}_{A=1}$.
\end{remark}

\subsection{Proofs}\label{app:AsecIII}

We begin by establishing the following identity for our target parameter $\psi_0^1$ defined in \eqref{eqn:psi}.

\begin{equation}\label{eqn:psi10_identity}
\psi^1_0 = \mu_y + (\bar{X}_0 - \upsilon_1)^\top \beta_1
\end{equation}
where $\mu_y = \mathbb{E}[Y_{sc} \mid A_{sc} = 1]$ and $\upsilon_1 = \mathbb{E}[X_{sc} \mid A_{sc} = 1]$.

\begin{proof}[Proof of (\ref{eqn:psi10_identity})]
Using our causal and modeling assumptions we have that:

\begin{align*}
\mathbb{E}[Y_{sc}^1 \mid X_{sc}, A_{sc} = 0] &= \mathbb{E}[Y_{sc}^1 \mid X_{sc}, A_{sc} = 1] \\
&= \mathbb{E}[Y_{sc} \mid X_{sc}, A_{sc} = 1] \\
&= \alpha_1 + X_{sc}^\top \beta_1 \\
&= \mu_y + (X_{sc} - \upsilon_1)^\top \beta \\
&\implies \psi_0^1 = \mu_y + (\bar{X}_0 - \upsilon_1)^\top \beta_1
\end{align*}
where the first equality follows from unconfoundedness, the second equality from consistency, the third from our parametric modeling assumptions, and the fourth by definition of $\alpha$. The final equation follows from averaging over the control units.
\end{proof}

\begin{proof}[Proof of Propositon \ref{cl1}]
It can be seen from \eqref{eqn:regcal} that for all $sc: A_{sc}=1$,

\begin{align*}
   X_{sc} &= v_1 + (W_{sc} - v_1)^\top \kappa + \nu_{sc}'
\end{align*}
where $\nu_{sc}' = X_{sc} - \mathbb{E}[X_{sc}|W,A=1]$ may be viewed as an independent zero-mean noise term. Plugging into \eqref{eqn:linmod} yields 

\begin{align*}
   Y_{sc} & = \alpha_1 + v_1^\top (I - \kappa)\beta_1 + W_{sc}^\top \kappa \beta_1 + \epsilon_{sc}'
\end{align*}
for $\epsilon_{sc}' = \beta_1^\top \nu_{sc}' + \epsilon_{sc}$. It follows that the OLS estimate $\hat{\beta}$ given by \eqref{eqn:prop1.beta} satisfies \citep{gleser1992importance},

\begin{equation}\label{eqn:prop1.0}
\mathbb{E}[\hat{\beta}|W_{A=1}] = \kappa \beta_1, \qquad \text{and} \qquad \mathbb{E}[\bar{W}_1 \hat{\beta}] = v_1 \kappa \beta_1
\end{equation}
To show that $\hat{\psi}_0^{1,\textup{ols}}$ and $\hat{\psi}_0^{1, \textup{sbw}}$ have identical bias, we compute their expectations:

\begin{align}
\nonumber	\mathbb{E}[\hat{\psi}_0^{1,\textup{ols}}] &= \mathbb{E}[ \bar{Y}_1 + (\bar{W}_0 - \bar{W}_1)^\top \hat{\beta}] \\
	& = \bar{\mu}_y + (\bar{X}_0 - \upsilon_1)^\top\kappa\beta_1 \label{eqn:prop1.1}\\
	& = \psi_0^1 + (\bar{X}_0 - \upsilon_1)^\top(\kappa - I_q)\beta_1 \label{eqn:prop1.2}
\end{align}
where \eqref{eqn:prop1.1} holds by \eqref{eqn:prop1.0}, and \eqref{eqn:prop1.2} holds by \eqref{eqn:psi10_identity}. We next derive the expected value of $\hat{\psi}^{1, \textup{sbw}}$:

\begin{align}
\nonumber	\mathbb{E}[\hat{\psi}_0^{1, \textup{sbw}}] & = \mathbb{E}\left[ \sum_{A_{sc} = 1} {\gamma}_{sc} Y_{sc}\right] \\
	& = \mathbb{E}\left[ \sum_{A_{sc}=1} {\gamma}_{sc} \left(\alpha_1 + (W_{sc} - W_{sc} + X_{sc})^\top \beta_1 + \epsilon_{sc}\right)\right] \label{eqn:prop1.4}\\
\nonumber	& = \mathbb{E}\left[ \alpha_1 + \sum_{A_{sc} = 1} {\gamma}_{sc} W_{sc}^\top \beta_1 + \sum_{A_{sc}=1} {\gamma}_{sc} (X_{sc} - W_{sc})^\top \beta_1 + \sum_{A_{sc}=1} {\gamma}_{sc} \epsilon_{sc} \right] \\
	& = \alpha_1 + \bar{X}_0^\top \beta_1 + \mathbb{E}\left[ \sum_{A_{sc} = 1} {\gamma}_{sc}(X_{sc} - W_{sc})^\top \beta_1\right] \label{eqn:prop1.5}\\
	& = \psi_0^1 + \mathbb{E} \left[ \sum_{A_{sc} = 1} {\gamma}_{sc}(X_{sc} - W_{sc})^\top \beta_1 \right] \label{eqn:prop1.6} \\
	& = \psi_0^1 + \mathbb{E} \left[ \sum_{A_{sc} = 1} \mathbb{E}\left[ {\gamma}_{sc}(X_{sc} - W_{sc})^\top \beta_1 | W \right] \right] \label{eqn:prop1.7} \\
	& = \psi_0^1 + \mathbb{E} \left[ \sum_{A_{sc} = 1}  {\gamma}_{sc} (\mathbb{E}[X_{sc}|W] - W_{sc})^\top \beta_1 \right] \label{eqn:prop1.8} \\
	& = \psi_0^1 + \mathbb{E} \left[ \sum_{A_{sc} = 1}  {\gamma}_{sc} (\upsilon_1 + \kappa^\top(W_{sc} - \upsilon_1) - W_{sc})^\top \beta_1 \right] \label{eqn:prop1.9} \\
\nonumber	& = \psi_0^1 + \mathbb{E} \left[ \sum_{A_{sc} = 1}  {\gamma}_{sc} (W_{sc} - \upsilon_1)^\top(\kappa - I)\beta_1 \right] \\
\nonumber	& = \psi_0^1 + \left(\mathbb{E}\left[\sum_{A_{sc} = 1} {\gamma}_{sc} W_{sc}\right] - \upsilon_1\right)^\top(\kappa - I)\beta_1  \\
	& = \psi_0^1 + \left(\bar{X}_0 - \upsilon_1\right)^\top(\kappa - I_q)\beta_1  \label{eqn:prop1.10}
\end{align}
where \eqref{eqn:prop1.4} holds by the assumed linear model for $Y_{sc}$ given by  \eqref{eqn:linmod}; \eqref{eqn:prop1.5} and \eqref{eqn:prop1.10} hold because the SBW algorithm enforces that $\sum \gamma_{sc} W_{sc} = \bar{W}_0$, which has expectation $\bar{X}_0$, and because $\epsilon_{sc}$ is zero-mean and independent of $W_{sc}$ and hence independent of $\gamma_{sc}$; \eqref{eqn:prop1.6} holds by definition of $\psi_0^1$ and the assumed linear model in \eqref{eqn:linmod}; \eqref{eqn:prop1.7} is the tower property of expectations; \eqref{eqn:prop1.8} follows because $\gamma_{sc}$ and $W_{sc}$ are deterministic given $W$; and \eqref{eqn:prop1.9} uses the expression for the conditional expectation given by \eqref{eqn:regcal}. It can be seen that \eqref{eqn:prop1.2} and \eqref{eqn:prop1.10} are equal, and hence show that $\hat{\psi}_0^{1,\textup{ols}}$ and $\hat{\psi}_0^{1, \textup{sbw}}$ have equal bias.

It remains to show that the bias of the OLS estimator is unchanged if the gaussian assumption is relaxed so that \eqref{eqn:regcal} no longer holds. It follows from \eqref{eqn:prop1.beta} that $\hat{\beta}$ is asymptotically given by

    \begin{align*}
    \hat{\beta} &= \left(\sum_{A_{sc}=1} (W_{sc} - \bar{W}_1)(W_{sc} - \bar{W}_1)^\top \right)^{-1} \left(\sum_{A_{sc}=1} (W_{sc} - \bar{W}_1)(Y_{sc} - \bar{Y}_1)^\top\right) \\
     & \rightarrow^p  (S_X + \Sigma_\nu)^{-1}S_X \beta_1 = \check{\kappa} \beta_1
    \end{align*}
where we have used \eqref{eqn:limitW} and \eqref{eqn:limitWY}. Plugging into $\psi_0^{1,\textup{ols}}$ yields

\begin{align*}    
    \mathbb{E}[\hat{\psi}_0]^{1,\textup{ols}} = \mathbb{E}[ \bar{Y}_1 + (\bar{W}_0 - \bar{W}_1)^\top \hat{\beta}] \\
    \rightarrow^p  \bar{\mu}_y + (\bar{X}_0 - \bar{X}_1)\check{\kappa} \beta_1
\end{align*}
from which the result follows by the same steps used to show \eqref{eqn:prop1.2}.

\end{proof}

\begin{proof}[Proof of Proposition \ref{cl2}]
Assuming $\epsilon_{sc} = 0$, by linearity we know that

\begin{equation}\label{eqn:outcomerevised}
Y_{sc} = \alpha_1 + \tilde{X}_{sc}^\top \beta_1 + (X_{sc} - \tilde{X}_{sc})^\top \beta_1 \qquad \forall sc: A_{sc} = 1
\end{equation}

We then have that:

\begin{align}\nonumber
    \hat{\psi}_0^{1,\textup{ideal}} - \psi_0^1 &= \sum_{sc: A_{sc} = 1}\gamma_{sc}^\star Y_{sc} - (\alpha_1 + \bar{X}_0^\top \beta_1) \\
    \nonumber &= \sum_{sc: A_{sc} = 1}\gamma_{sc}^\star\alpha_1 + \sum_{sc: A_{sc} = 1}\gamma_{sc}^\star\tilde{X}_{sc}^\top \beta_1 \\ 
    &+ \sum_{sc: A_{sc} = 1}\gamma_{sc}^\star(X_{sc} - \tilde{X}_{sc})^\top \beta_1 - (\alpha_1 + \bar{X}_0^\top \beta_1) \label{eqn:outcomerevised_proof1}\\
    &= \sum_{sc: A_{sc} = 1}\gamma_{sc}^\star(X_{sc} - \tilde{X}_{sc})^\top \beta_1\label{eqn:sbwregcalerror}
\end{align}
where \eqref{eqn:outcomerevised_proof1} follows from \eqref{eqn:outcomerevised}, and \eqref{eqn:sbwregcalerror} holds since $\sum \gamma_{sc}^\star = 1$ and $\sum \gamma_{sc}^\star \tilde{X}_{sc} = \bar{X}_0$. Conditioned on $W$, it can be seen that $\gamma^*$ is fixed and $X_{sc} - \tilde{X}_{sc}$ has expectation zero; therefore, \eqref{eqn:sbwregcalerror} implies that the estimator is unbiased.
\end{proof}

\begin{proof}[Proof of Proposition \ref{prop:variance_rate}] 
To derive $\operatorname{Var}\left(\hat{\psi}_0^{1,\textup{ideal}} | W\right)$, we use

\begin{align}
\operatorname{Var}\left(\hat{\psi}_0^{1,\textup{ideal}} - \psi_0^1 | W\right) &= \operatorname{Var}\left[\sum_{sc: A_{sc} = 1}\gamma_{sc}^\star(X_{sc} - \tilde{X}_{sc})^\top\beta_1 \mid W\right] \label{eqn:prop:variance.1}\\
 &= \sum_{sc: A_{sc} = 1} \operatorname{Var}(\gamma_{sc}^\star(X_{sc} - \tilde{X}_{sc})^\top\beta_1 \mid W) \label{eqn:prop:variance.2}\\
 &= \sum_{sc: A_{sc} = 1} \gamma_{sc}^{\star^2}\beta_1^\top (\Sigma_{X} - \Sigma_{X}\Sigma_{W}^{-1}\Sigma_{X})\beta_1  \label{eqn:prop:variance.3}\\
& = \|\gamma^*\|^2 \cdot \beta_1^\top (\Sigma_{X} - \Sigma_{X}\Sigma_{W}^{-1}\Sigma_{X})\beta_1 \label{eqn:variance}
\end{align}
where \eqref{eqn:prop:variance.1} follows from \eqref{eqn:sbwregcalerror}, \eqref{eqn:prop:variance.2} holds because the tuples $(X_{sc}, W_{sc})$ are i.i.d, and \eqref{eqn:prop:variance.3} holds because $\gamma_{sc}^*$ is fixed given $W$ and $(X_{sc}, W_{sc})$ are jointly normal. 

To upper bound the conditional variance given by \eqref{eqn:variance}, we will construct a feasible solution $\gamma'$ to the SBW objective over the constraint set $\Gamma(\tilde{X}, \bar{X}_0, 0)$ such that $\|\gamma'\|^2 = O_P(n_1^{-1})$. As the optimal solution $\gamma^*$ satisfies $\|\gamma^*\|^2 \leq \|\gamma'\|^2$, the result follows.

Our construction is the following. Divide the $n_1$ treated units into $L = \lfloor n_1/n^{\text{sub}} \rfloor$ subsets of size $n^{\text{sub}}$, and a remainder subset. For the subsets $\ell=1,\ldots,L$, let $X^{(\ell)}$ denote its covariates, $\tilde{X}^{(\ell)}$ the conditional expectation $\mathbb{E}[X^{(\ell)}|W, A]$, and  $\gamma^{(\ell)}$ the solution to the SBW objective over the constraint set $\Gamma(\tilde{X}^{(\ell)}, \bar{X}_0, 0)$, with $\gamma^{(\ell)}=0$ if the constraint set is infeasible. As the units are assumed to be i.i.d., it follows that $\gamma^{(1)}, \ldots, \gamma^{(L)}$ are also i.i.d. Let $n^{\text{sub}}$ be large enough so that each $\gamma^{(\ell)}$ has positive probability of being non-zero. 

Let $L'$ denote the number of subsets whose $\gamma^{(\ell)}$ is non-zero. As each non-zero weight vector $\gamma^{(\ell)}$ is feasible for $\Gamma(\tilde{X}^{(\ell)}, \bar{X}_0, 0)$, it can be seen that the concatenated vector $\gamma' = (\gamma^{(1)}/L', \ldots, \gamma^{(L)}/L', 0)$ is feasible for $\Gamma(\tilde{X},\bar{X}_0,0)$. As the weights $\gamma^{(\ell)}$ are i.i.d, it follows that $\| \gamma'\|^2$ which equals $\frac{1}{(L')^2} \sum_\ell \|\gamma_\ell\|^2$  converges in probability to $\frac{1}{L'} \mathbb{E}\|\gamma^{(1)}\|^2 = O_P(n_1^{-1})$, proving the bound on $\operatorname{Var}\left(\hat{\psi}_0^{1,\textup{ideal}} - \psi_0^1 | W\right)$.

To show this rate also holds for $\operatorname{Var}\left(\hat{\psi}_0^{1,\textup{ideal}}\right)$, we can apply the law of total variance to $f = \hat{\psi}_0^{1,\textup{ideal}} - \psi_0^1$, %= \hat{\psi}_0^{1,\textup{ideal}},$
\[ \operatorname{Var}(f) = \underbrace{\mathbb{E}[\operatorname{Var}(f|W)]}_{(i)} + \underbrace{\operatorname{Var}(\mathbb{E}[f|W])}_{(ii)} \]
observing that $\mathbb{E}[f|W] = 0$ by Proposition \ref{cl2}, so that (ii) is zero. To show that term (i) is $O(n_1^{-1})$, we observe that as $\operatorname{Var}(f|W) = O_P(n_1^{-1})$ and is bounded (since $\gamma^*$ is non-negative and sums to 1), it follows that  $\mathbb{E}[ \operatorname{Var}(f|W)]$ must be $O(n_1^{-1})$ as well.
\end{proof}

\begin{proof}[Proof of Proposition \ref{cl3}]
Following Proposition~\ref{cl2}, assuming $\epsilon_{sc}=0$ we can decompose the error of the estimator as follows:

\begin{align}
\nonumber    \hat{\psi}^{1,\textup{adjusted}}_0 - \psi_0^1 &= \sum_{A_{sc}=1} \hat{\gamma}_{sc} Y_{sc} - \psi_0^1 \\
    & = \sum_{A_{sc}=1} \hat{\gamma}_{sc} (\alpha_1 + X_{sc}^\top \beta_1) - \psi_0^1 \label{eqn:cl3.1}\\
    \nonumber & = \sum_{A_{sc}=1} \hat{\gamma}_{sc} (\alpha_1 + \hat{X}_{sc}^\top \beta_1 + (X_{sc} - \hat{X}_{sc})^\top \beta_1 ) - \psi_0^1 \\
    \nonumber & = \alpha_1 + \sum_{A_{sc}=1} \hat{\gamma}_{sc} \hat{X}_{sc}^\top \beta_1 + \sum_{A_{sc}=1} \hat{\gamma}_{sc}(X_{sc} - \hat{X}_{sc})^\top \beta_1  - \psi_0^1 \\
    & = \alpha_1 + \bar{W}_0^\top \beta_1 + \sum_{A_{sc}=1} \hat{\gamma}_{sc}(X_{sc} - \hat{X}_{sc})^\top \beta_1  - \psi_0^1 \label{eqn:cl3.4}\\
    & = \underbrace{(\bar{X}_0 - \bar{W}_0)^\top \beta_1}_{(i)} + \underbrace{\sum_{A_{sc}=1} \hat{\gamma}_{sc}(X_{sc} - \tilde{X}_{sc})^\top \beta_1}_{(ii)} + \underbrace{\sum_{A_{sc}=1} \hat{\gamma}_{sc} (\tilde{X}_{sc} - \hat{X}_{sc})^\top \beta_1}_{(iii)} \label{eqn:cl3.5}
\end{align}

where \eqref{eqn:cl3.1} holds by \eqref{eqn:linmod}, \eqref{eqn:cl3.4} uses that $\sum \hat{\gamma}_{sc} \hat{X}_{sc} = \bar{W}_0$, and \eqref{eqn:cl3.5} uses that $\psi_{0}^1 = \alpha_1 + \bar{X}_0^\top \beta_1$.

We observe that term (i) goes to zero by the law of large numbers. Term (iii) goes to zero because $\|\hat{\gamma}\| \leq 1$ (since $\hat{\gamma} \geq 0$ and sums to 1), and because $\hat{X}$ converges to $\tilde{X}$ uniformly over all units as $\hat{\Sigma}_X$ and $\hat{\Sigma}_{\nu}$ converge. 

To show that (ii) goes to zero, we will show that conditioned on $W$, (ii) is zero mean and has variance going to zero. Conditioned on $W$, $\hat{\gamma}$ is fixed; this implies that conditioned on $W$, term (ii) is zero-mean, and has conditional variance 

\[ \|\hat{\gamma}\|^2 \cdot \beta_1^\top (\Sigma_X - \Sigma_X\Sigma_W^{-1}\Sigma_X) \beta_1\]
By a construction argument identical to one used in the proof of Proposition \ref{prop:variance_rate}, it can be shown that $\|\hat{\gamma}\|^2\rightarrow 0$, implying that the conditional variance goes to zero as well.
\end{proof}

\begin{proof}[Proof of Proposition \ref{prop:jackknife}]
Let $U_s = \{(J_{sc}, W_{sc}): c = 1,\ldots,p_s\}$ denote the observed outcomes and covariates corresponding to the CPUMAs in state $s$. Under our assumptions, it holds that $U_s$ is i.i.d. for the treated states. It can also be seen that $\hat{\psi}_0^{1,\textup{adjusted}}$ is a symmetric function of the treated state observations $\{U_s: A_s=1\}$. It follows that equation (1.6) of  \cite{efron1981jackknife} can be seen to apply to our setting; as this equation is equal to (\ref{eqn:prop.jackknife}), this proves the result.
\end{proof}

\begin{proof}[Proof of Proposition \ref{cl8}]
Let $\mu$ denote 
\[ \mu = \frac{1}{n_1} \sum_{A_{sc} = 1} \check{X}_{sc}\]
so that
\[ \check{X}_{sc} - \mu = \check{\kappa}^\top(W_{sc} - \bar{W}_1)\]
and hence that 
\begin{align}
 \nonumber \check{\beta} &=  \left(\sum_{A_{sc}=1} (\check{X}_{sc} - \mu)(\check{X}_{sc} - \mu)^\top\right)^{-1} \sum_{A_{sc}=1} (\check{X}_{sc} - \mu)(Y_{sc} - \bar{Y}_1) \\
 \nonumber & = \left(\sum_{A_{sc}=1} \check{\kappa}^\top (W_{sc} - \bar{W}_1)(W_{sc} - \bar{W}_1)^\top \check{\kappa}\right)^{-1} \sum_{A_{sc}=1} \check{\kappa}^\top(W_{sc} - \bar{W}_1)(Y_{sc} - \bar{Y}_1) \\
 & \to^p (\check{\kappa}^\top (S_X + \Sigma_\nu) \check{\kappa})^{-1} \check{\kappa}^\top S_X \beta_1  \label{eqn:cl8.1}\\
 \nonumber 
 & = \check{\kappa}^{-1} (S_X + \Sigma_{\nu})^{-1} S_X \beta_1 \\
 \nonumber 
 & = \beta_1
 \end{align}
 where \eqref{eqn:cl8.1} follows by \eqref{eqn:limitW} and \eqref{eqn:limitWY}, and the last step follows from the definition of $\check{\kappa}$. It then follows that 
 \[ \bar{Y}_1 - (\bar{W}_0 - \bar{W}_1)^\top \check{\beta} \to^p \alpha_1 + \bar{X}_0^\top \beta_1 = \psi_0^1\]
 proving consistency.
\end{proof}

\begin{proof}[Proof of Proposition \ref{cl9}]

We will use Proposition \ref{cl56} which is proved later in this section. To apply it, we let $\Omega = I$, $Z = \check{X}_{A=1}$, and $v = \bar{W}_0$. Using  $\check{X}_{sc} = \bar{W}_1 + \check{\kappa}^\top(W_{sc} - \bar{W}_1)$, we find that 
\begin{align}
    \nonumber \mu & = \frac{1}{n_Q} \sum_{sc \in Q} \check{X}_{sc} \\
    \label{eqn:cl9.mu} & = \bar{W}_1 + \check{\kappa}^\top(\bar{W}_Q - \bar{W}_1)
\end{align}
and hence that $\check{X}_{sc} - \mu = \check{\kappa}^\top(W_{sc} - \bar{W}_Q)$. Plugging into \eqref{eqn:a1.3} yields 

\begin{align}
 \nonumber \check{\gamma}_{sc} & = \frac{1}{n_Q}(W_{sc} - \bar{W}_Q)^\top \check{\kappa} (\check{\kappa}^\top S_{W_Q} \check{\kappa})^{-1}(\bar{W}_0 - \mu) + \frac{1}{n_Q} \\
 \label{eqn:cl9.proof.gamma}& = \frac{1}{n_Q}(W_{sc} - \bar{W}_Q)^\top S_{W_Q}^{-1} \check{\kappa}^{-\top}(\bar{W}_0 - \mu) + \frac{1}{n_Q} \qquad \forall \ sc \in Q
\end{align}
As $Y_{sc} = \alpha_1 + \beta_1^\top (\bar{X}_Q + X_{sc} - \bar{X}_Q)$ for the treated units, the SBW estimator of $\psi_0^1$ can be seen to equal
\begin{align}
    \nonumber \sum_{A_{sc}=1} Y_{sc}\check{\gamma}_{sc} & = \sum_{A_{sc}=1} (\alpha_1 + \beta_1^\top \bar{X}_Q) \check{\gamma}_{sc} + \sum_{A_{sc}=1}  \beta_1^\top (X_{sc} - \bar{X}_Q)\check{\gamma}_{sc} \\
 \label{eqn:cl9.proof1}    & = (\alpha_1 + \beta_1^\top \bar{X}_Q)\\
    \nonumber & \hskip.5cm {} + \sum_{sc \in Q}  \beta_1^\top (X_{sc} - \bar{X}_Q)\left[\frac{1}{n_Q} (W_{sc} - \bar{W}_Q)^\top S_{W_Q}^{-1} \check{\kappa}^{-\top}(\bar{W}_0 - \mu) +  \frac{1}{n_Q}\right] \\
\label{eqn:cl9.proof2}    & = (\alpha_1 + \beta_1^\top \bar{X}_Q)\\
    \nonumber & \hskip.5cm {} +  \beta_1^\top S_{XW_Q} S_{W_Q}^{-1} \check{\kappa}^{-\top}(\bar{W}_0 - \mu) + \underbrace{\frac{1}{n_Q}\sum_{sc \in Q}   \beta_1^\top (X_{sc} - \bar{X}_Q) }_{= 0} \\ 
\label{eqn:cl9.proof3}    & = \alpha_1 + \beta_1^\top \bar{X}_0 - \underbrace{(\beta_1^\top \bar{X}_0 -  \beta_1^\top S_{XW_Q} S_{W_Q}^{-1} \check{\kappa}^{-\top}\bar{W}_0)}_{(i)} \\
    \nonumber & \hskip.5cm {} + \underbrace{\beta_1^\top (\bar{X}_Q - S_{XW_Q} S_{W_Q}^{-1} \check{\kappa}^{-\top}\bar{W}_1)}_{(ii)} - \underbrace{\beta_1^\top S_{X_Q}S_{W_Q}^{-1}(\bar{W}_Q - \bar{W}_1))}_{(iii)}\\
\label{eqn:cl9.proof4}    & \to^p \psi_0^1 - \underbrace{\beta_1^\top (I - S_{XW_Q}S_{W_Q}^{-1}S_WS_X^{-1})\bar{X}_0}_{(i)} \\
    \nonumber & \hskip.5cm {} + \underbrace{\beta_1^\top (\bar{X}_Q - S_{XW_Q}S_{W_Q}^{-1} S_W S_X^{-1} \bar{X}_1)}_{(ii)} - \underbrace{\beta_1^\top S_{XW_Q}S_{W_Q}^{-1}(\bar{X}_Q - \bar{X}_1)}_{(iii)}
\end{align}
where \eqref{eqn:cl9.proof1} uses the expression for $\check{\gamma}$ given by (\ref{eqn:cl9.proof.gamma}; \eqref{eqn:cl9.proof2} follows by algebraic manipulations, and notes that $n_Q^{-1}\sum_{sc \in Q} (X_{sc} - \bar{X}_Q) = 0$; \eqref{eqn:cl9.proof3} adds and subtracts $\beta_1^\top \bar{X}_0$,  substitutes for $\mu$ using (\ref{eqn:cl9.mu}), and groups the terms into (i), (ii), and (iii); and \eqref{eqn:cl9.proof4} substitutes for $\check{\kappa}$ and uses $\bar{W}_0 \to^p \bar{X}_0$ and $\bar{W}_1 \to^p \bar{X}_1$.

It can be seen that terms (i), (ii), and (iii) each go to zero if $S_{XW_Q} \to S_X$, $S_{W_Q} \to S_W$, and $\bar{X}_Q \to \bar{X}_1$, proving the result. 
\end{proof}

\begin{proof}[Proof of Propositions \ref{cl7} and \ref{cl7hsbw}]
    We prove the result for H-SBW only, as letting $\rho=0$ includes SBW as a special case. Following the derivation of \eqref{eqn:sbwregcalerror} with the H-SBW weights $\tilde{\gamma}^{\textup{hsbw}*}$ in place of $\gamma^*$, it can be shown that: 
    
    \begin{align*}
        \sum_{A_{sc}=1} Y_{sc} \tilde{\gamma}_{sc}^{\textup{hsbw}*} - \psi^1_0 = \sum_{sc: A_{sc} = 1}\tilde{\gamma}^{\textup{hsbw}*}_{sc}(X_{sc} - \tilde{X}_{sc}^\star)^\top \beta_1
    \end{align*}
Conditional on $W$ (and assuming that the correspondence between states and CPUMAs is known), $\tilde{\gamma}_{sc}^{\textup{hsbw}*}$ is fixed and $X_{sc} - \tilde{X}_{sc}^*$ equals $X_{sc} - \mathbb{E}[X_{sc}|W, A=1]$ which has mean zero, proving the result.
\end{proof}

\begin{proof}[Proof of Proposition \ref{cl4}]
\begin{align*}
    \operatorname{Var}\left( n_t^{-1}\sum_{s: A_s = 1}\sum_{c = 1}^{p_s}\gamma_{sc}Y_{sc} \mid X, A\right) &= n_t^{-2}\sum_{s: A_s = 1}\sum_{c = 1}^{p_s}\gamma_{sc}^2(\sigma^2_{\epsilon} + \sigma^2_{\varepsilon}) + \sum_{c \ne d}\gamma_{sc}\gamma_{sd}\sigma^2_{\varepsilon} \\
    &\propto \sum_{s: A_s = 1}\sum_{c = 1}^{p_s}\gamma_{sc}^2 + \sum_{c \ne d}\rho \gamma_{sc}\gamma_{sd}
\end{align*}
where the second line follows by dividing by $\sigma^2_{\epsilon} + \sigma^2_{\varepsilon}$. By definition of the H-SBW objective, which minimizes this function for known $\rho$, the H-SBW estimator must produce the minimum conditional-on-X variance estimator within the constraint set.
\end{proof}

\begin{proof}[Proof of Proposition \ref{cl56}]

    To show that $\gamma^*$ solves (\ref{eqn:a1.2}), we first observe that it is a feasible solution, by definition of $Q$. The result can then be proven by contradiction: if $\gamma^*$ is feasible but does not solve (\ref{eqn:a1.2}), then a feasible $\tilde{\gamma}$ must exist with lower objective value. Then for some convex combination $\gamma_\lambda = \lambda \tilde{\gamma} + (1-\lambda)\gamma^*$ with $\lambda > 0$, we can show that that $\gamma_\lambda$ is both feasible for (\ref{eqn:a1.1}), and has lower objective value than $\gamma^*$:
    \begin{enumerate}
        \item     To establish that $\gamma_\lambda$ is feasible, we observe that $\tilde{\gamma}$ is feasible for (\ref{eqn:a1.2}). This implies that if $\gamma^*_i=0$, then $\tilde{\gamma}_i=0$ as well; as a result, there exists $\lambda > 0$ such that the convex combination $\gamma_\lambda$ satisfies $\gamma_\lambda \geq 0$ and hence is feasible for (\ref{eqn:a1.1}).
    \item     To show that this $\gamma_\lambda$ has lower objective value than $\gamma^*$, we observe that if $\tilde{\gamma}$ has lower objective value than $\gamma^*$, then by strict convexity of the objective any convex combination with $\lambda > 0$ must have lower objective value than $\gamma^*$ as well.
    \end{enumerate}
    This shows that if $\gamma^*$ is not optimal for (\ref{eqn:a1.2}), then it is not optimal for (\ref{eqn:a1.1}) either. But as $\gamma^*$ is the optimal solution to (\ref{eqn:a1.1}), this is a contradiction; hence by taking the contrapositive it follows that $\gamma^*$ must solve (\ref{eqn:a1.2}).
    
To show \eqref{eqn:a1.3}, we observe that \eqref{eqn:a1.2} can be written as

\[ \min_{\gamma_Q} \gamma_Q^\top \Omega_Q \gamma_Q \quad \text{subject to} \quad Z_Q \gamma_Q = v \ \text{ and } \ {\bf 1}^\top \gamma_Q = 1\] 
which can be rewritten as

\[ \min_{\gamma_Q} \gamma_Q^\top \Omega_Q \gamma_Q \quad \text{subject to} \quad \left[ \begin{array}{cr} Z_Q - \mu {\bf 1}^\top\\ {\bf 1}^\top \end{array}\right] \gamma_Q = \left[\begin{array}{cr} v - \mu & 1 \end{array}\right]\] 
where we have subtracted $\mu{\bf 1}^\top \gamma_Q$ (which equals $\mu$) from both sides of the constraint. This is a least norm problem, and when feasible has solution

\begin{equation}\label{eq:a1.least_norm}
 \gamma_Q^* = \Omega^{-1}_Q A^\top (A\Omega^{-1}_QA^\top)^{-1} b
\end{equation}
where $A = \left[ \begin{array}{cr} Z_Q - \mu {\bf 1}^\top\\ {\bf 1}^\top \end{array}\right]$ and $b = \left[\begin{array}{cr} v - \mu & 1 \end{array}\right]$. As $(Z_{Q} - \mu{\bf 1}^\top)\Omega_Q^{-1} {\bf 1} = 0$, it follows that $A\Omega^{-1}A^\top$ is block diagonal

\[ A\Omega_Q^{-1}A^\top = \left[\begin{array}{cc} (Z_Q - \mu)\Omega_Q^{-1} (Z_Q- \mu)^\top & 0  \\ 0 & {\bf 1}^\top \Omega^{-1}_Q{\bf 1}\end{array}\right] \]
so that plugging into \eqref{eq:a1.least_norm} yields
 \begin{equation*}
 \gamma^*_{Q} = \Omega_{Q}^{-1} (Z_{Q} - \mu)^\top\left[ (Z_Q - \mu) \Omega_{Q}^{-1} (Z_Q - \mu)\right]^{-1} (v - \mu) + \frac{\Omega^{-1}_Q {\bf 1} }{{\bf 1}^\top \Omega^{-1}_Q {\bf 1}}
 \end{equation*}

This is equivalent to the implied regression weights for the estimate of $\beta^\top b$ using estimate of $\beta$ solving the GLS problem on the set $Q$ given an outcome vector $Y$, 
\begin{align*}
    \min_{\beta} (Y_Q - A^\top \beta)\Omega_Q^{-1}(Y_Q - A^\top \beta)^\top
\end{align*}
where $\hat{\beta} = (A\Omega_Q^{-1}A^\top)^{-1}A\Omega_Q^{-1}Y_Q$, proving the result.

%\clearpage
\end{proof}

\clearpage

\section{Covariate and Adjustment Details}\label{app:adjustmentdetails}

In this section we detail our estimation of the observed covariates $W$ and our adjusted covariates $\hat{X}$. 

\subsection{Unadjusted covariate estimates}

To estimate our CPUMA-level covariates $W$ using the ACS microdata, we require estimating both numerator and denominator counts given that we are ultimately interested in calculating rates. For each CPUMA we estimate: the total non-elderly adult population for each year 2011-2014; the total labor force population (among non-elderly adults) for each year 2011-2013; and the total number of households averaged from 2011-2013. We also construct an average of the total non-elderly adult population from 2011-2013. These are our denominator variables. For our numerator counts, we estimate the total number of: females; White individuals; people of Hispanic ethnicity; people born outside of the United States; citizens; people with disabilities; married individuals; students; people with less than a high school education, high school degrees, some college, or college graduates or higher; people living under 138 percent of the FPL, between 139 and 299 percent, 300 and 499 percent, more than 500 percent, and who did not respond to the income survey question; people aged 19-29, 30-39, 40-49, 50-64; households with one, two, or three or more children, and households that did not respond about the number of children. We average these estimated counts from 2011-2013. For each individual year from 2011-2013, we then estimate the total number of people who were unemployed and uninsured at the time of the survey (calculated among all non-elderly adults and all non-elderly adults within the labor force, respectively). We divide the numerator totals by the corresponding denominator totals to estimate the percentage in each category. For the demographics, these include the average number of non-elderly adults from 2011-2013. For the time-varying variables, we use the corresponding year (where uninsurance rates are calculated as a fraction of the labor force rather than the non-elderly adult population). We also calculate the average non-elderly adult population growth and the average number of households to adults across 2011-2013. All estimates are calculated using the associated set of survey weights provided in the public use microdata files.

\subsection{Covariate adjustment}

We provide additional details about estimating $\tilde{X}$. We begin by estimating the unpooled unit-level covariance matrices $\Sigma_{\nu, sc}^{\text{raw}}$, the sampling variability for each CPUMA among the treated units, by using the individual replicate survey weights to generate $b = 80$ additional CPUMA-level datasets. We then compute:

\begin{equation}
\hat{\Sigma}_{\nu, sc}^{\text{raw}} = \frac{4}{80}\sum_{b=1}^{80}(W_{b, sc} - \bar{W}_{sc})(W_{b, sc} - \bar{W}_{sc})^\top
\end{equation}
where the $4$ in the numerator comes from the process used to generate the replicate survey weights and $\bar{W}_{sc}$ is the vector of covariate values estimated using the original ACS weights.

We let $\hat{\Sigma}_{W} = n_1^{-1}\sum_{sc: A_s = 1} (W_{sc} - \bar{W}_1)(W_{sc} - \bar{W}_1)^\top$. This estimate is calculated on the original observed dataset. We then estimate $\Sigma_{X}$ using:

\begin{equation}
\hat{\Sigma}_X = \hat{\Sigma}_W - n_1^{-1}\sum_{sc: A_{sc} = 1} \hat{\Sigma}_{\nu, sc}^{\text{raw}}
\end{equation}

Define

\begin{equation}
\hat{\kappa} = \hat{\Sigma}_W^{-1}\hat{\Sigma}_X
\end{equation}

Notice that $\hat{\kappa}$ is a matrix of estimated coefficients of a linear regressions of the (unobserved) matrix $X_{sc}$ on (observed) matrix $W_{sc}$. We can then estimate $\mathbb{E}[X_{sc} \mid W_{sc}, A_{sc} = 1]$ using: 

\begin{equation}\label{eqn:homogeneousadjustment}
\hat{X}_{sc} = \hat{\mathbb{E}}[X_{sc} \mid W_{sc}, A_{sc} = 1] = \bar{W}_1 + \hat{\kappa}^\top(W_{sc} - \bar{W}_1) \forall sc: A_{sc} = 1
\end{equation}

We call this the ``homogeneous adjustment'' and denote the corresponding set of units in \eqref{eqn:homogeneousadjustment} $\hat{X}_{A=1}^{hom}$. This adjustment approximately aligns with the adjustments suggested by \cite{carroll2006measurement} and \cite{gleser1992importance}. This estimator for $\tilde{X}$ is consistent for $\mathbb{E}[X \mid W, A]$ if we assume, for example, that the measurement errors are homoskedastic (see, e.g., $\Sigma_{\nu, sc} = \Sigma_{\nu}$).\footnote{We can theoretically weaken this assumption to still obtain a consistent estimate (see, e.g., \cite{buonaccorsi2010measurement}).}

To potentially improve upon this procedure, we also consider a second estimate that we call the ``heterogeneous adjustment.'' This adjustment accounts for the fact that some regions with large populations are estimated quite precisely, while regions with small populations are estimated much less precisely (additionally, for a given CPUMA, some covariates are measured using three years of data, and others only one). For this adjustment we model an individual-level $\Sigma_{\nu, sc}$ as a function of the sample sizes used to estimate each covariate. 

Specifically, let $s_{sc}$ be the q-dimensional vector of the sample sizes used to estimate each covariate value for a given CPUMA. Let $\odot$ reflect the Hadamard product, and $\oslash$ reflect Hadamard division. We assume that $\sqrt{s_{sc}} \odot \nu_{sc} \mid A_{sc} = 1 \sim N(0, \Sigma_{\nu})$. Let $S_{sc} = \sqrt{s_{sc}}\sqrt{s_{sc}}^\top$. We then know that $\Sigma_{\nu, sc} = \Sigma_{\nu} \oslash S_{sc}$.

To estimate $\Sigma_{\nu, sc}$, we first pool our initial estimates of the CPUMA-level covariance matrices $\hat{\Sigma}_{\nu, sc}^{\text{raw}}$ to generate $\hat{\Sigma}_{\nu} = n_1^{-1}\sum_{sc: A_s = 1} \matr{S}_{sc} \circ \hat{\Sigma}_{\nu, sc}^{\text{raw}}$. We estimate $\hat{\Sigma}_{\nu, sc} = \hat{\Sigma}_{\nu} \oslash S_{sc}$. Using the same estimate of $\hat{\Sigma}_X$ as before, we calculate $\hat{\kappa}_{sc} = (\hat{\Sigma}_{X} + \hat{\Sigma}_{\nu, sc})^{-1}\hat{\Sigma}_X$, which we use to estimate the heterogeneous adjustment. We call the corresponding set of estimates that use this adjustment $\hat{X}_{A=1}^{het}$.

This adjustment accounts for CPUMA-level variability in the measurement error, and should more greatly affect outlying values of imprecisely estimated covariates, while leaving precisely estimated covariates closer to their observed value in the dataset. Moreover, unlike using the original individual-level CPUMA estimates $\hat{\Sigma}_{\nu, sc}^{\text{raw}}$, we are able to use the full efficiency of using all units in the modeling. On the other hand, this model assumes that all differences in the sampling variability are due to the sample sizes, and assumes away heterogeneity due to heteroskedasticity. In our simulation results in Appendix~\ref{app:simstudy}, we also find that this estimator leads to bias if the measurement errors are in truth homoskedastic. 

As a final adjustment we can also build on the homogeneous adjustment to account for the correlation structure of the data. To motivate this adjustment we view the entire data among the treated states $X = (X_{11}, ..., X_{1p_1}, ..., X_{sp_s})$ as reflecting a single draw from the distribution $MVN(\boldsymbol{\upsilon}_1, \mathbf{\Sigma}_X)$, where $\boldsymbol{\upsilon}_1$ is a $qn_1$ dimensional vector that repeats the $q$ dimensional vector of means $\upsilon_1$ (defined previously) $n_1$ times. Conditional on $X$, $W = (W_{11}, ..., W_{1p_1}, ..., W_{sp_s})$ reflects a single draw from the distribution $MVN(X, \boldsymbol{\Sigma}_{\nu})$, where $\boldsymbol{\Sigma}_{\nu}$ is a diagonal matrix with $\Sigma_{\nu}$ in the diagonals. That is, we assume that the measurement errors are drawn independently across units with constant covariance matrix $\Sigma_{\nu}$.

We then let $\Sigma_B$ represent the covariance between CPUMAs that share a state, which we assume is constant across all CPUMAs and states. 

Let $\boldsymbol{\Sigma}_W = \boldsymbol{\Sigma}_X + \boldsymbol{\Sigma}_{\nu}$. $(X, W)$ are then jointly normal with common mean $\boldsymbol{\upsilon}_1$ and covariance matrix $\boldsymbol{\Sigma}$: 

\begin{align}\label{eqn:xcormodel}
\boldsymbol{\Sigma} &= \begin{pmatrix}
\mathbf{\Sigma}_X & \mathbf{\Sigma}_X \\
\mathbf{\Sigma}_X & \mathbf{\Sigma}_W
\end{pmatrix} 
\end{align}

Notice that $\mathbf{\Sigma}_X$ and $\mathbf{\Sigma}_W$ are block-diagonal matrices with $qp_s$ by $qp_s$ blocks indexed by $s$ that we denote $\Sigma_{X, s}$ and $\Sigma_{W, s}$. Each element of these matrices contains the $q$ by $q$ matrices defined previously: 

\begin{align}\label{eqn:xcormodel2}
\Sigma_{X, s} &= \begin{pmatrix}
\Sigma_{X}, \Sigma_B, ..., \Sigma_B \\
\Sigma_B, \Sigma_{X}, ..., \Sigma_B \\
... \\
\Sigma_B, \Sigma_B, ..., \Sigma_{X}
\end{pmatrix};
\Sigma_{W, s} = \begin{pmatrix}
\Sigma_{W}, \Sigma_B, ..., \Sigma_B \\
\Sigma_B, \Sigma_{W}, ..., \Sigma_B \\
... \\
\Sigma_B, \Sigma_B, ..., \Sigma_{W} \\
\end{pmatrix}
\end{align}
where we have defined $\Sigma_{X}$, $\Sigma_{W}$, and $\Sigma_B$ previously. 
Let $\boldsymbol{\kappa} = \mathbf{\Sigma_W}^{-1}\mathbf{\Sigma_X}$. We then have that by the conditional normal distribution:

\begin{align}\label{eqn:xcoradjust}
    \mathbb{E}[X \mid W, A = 1] &=  \boldsymbol{\upsilon}_1 + \boldsymbol{\kappa}^\top(W - \boldsymbol{\upsilon}_1)  
\end{align}

To estimate this quantity we need only additionally generate an estimate of $\Sigma_B$ in addition to the previous estimates. We first generate state-specific estimates $\hat{\Sigma}_{B, s}= \frac{2}{p_s(p_s - 1)}\sum_{c < d}(W_{sc} - \bar{W})(W_{sd} - \bar{W})^\top$ and then generate $\hat{\Sigma}_B$ as an average of these estimates. We then replace the other quantities by their empirical counterparts to generate the adjustment set $\hat{X}_{A=1}^{cor}$.

We do not use this adjustment for our application: in practice we find that it adds substantial variability to the observed data, predicting values that frequently fall outside of the support of the original dataset. In simulations in Appendix~\ref{app:simstudy} we find that this adjustment adds quite a bit of variability to the final estimates given a $m_1 = 25$ states. 

\clearpage

\section{Summary Statistics and Covariates}\label{app:sumstats}

In this section we display summary statistics about the CPUMA-level datasets. The first two tables pertain to treatment assignment classifications. Table~\ref{tab:txassign} lists the states that are assigned to each group: the first two columns include the treatment states and control states in our primary analysis. The third column lists the treatment states for our sensitivity analysis that excludes ``early expansion'' states. The final column indicates states that were always excluded from the analysis. Table~\ref{tab:cpumasperstate} displays the total number of CPUMAs per state, as well as a column reiterating the state's treatment assignment and whether it was an early expansion state.

The subsequent tables and figure display summary information about the expansion state data and the homogeneous and heterogeneous covariate adjustments detailed in Appendix~\ref{app:adjustmentdetails}. 

Table~\ref{tab:summarytab1} displays univariate summary statistics for the treated CPUMAs. Specifically, the table displays mean, interquartile range, and the range (as defined by the maximum value minus the minimum value) for the unadjusted dataset, the heterogeneous adjustment, and the homogeneous adjustment. We see that the covariate adjustments generally reduce the variability relative to the unadjusted data. 

Table~\ref{tab:extreme1} displays the frequency that the adjusted covariates fell outside of the support of the unadjusted dataset on our primary dataset. The frequency is comparable for either adjustment and the counts are low, supporting the use of the linear model outlined in \eqref{eqn:regcal}. We also calculate these adjustments excluding early expansion states, and recalculate these adjustments excluding each state one at a time to calculate our variance estimates, yielding different results with respect to the quality of the resulting adjustments. These results are available on request.

Table~\ref{tab:timetrends} displays the trends in the outcome over time by treatment group. We use these estimates to compute the difference-in-differences estimator of the ETT in footnote \ref{footnote_did}. 

\begin{figure}[h!]
\begin{center}
    \caption{Correlation matrix: full data, unadjusted covariates \newline (primary dataset)}
    \label{fig:corrmatrix}
    \includegraphics[scale=0.25]{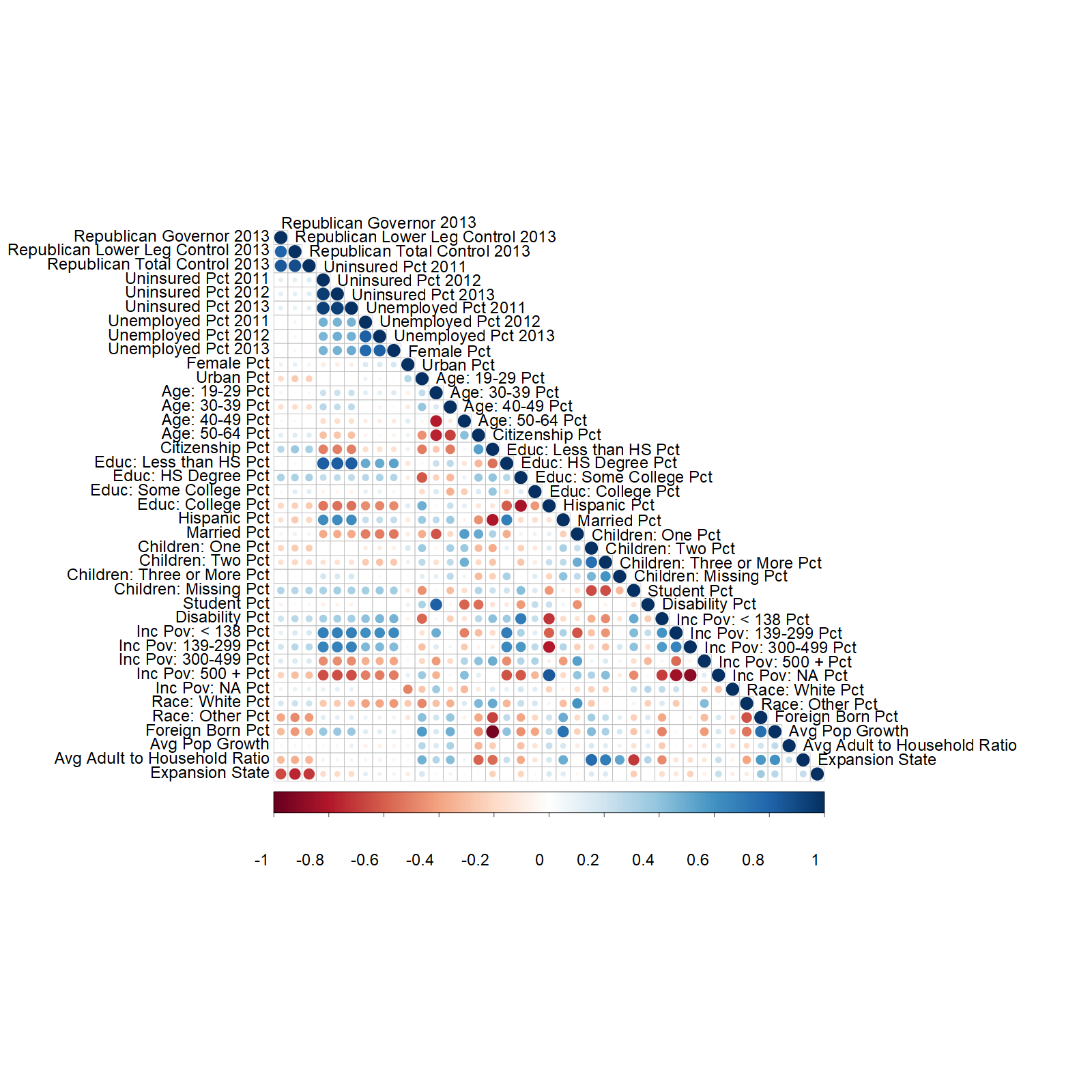}
\end{center}
\end{figure}

Figure~\ref{fig:corrmatrix} displays the Pearson's correlation coefficients for the bivariate relationships between the covariates on the unadjusted dataset (including both treated and untreated units). These point estimates may be biased due to the measurement error in the covariates. Nevertheless, this matrix is useful for at least two reasons: first, assuming the correlations among the treated and untreated units are similar, the more heavily correlated the data the easier it should be to attain covariate balance (see, e.g., \cite{d2021overlap}). This matrix gives a general sense of how correlated the data are, even if the estimates are biased. Second, these correlations can suggest potential confounders by revealing which variables are most heavily associated with treatment assignment and the pre-treatment outcomes. For example, the plot shows a strong association between Republican governance and treatment assignment, and a smaller association between these variables with pre-treatment outcomes. The plot also illustrates strong associations between the pre-treatment uninsurance rates, though they are more weakly associated with treatment assignment. 

\begin{table}[h!]\caption{Treatment assignment classification}\label{tab:txassign}
\centering
%\hline 
\begin{tabularx}{\textwidth}{XXXX} \\ 
Treated states & Control states & Early expansion states & Always excluded \\ 
\hline
AR, AZ, CA, CO, CT, HI, IA, IL, KY, MD, MI$^\textrm{a}$, MN, ND, NJ, NM, NV, OH, OR, RI, WA, WV & AK, AL, FL, GA, ID, IN, KS, LA, ME, MO, MS, MT, NC, NE, OK, PA, SC, SD, TN, TX, UT, VA, WI, WY & CA, CT, MN, NJ, WA & DE$^\textrm{c}$, MA$^\textrm{c}$, NH$^\textrm{b}$, NY$^\textrm{c}$, VT$^\textrm{c}$, DC$^\textrm{c}$\\ 
\hline 
\end{tabularx} {
     \vspace{1ex} }
     {\par \raggedright $^\textrm{a}$ Expanded April 2014 \par 
     \raggedright $^\textrm{b}$ Expanded September 2014; included for covariate adjustment estimates but not as a possible weight donor for treatment effect estimates \par 
    \raggedright $^\textrm{c}$ Comparable coverage policies prior to 2014 
    \par}
\end{table}

\begin{table}[ht]
\centering
\caption{Number of CPUMAs per state}\label{tab:cpumasperstate}
\begin{tabular}{lllrl}
  \hline
State Full & State & Treatment & Number CPUMAs & Early Expansion \\ 
  \hline
Delaware & DE & Excluded &   4 & No \\ 
  Massachusetts & MA & Excluded &  15 & No \\ 
    New Hampshire & NH & Excluded $^\textrm{a}$ &   4 & No \\ 
  New York & NY & Excluded & 123 & No \\ 
  Vermont & VT & Excluded &   4 & No \\ 
  Arizona & AZ & Expansion &  11 & No \\ 
  Arkansas & AR & Expansion &  15 & No \\ 
  Colorado & CO & Expansion &  15 & No \\ 
  Hawaii & HI & Expansion &   8 & No \\ 
  Illinois & IL & Expansion &  47 & No \\ 
  Iowa & IA & Expansion &   7 & No \\ 
  Kentucky & KY & Expansion &  23 & No \\ 
  Maryland & MD & Expansion &  36 & No \\ 
  Michigan & MI & Expansion &  44 & No \\ 
  Nevada & NV & Expansion &   7 & No \\ 
  New Mexico & NM & Expansion &   6 & No \\ 
  North Dakota & ND & Expansion &   2 & No \\ 
  Ohio & OH & Expansion &  44 & No \\ 
  Oregon & OR & Expansion &  17 & No \\ 
  Rhode Island & RI & Expansion &   6 & No \\ 
  West Virginia & WV & Expansion &   4 & No \\ 
  California & CA & Expansion & 110 & Yes \\ 
  Connecticut & CT & Expansion &  22 & Yes \\ 
  Minnesota & MN & Expansion &  27 & Yes \\ 
  New Jersey & NJ & Expansion &  38 & Yes \\ 
  Washington & WA & Expansion &  22 & Yes \\ 
  Alabama & AL & Non-expansion &  18 & No \\ 
  Alaska & AK & Non-expansion &   4 & No \\ 
  Florida & FL & Non-expansion &  59 & No \\ 
  Georgia & GA & Non-expansion &  20 & No \\ 
  Idaho & ID & Non-expansion &   1 & No \\ 
  Indiana & IN & Non-expansion &  24 & No \\ 
  Kansas & KS & Non-expansion &   9 & No \\ 
  Louisiana & LA & Non-expansion &  15 & No \\ 
  Maine & ME & Non-expansion &   5 & No \\ 
  Mississippi & MS & Non-expansion &   7 & No \\ 
  Missouri & MO & Non-expansion &  16 & No \\ 
  Montana & MT & Non-expansion &   1 & No \\ 
  Nebraska & NE & Non-expansion &  11 & No \\ 
  North Carolina & NC & Non-expansion &  27 & No \\ 
  Oklahoma & OK & Non-expansion &   8 & No \\ 
  Pennsylvania & PA & Non-expansion &  55 & No \\ 
  South Carolina & SC & Non-expansion &  10 & No \\ 
  South Dakota & SD & Non-expansion &   1 & No \\ 
  Tennessee & TN & Non-expansion &  28 & No \\ 
  Texas & TX & Non-expansion &  49 & No \\ 
  Utah & UT & Non-expansion &   8 & No \\ 
  Virginia & VA & Non-expansion &  15 & No \\ 
  Wisconsin & WI & Non-expansion &  21 & No \\ 
  Wyoming & WY & Non-expansion &   2 & No \\ 
   \hline
\end{tabular}
     \vspace{1ex}
     \newline
     {\raggedright $^\textrm{a}$ Included for covariate adjustment estimates but not as a possible weight donor for treatment effect estimates \par }
\end{table}

\begin{table}[h!]
\centering
\caption{Univariate summary statistics on adjusted data, primary dataset \\ (Mean, IQR, Range)}\label{tab:summarytab1}
\begin{tabular}{rllll}
  \hline
Variable & Unadjusted & Heterogeneous & Homogeneous \\ 
  \hline
  Age: 19-29 Pct & (24.5, 6, 30.9) & (24.5, 5.9, 29) & (24.5, 5.9, 29) \\ 
  Age: 30-39 Pct & (20.9, 3.4, 20.9) & (20.9, 3.1, 19.1) & (20.9, 3.1, 19.4) \\ 
  Age: 40-49 Pct & (22.2, 2.5, 15.4) & (22.2, 2.3, 13.7) & (22.2, 2.2, 13.7) \\ 
  Avg Adult to Household Ratio & (151, 27.2, 174.3) & (151, 27.2, 173.8) & (151, 27.2, 173.3) \\ 
  Avg Pop Growth & (100.3, 1.9, 13.7) & (100.3, 1.2, 6.2) & (100.3, 1.2, 6.5) \\ 
  Children: Missing Pct & (10.5, 6.6, 41) & (10.5, 6.5, 40.8) & (10.5, 6.5, 40.5) \\ 
  Children: One Pct & (11.1, 3.1, 14.3) & (11.1, 2.8, 12.3) & (11.1, 2.8, 12.5) \\ 
  Children: Three or More Pct & (5.2, 2, 14.1) & (5.2, 1.7, 13.5) & (5.2, 1.7, 13.3) \\ 
  Children: Two Pct & (9.7, 3.5, 15) & (9.7, 3.3, 13.5) & (9.7, 3.2, 13.6) \\ 
  Citizenship Pct & (90, 11.9, 57.1) & (90, 11.8, 55.4) & (90, 11.7, 55.7) \\ 
  Disability Pct & (10.5, 5.3, 28.6) & (10.4, 5.3, 26.7) & (10.5, 5.4, 27.2) \\ 
  Educ: HS Degree Pct & (26.3, 10.7, 43.2) & (26.3, 10.6, 41.8) & (26.3, 10.6, 42) \\ 
  Educ: Less than HS Pct & (11.4, 7.7, 45.3) & (11.4, 7.5, 45) & (11.4, 7.4, 44.6) \\ 
  Educ: Some College Pct & (33.5, 7.9, 34.2) & (33.5, 7.5, 32.9) & (33.5, 7.4, 33) \\ 
  Female Pct & (50.1, 1.6, 15.4) & (50.1, 1.4, 14.2) & (50.1, 1.5, 14.2) \\ 
  Foreign Born Pct & (18.1, 22.4, 76) & (18.1, 22.2, 75.2) & (18.1, 22.2, 75.4) \\ 
  Hispanic Pct & (15.9, 17.7, 97.2) & (15.9, 17.7, 97) & (15.9, 17.7, 97) \\ 
  Inc Pov: $<$ 138 Pct & (20, 11.9, 45.6) & (20, 11.8, 44.8) & (20, 11.8, 43.9) \\ 
  Inc Pov: 139-299 Pct & (24.9, 8.4, 34.2) & (24.9, 7.9, 34.2) & (24.9, 7.8, 34.1) \\ 
  Inc Pov: 300-499 Pct & (23.6, 5.5, 23) & (23.6, 4.9, 22.2) & (23.6, 4.9, 22.2) \\ 
  Inc Pov: 500 + Pct & (29.3, 18.5, 69.1) & (29.3, 18.5, 68.1) & (29.3, 18.5, 68) \\ 
  Married Pct & (50.7, 9.4, 45.1) & (50.7, 9, 44.1) & (50.7, 9.1, 44.2) \\ 
  Race: White Pct & (73.8, 25.4, 91.9) & (73.8, 25.5, 91.7) & (73.8, 25.5, 91.8) \\ 
  Republican Governor 2013 $^\textrm{a}$& (31.1, 100, 100) & (31.1, 100, 100) & (31.1, 100, 100) \\ 
  Republican Lower Leg Control 2013 $^\textrm{a}$ & (24.1, 0, 100) & (24.1, 0, 100) & (24.1, 0, 100) \\ 
  Republican Total Control 2013 $^\textrm{a}$ & (19.8, 0, 100) & (19.8, 0, 100) & (19.8, 0, 100) \\ 
Student Pct & (11.7, 3.4, 29.5) & (11.7, 3.5, 28.1) & (11.7, 3.5, 28) \\ 
  Unemployed Pct 2011 & (10.2, 4.6, 25.5) & (10.2, 3.9, 23.8) & (10.2, 3.9, 22.5) \\ 
  Unemployed Pct 2012 & (9.4, 4.5, 28.3) & (9.4, 4.3, 23.6) & (9.4, 4.3, 23.5) \\ 
  Unemployed Pct 2013 & (8.4, 3.6, 23.4) & (8.4, 3.5, 20.1) & (8.4, 3.5, 20.5) \\ 
  Uninsured Pct 2011 & (19.6, 11.2, 59) & (19.7, 10.9, 51.8) & (19.6, 10.9, 52.5) \\ 
  Uninsured Pct 2012 & (19.4, 9.9, 50.6) & (19.4, 10.1, 49.7) & (19.4, 10.3, 50.2) \\ 
  Uninsured Pct 2013 & (19, 11.2, 49.9) & (19, 10.3, 48.2) & (19, 10.5, 48.7) \\ 
  Urban Pct $^\textrm{b}$ & (82.9, 31.3, 91.3) & (82.9, 31.3, 91.3) & (82.9, 31.3, 91.3) \\ 
   \hline
\end{tabular}
     \vspace{1ex}
     
     {\raggedright $^\textrm{a}$ Derived from data obtained from National Conference of State Legislatures \par
     $^\textrm{b}$ Derived from 2010 Census \par
     }
\end{table}

\begin{table}[h!]
\centering
    \caption{Frequency of covariate adjustments falling outside the support of the unadjusted data \newline (primary dataset)}
    \label{tab:extreme1}
\begin{tabular}{lll}
  \hline
Variables & Heterogeneous & Homogeneous \\ 
  \hline
Age: 19-29 Pct & 0 & 0 \\ 
  Age: 30-39 Pct & 0 & 0 \\ 
  Age: 40-49 Pct & 0 & 0 \\ 
  Avg Adult to Household Ratio & 0 & 0 \\ 
  Citizenship Pct & 1 & 0 \\ 
  Disability Pct & 2 & 2 \\ 
  Educ: HS Degree Pct & 0 & 0 \\ 
  Educ: Less than HS Pct & 1 & 1 \\ 
  Educ: Some College Pct & 0 & 0 \\ 
  Female Pct & 0 & 0 \\ 
  Foreign Born Pct & 1 & 1 \\ 
  Uninsured Pct 2011 & 0 & 0 \\ 
  Uninsured Pct 2012 & 1 & 1 \\ 
  Uninsured Pct 2013 & 0 & 1 \\ 
  Hispanic Pct & 0 & 0 \\ 
  Inc Pov: $<$ 138 Pct & 1 & 1 \\ 
  Inc Pov: 139-299 Pct & 1 & 1 \\ 
  Inc Pov: 300-499 Pct & 0 & 0 \\ 
  Inc Pov: 500 + Pct & 0 & 0 \\ 
  Married Pct & 0 & 0 \\ 
  Children: Missing Pct & 1 & 1 \\ 
  Children: One Pct & 0 & 0 \\ 
  Avg Pop Growth & 0 & 0 \\ 
  Race: White Pct & 1 & 1 \\ 
  Student Pct & 0 & 0 \\ 
  Children: Three or More Pct & 2 & 1 \\ 
  Children: Two Pct & 1 & 1 \\ 
  Unemployed Pct 2011 & 0 & 0 \\ 
  Unemployed Pct 2012 & 0 & 0 \\ 
  Unemployed Pct 2013 & 0 & 0 \\ 
   \hline
\end{tabular}
\end{table}

\begin{table}[ht]
\centering
\caption{Mean non-elderly adult uninsurance rates, 2009-2014}\label{tab:timetrends}
\begin{tabular}{lrrrrrr}
  \hline
Treatment Group & 2009 & 2010 & 2011 & 2012 & 2013 & 2014 \\ 
  \hline
Non-expansion & 21.84 & 22.97 & 22.72 & 22.41 & 22.01 & 19.07 \\ 
  Expansion (primary dataset) & 19.52 & 20.20 & 19.63 & 19.42 & 19.01 & 14.02 \\ 
  Expansion (early excluded) & 19.40 & 20.08 & 19.21 & 19.01 & 18.55 & 13.64 \\ 
   \hline
\end{tabular}
\end{table}

\clearpage

\clearpage

\section{Weight Diagnostics}\label{app:weightdiagnostics}

This section contains additional information pertaining to the generation of the balancing weights and their properties.

Table~\ref{tab:toltable} displays each covariate with the targeted level of maximal imbalance $\delta$ when generating approximate balancing weights. All variables and tolerances are measured in percentage points.

Table~\ref{tab:baltab1} displays the differences between the weighted mean covariate values of the expansion region and the mean of the non-expansion region for our primary dataset and with the early expansion states excluded (calculated using our the homogeneous covariate adjustments). The weights presented here are for the H-SBW estimator. The values under each column are in the following format: (unweighted difference, weighted difference). ``Primary'' and ``Early excluded'' refer to the primary dataset and those that exclude the early expansion states. ``Percent'' indicates that the differences displayed are in percentage points while ``Standardized'' indicates that the standardized mean differences are displayed. Additional results are available on request.

\begin{figure}[H]
\begin{center}
    \caption{Total weights summed by state, primary dataset}
    \label{fig:weightsbystatec1all}
    \includegraphics[scale=0.5]{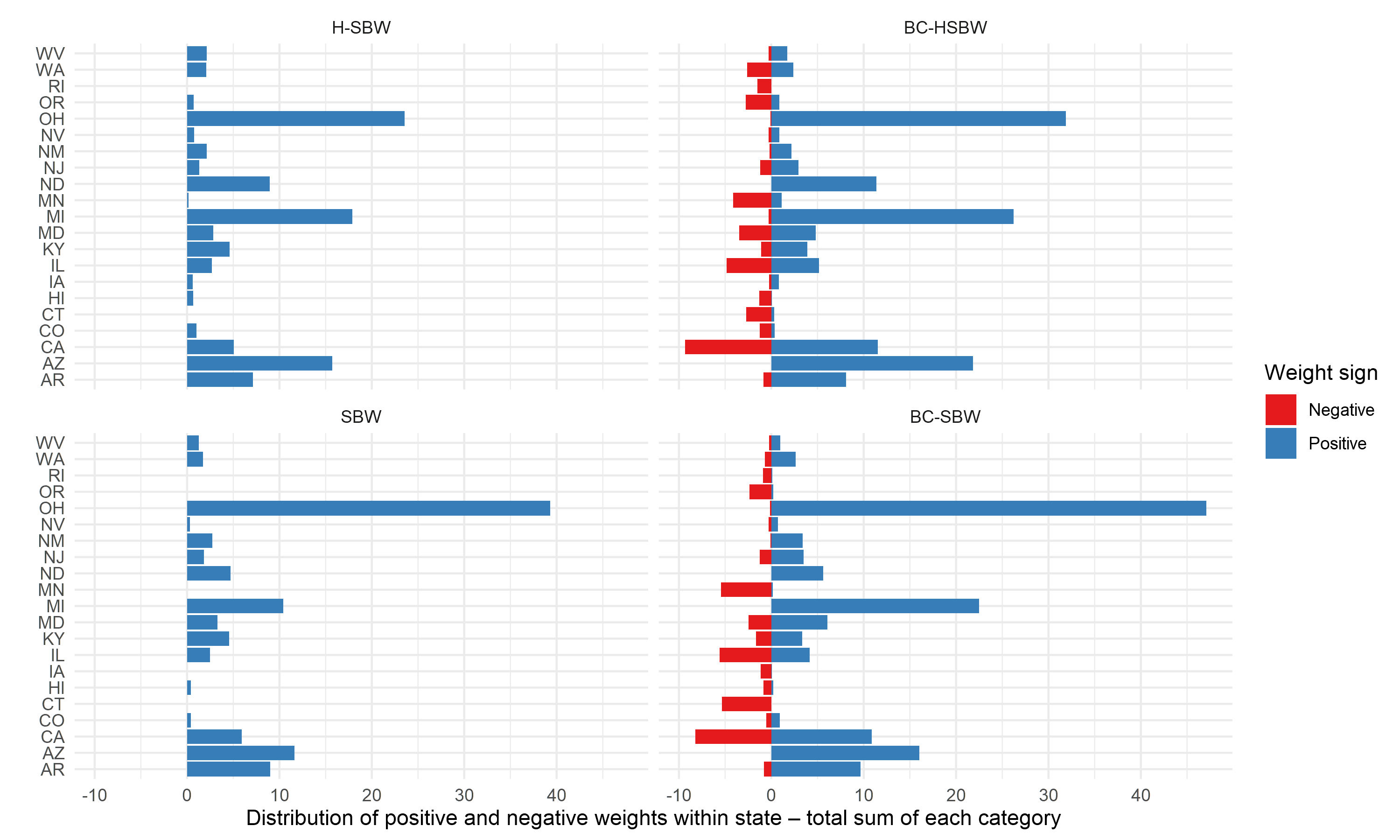}
\end{center}
\end{figure}

\begin{figure}[H]
\begin{center}
    \caption{Total weights summed by state, early expansion excluded}
    \label{fig:weightsbystatec2all}
    \includegraphics[scale=0.5]{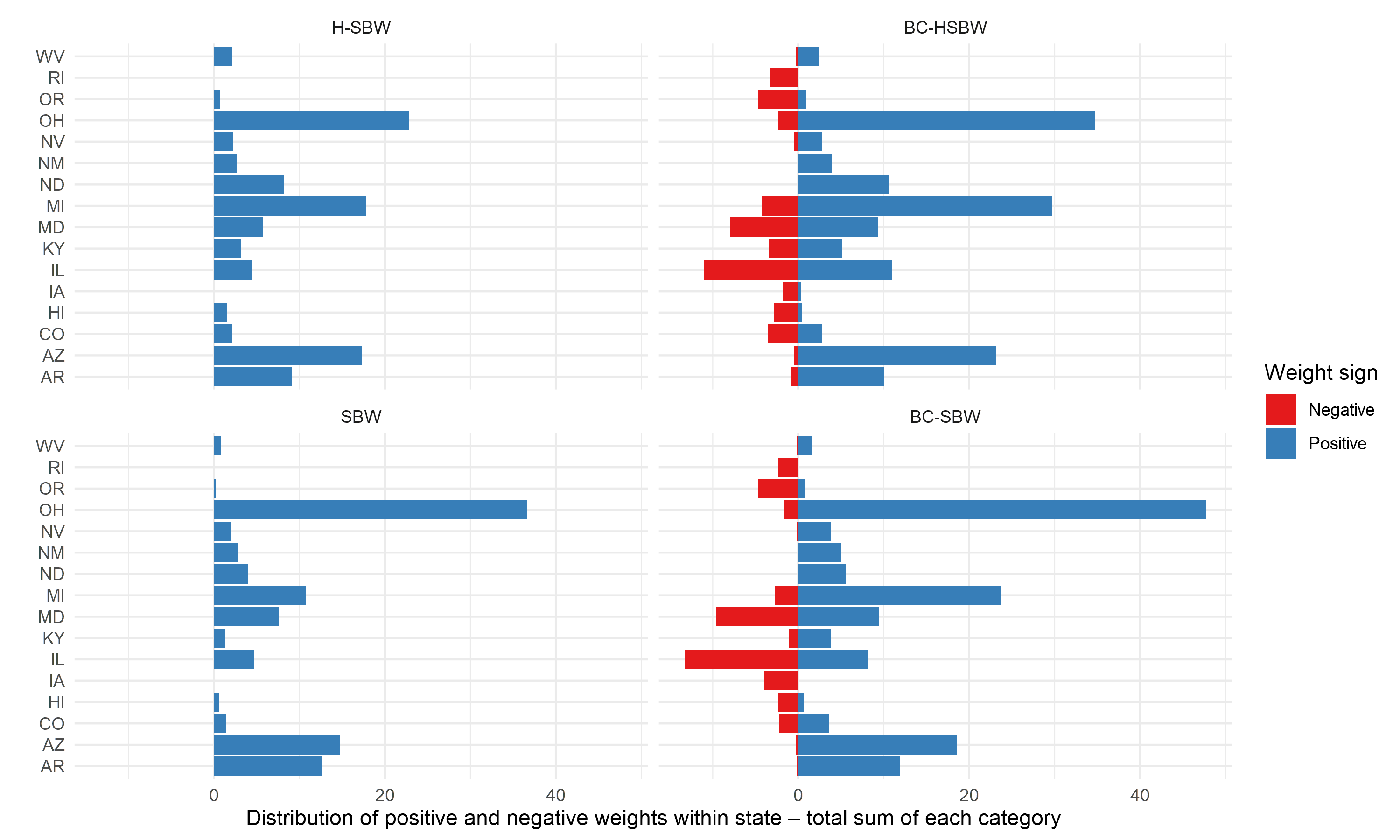}
\end{center}
\end{figure}

Figure~\ref{fig:weightsbystatec1all} shows the weights summed by states for the H-SBW, BC-HSBW, SBW, and BC-HSBW weights on the primary dataset. The positive and negative weights are displayed separately and we standardize the weights to sum to 100. Figure~\ref{fig:weightsbystatec2all} displays the same plot excluding the early expansion states. These plots again show that H-SBW and BC-HSBW more evenly disperses the weights across states relative to SBW and BC-SBW, and also highlights the extent to which the weights extrapolate from each state for the bias-corrected estimators.

We conclude by examining whether the H-SBW weights generated using the unadjusted data balance the adjusted covariates. While these metrics do not reflect the ``true'' imbalances, the comparison can provide some indication of whether the unadjusted weights are overfitting to noisy covariate measurements. Table~\ref{tab:balcomp} compares the imbalances among our pre-treatment outcomes using H-SBW weights generated on our unadjusted dataset applied to the adjusted (homogeneous) dataset. The ``Unweighted Difference'' column represents the raw difference in means, while the ``Weighted Difference'' column reflects the weighted difference that we calculate on the unadjusted dataset. The ``Homogeneous Diff'' column displays the weighted imbalance when applying the H-SBW weights to the dataset using the homogeneous adjustment, and similarly for ``Heterogeneous Diff.'' The weighted pre-treatment outcomes are approximately one percentage point lower than we desired in the two years prior to treatment using the heterogeneous adjustment, and -0.2 percentage points lower on average using the homogeneous adjustment. On the other hand, the naive difference suggests that the imbalance is only -0.05 percentage points. This result suggests that the unadjusted weights are overfitting to noisy covariates and may give an overly optimistic view of the covariate balance. 

\begin{table}[ht]
\centering
\caption{Variables and maximal level of targeted imbalances ($\delta$)}
\begin{tabular}{lr}\label{tab:toltable}
Variable & $\delta$ \\ 
  \hline
Uninsured Pct 2011 & 0.05 \\ 
  Uninsured Pct 2012 & 0.05 \\ 
  Uninsured Pct 2013 & 0.05 \\ 
  Unemployed Pct 2011 & 0.15 \\ 
  Unemployed Pct 2012 & 0.15 \\ 
  Unemployed Pct 2013 & 0.15 \\ 
  Avg Pop Growth & 0.50 \\ 
  Avg Adult to Household Ratio & 0.50 \\ 
  Female Pct & 1.00 \\ 
  Age: 19-29 Pct & 1.00 \\ 
  Age: 30-39 Pct & 1.00 \\ 
  Age: 40-49 Pct & 1.00 \\ 
  Age: 50-64 Pct & 1.00 \\ 
  Married Pct & 1.00 \\ 
  Disability Pct & 1.00 \\ 
  Hispanic Pct & 1.00 \\ 
  Race: White Pct & 1.00 \\ 
  Children: One Pct & 1.00 \\ 
  Children: Two Pct & 1.00 \\ 
  Children: Three or More Pct & 1.00 \\ 
  Children: Missing Pct & 1.00 \\ 
  Urban Pct & 2.00 \\ 
  Citizenship Pct & 2.00 \\ 
  Educ: Less than HS Pct & 2.00 \\ 
  Educ: HS Degree Pct & 2.00 \\ 
  Educ: Some College Pct & 2.00 \\ 
  Educ: College Pct & 2.00 \\ 
  Student Pct & 2.00 \\ 
  Inc Pov: $<$ 138 Pct & 2.00 \\ 
  Inc Pov: 139-299 Pct & 2.00 \\ 
  Inc Pov: 300-499 Pct & 2.00 \\ 
  Inc Pov: 500 + Pct & 2.00 \\ 
  Inc Pov: NA Pct & 2.00 \\ 
  Race: Other Pct & 2.00 \\ 
  Foreign Born Pct & 2.00 \\ 
  Republican Governor 2013 & 25.00 \\ 
  Republican Lower Leg Control 2013 & 25.00 \\ 
  Republican Total Control 2013 & 25.00 \\ 
   \hline
\end{tabular}
\end{table}

\newpage

\begin{landscape}
\begin{table}[h!]\caption{Balance table: percent and standardized mean differences (H-SBW weights on primary dataset, homogeneous adjustment) \\ (Unweighted difference, Weighted difference)}\label{tab:baltab1}
\centering
\begin{threeparttable}\begin{tabular}{lllll}
  \hline
Variables & Preferred (Percent) & Preferred (Standardized) & Early excluded (Percent) & Early excluded (Standardized) \\ 
  \hline
Age: 19-29 Pct & (-0.34, -0.34) & (-0.05, -0.05) & (-0.62, -0.21) & (-0.09, -0.03) \\ 
  Age: 30-39 Pct & (0.36, 0.17) & (0.1, 0.05) & (-0.04, 0.32) & (-0.01, 0.09) \\ 
  Age: 40-49 Pct & (0.19, -0.3) & (0.06, -0.1) & (-0.01, -0.44) & (0, -0.15) \\ 
  Avg Adult to Household Ratio & (11.29, -0.04) & (0.37, 0) & (3.37, 0.1) & (0.13, 0) \\ 
  Citizenship Pct & (-3.61, -1.59) & (-0.33, -0.15) & (-0.24, -1.45) & (-0.03, -0.16) \\ 
  Disability Pct & (-1.45, 0.52) & (-0.27, 0.1) & (-0.17, 0.63) & (-0.03, 0.11) \\ 
  Educ: HS Degree Pct & (-3.37, 0.54) & (-0.32, 0.05) & (-1.02, 0.64) & (-0.1, 0.06) \\ 
  Educ: Less than HS Pct & (-0.37, 0.83) & (-0.04, 0.1) & (-1.22, 0.76) & (-0.16, 0.1) \\ 
  Educ: Some College Pct & (-0.35, 0.4) & (-0.05, 0.06) & (0.36, 0.57) & (0.05, 0.08) \\ 
  Female Pct & (-0.34, -0.64) & (-0.16, -0.3) & (-0.25, -1) & (-0.12, -0.48) \\ 
  Foreign Born Pct & (7.6, 2) & (0.42, 0.11) & (1.02, 2) & (0.07, 0.13) \\ 
  Uninsured Pct 2011 & (-3.08, 0.05) & (-0.28, 0) & (-3.51, -0.05) & (-0.34, 0) \\ 
  Uninsured Pct 2012 & (-3, -0.05) & (-0.27, 0) & (-3.4, 0.05) & (-0.33, 0) \\ 
  Uninsured Pct 2013 & (-2.99, -0.05) & (-0.27, 0) & (-3.45, -0.05) & (-0.34, 0) \\ 
  Hispanic Pct & (4.46, 1) & (0.2, 0.04) & (-1.35, 1) & (-0.07, 0.05) \\ 
  Inc Pov: $<$ 138 Pct & (-2.05, 0.63) & (-0.19, 0.06) & (-1.33, 0.12) & (-0.12, 0.01) \\ 
  Inc Pov: 139-299 Pct & (-2.45, 0.65) & (-0.35, 0.09) & (-1.53, 0.5) & (-0.23, 0.08) \\ 
  Inc Pov: 300-499 Pct & (-0.59, -0.18) & (-0.12, -0.04) & (0.28, -0.18) & (0.06, -0.04) \\ 
  Inc Pov: 500 + Pct & (5.58, -1.3) & (0.35, -0.08) & (2.9, -1.23) & (0.2, -0.08) \\ 
  Married Pct & (-0.76, -0.43) & (-0.07, -0.04) & (-0.21, -0.53) & (-0.02, -0.05) \\ 
  Children: Missing Pct & (-3.25, -1) & (-0.36, -0.11) & (-1.99, -0.1) & (-0.21, -0.01) \\ 
  Children: One Pct & (0.7, -0.14) & (0.25, -0.05) & (0.11, -0.31) & (0.04, -0.12) \\ 
  Avg Pop Growth & (-0.09, -0.21) & (-0.07, -0.18) & (-0.26, -0.19) & (-0.22, -0.16) \\ 
  Race: White Pct & (-4.02, 1) & (-0.16, 0.04) & (0.09, 1) & (0, 0.04) \\ 
  Republican Governor 2013 & (-64.78, -25) & (-1.28, -0.5) & (-54.46, -24.87) & (-1.02, -0.47) \\ 
  Republican Lower Leg Control 2013 & (-74.72, -25) & (-1.69, -0.57) & (-56.67, -23.6) & (-1.12, -0.47) \\ 
  Republican Total Control 2013 & (-71.3, -25) & (-1.45, -0.51) & (-56.47, -25) & (-1.02, -0.45) \\ 
  Student Pct & (0.25, -0.5) & (0.04, -0.08) & (0.11, -0.25) & (0.02, -0.04) \\ 
  Children: Three or More Pct & (0, -0.21) & (0, -0.08) & (-0.17, -0.26) & (-0.07, -0.11) \\ 
  Children: Two Pct & (0.76, -0.31) & (0.23, -0.09) & (0.17, -0.37) & (0.05, -0.12) \\ 
  Unemployed Pct 2011 & (0.82, 0.15) & (0.18, 0.03) & (0.68, 0.15) & (0.15, 0.03) \\ 
  Unemployed Pct 2012 & (0.63, -0.03) & (0.14, -0.01) & (0.47, -0.03) & (0.11, -0.01) \\ 
  Unemployed Pct 2013 & (0.42, -0.15) & (0.11, -0.04) & (0.22, -0.15) & (0.06, -0.04) \\ 
  Urban Pct & (8.28, -2) & (0.26, -0.06) & (2.79, -2) & (0.08, -0.06) \\ 
   \hline
\end{tabular}
    \begin{tablenotes}
      \item[] The values displayed in each cell are the (unweighted, weighted) differences. The columns containing ``Standardized'' reflect the standardized mean differences while ``percent'' indicates the mean differences in percentage points. The columns containing ``Preferred'' indicate that this is for our primary analysis while ``Early excluded'' is for our analysis that excludes the early expansion states.
    \end{tablenotes}
\end{threeparttable}
\end{table}

\end{landscape}

\begin{table}[ht]
\caption{Balance comparison: weights estimated on unadjusted data applied to adjusted data \newline (primary dataset)}\label{tab:balcomp}
\begin{tabular}{p{3cm}p{2cm}p{1.6cm}p{2cm}p{2cm}}
  \hline
Variable & Unweighted Difference & Weighted Difference & Homogeneous Difference & Heterogeneous Difference\\ 
  \hline
Uninsured Pct 2011 & -3.09 & -0.05 & -0.11 & 0.92 \\ 
  Uninsured Pct 2012 & -2.99 & -0.05 & -0.21 & -1.06 \\ 
  Uninsured Pct 2013 & -3.00 & -0.05 & -0.38 & -0.93 \\
  \hline
\end{tabular}
\end{table}
\clearpage

\clearpage

\section{Additional Results}\label{app:allresults}

This section contains additional results that were not displayed in the main paper.

Tables~\ref{tab:pretxpredfullcc} and~\ref{tab:pretxpredfull} presents the full model validation results that also include estimators using the Oaxaca-Blinder OLS and GLS weights (see, e.g, \cite{kline2011oaxaca}). The first table only presents the mean error and RMSE, with rows ordered by RMSE for each dataset, while the second table contains the errors for each individual year and the RMSE, with rows ordered by RMSE on the primary dataset. We find that the OLS and GLS estimators perform quite poorly, again showing that the cost of extrapolation is quite high in our application. 

Table~\ref{tab:confintmain} displays the point estimates and confidence intervals from all estimators we considered, including those using the ``heterogeneous adjustment.''

\begin{figure}[h!]
\begin{center}
    \caption{Leave-one-state-out point estimates minus original estimate, primary dataset}
    \label{fig:loostateplot}
    \includegraphics[scale=0.6]{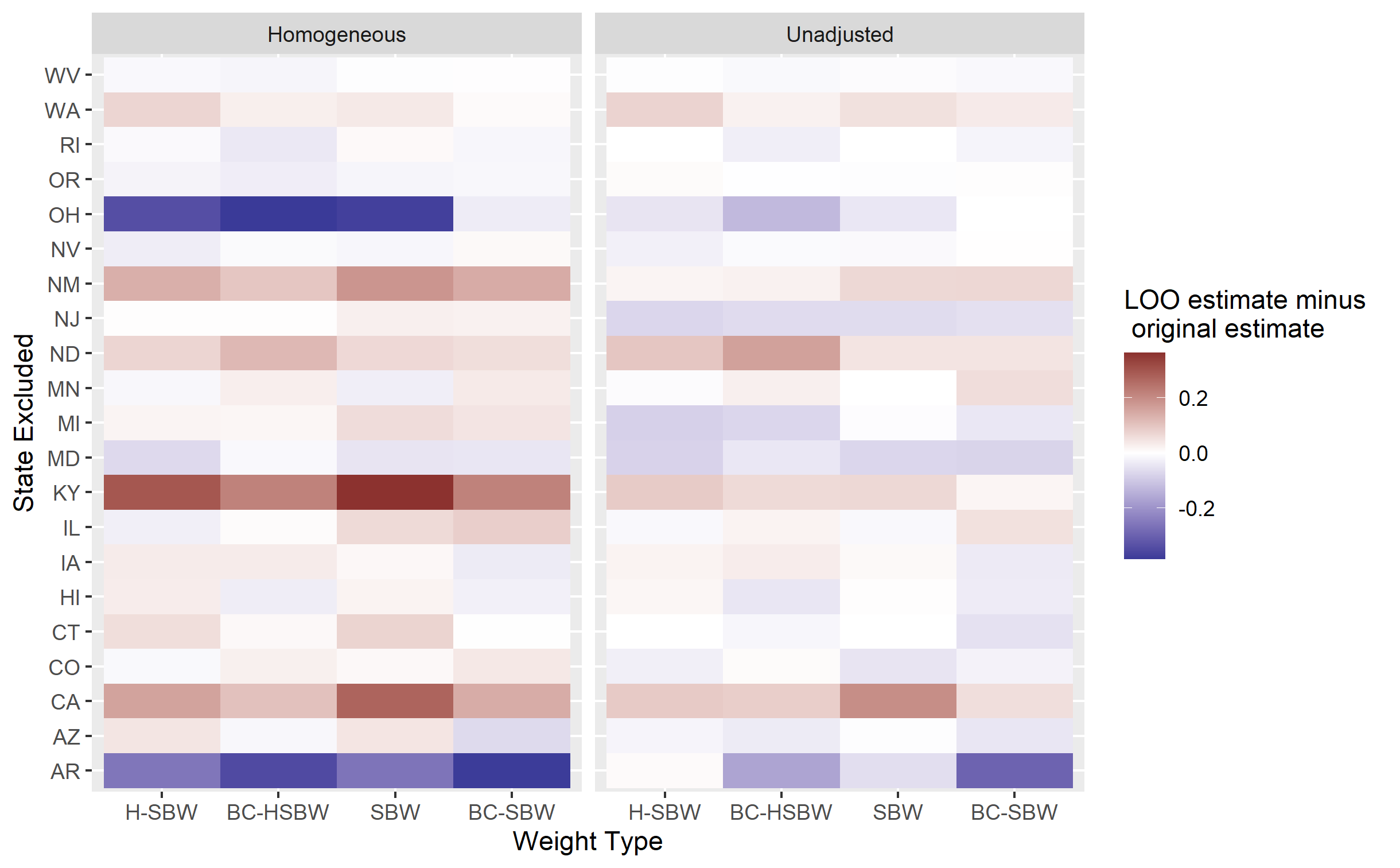}
\end{center}
\subcaption{Colors reflect the magnitude of the difference in the estimates when subtracting the original estimate from the estimate that excludes the specified state. The values in the right panel are on the unadjusted data.}
\end{figure}

\begin{figure}[h!]
\begin{center}
    \caption{Leave-one-state-out point estimates minus original estimate, early expansion excluded}
    \label{fig:loostateplotc2}
    \includegraphics[scale=0.6]{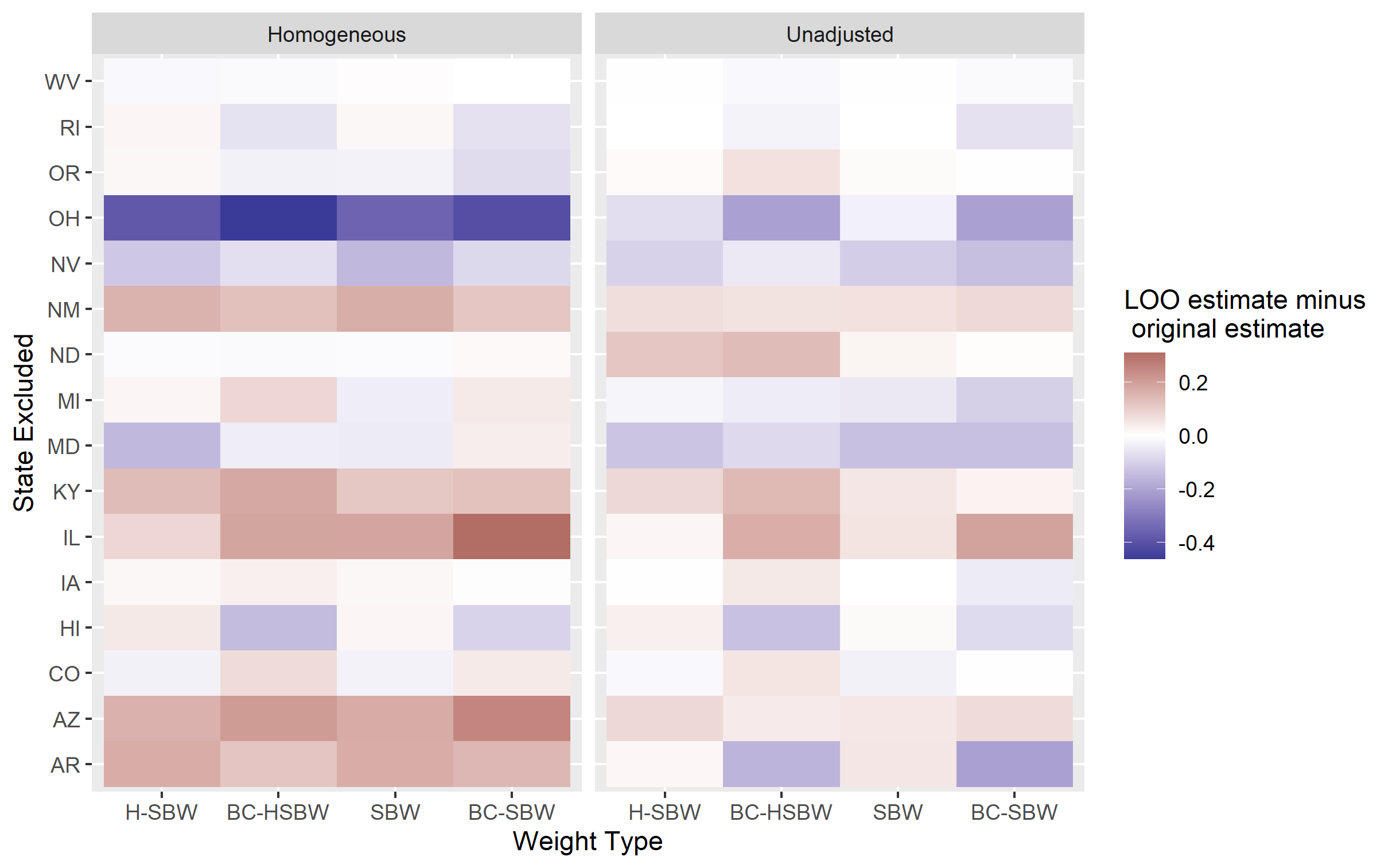}
\end{center}
\subcaption{Colors reflect the magnitude of the difference in the estimates when subtracting the original estimate from the estimate that excludes the specified state. The values in the right panel are on the unadjusted data.}
\end{figure}

Figure~\ref{fig:loostateplot} displays the change in our estimator when removing each state for all four of our estimators on the adjusted dataset (``homogeneous'') and the unadjusted dataset (``unadjusted'') for the primary dataset. Figure~\ref{fig:loostateplotc2} displays the same information when excluding the early expansion states. 

Our final two tables (Tables~\ref{tab:loojackknifec1} and \ref{tab:loojackknifec2}) display each point estimate associated with each estimator in the leave-one-state-out jackknife.

\begin{table}\caption{Estimator pre-treatment outcome mean prediction error and RMSE (in \% points)}\label{tab:pretxpredfullcc}
\centering
\begin{tabular}{llrrllrr}
\hline
\multicolumn{4}{c}{Primary dataset} & \multicolumn{4}{c}{Early expansion 
 excluded} \\ 
 \hline
% \cmidrule(lr){1-4} \cmidrule(lr){5-8}
Sigma estimate & Estimator & Mean Error & RMSE & Sigma estimate & Estimator & Mean Error & RMSE \\ 
\hline
Homogeneous & SBW & -0.20 & 0.20 & Homogeneous & BC-HSBW & -0.02 & 0.07 \\ 
Homogeneous & H-SBW & -0.23 & 0.23 & Homogeneous & BC-SBW & -0.03 & 0.12 \\ 
Heterogeneous & SBW & -0.27 & 0.27 & Heterogeneous & BC-HSBW & -0.08 & 0.14 \\ 
Heterogeneous & H-SBW & -0.35 & 0.36 & Heterogeneous & BC-SBW & -0.07 & 0.15 \\ 
Homogeneous & BC-SBW & -0.39 & 0.39 & Homogeneous & H-SBW & 0.01 & 0.25 \\ 
Heterogeneous & BC-SBW & -0.42 & 0.42 & Homogeneous & SBW & 0.07 & 0.26 \\ 
Unadjusted & SBW & -0.56 & 0.56 & Heterogeneous & SBW & 0.04 & 0.28 \\ 
Unadjusted & H-SBW & -0.57 & 0.57 & Heterogeneous & H-SBW & -0.04 & 0.29 \\ 
Homogeneous & BC-HSBW & -0.58 & 0.58 & Heterogeneous & GLS & -0.02 & 0.29 \\ 
Heterogeneous & BC-HSBW & -0.63 & 0.63 & Heterogeneous & OLS & -0.14 & 0.39 \\ 
Unadjusted & BC-SBW & -0.88 & 0.88 & Unadjusted & SBW & -0.37 & 0.42 \\ 
Unadjusted & OLS & -0.88 & 0.88 & Unadjusted & H-SBW & -0.43 & 0.46 \\ 
Unadjusted & GLS & -0.89 & 0.89 & Unadjusted & BC-HSBW & -0.60 & 0.60 \\ 
Unadjusted & BC-HSBW & -0.96 & 0.96 & Unadjusted & GLS & -0.61 & 0.63 \\ 
Homogeneous & OLS & -1.48 & 1.50 & Unadjusted & OLS & -0.70 & 0.71 \\ 
Homogeneous & GLS & -1.60 & 1.61 & Unadjusted & BC-SBW & -0.70 & 0.71 \\ 
Heterogeneous & GLS & -9.39 & 12.47 & Homogeneous & GLS & 0.28 & 0.98 \\ 
Heterogeneous & OLS & -11.88 & 16.21 & Homogeneous & OLS & 0.02 & 1.10 \\ 
 \hline
\end{tabular}
\end{table}

\begin{table}
    \centering
\begin{threeparttable}\caption{Estimator pre-treatment outcome prediction error (in \% points)}\label{tab:pretxpredfull}
\begin{tabular}{llrrr|rrr}\\ \hline
 &  & \multicolumn{3}{c}{Primary dataset} & \multicolumn{3}{c}{Early expansion 
 excluded} \\
 \hline
 %\cmidrule(lr){3-5} \cmidrule(lr){6-8}
Sigma estimate & Estimator & 2012 error & 2013 error & RMSE & 2012 error & 2013 error & RMSE \\ 
\hline
Homogeneous & SBW & -0.18 & -0.22 & 0.20 & 0.32 & -0.18 & 0.26 \\ 
Homogeneous & H-SBW & -0.24 & -0.21 & 0.23 & 0.26 & -0.24 & 0.25 \\ 
Heterogeneous & SBW & -0.25 & -0.30 & 0.27 & 0.31 & -0.24 & 0.28 \\ 
Heterogeneous & H-SBW & -0.32 & -0.39 & 0.36 & 0.25 & -0.32 & 0.29 \\ 
Homogeneous & BC-SBW & -0.42 & -0.35 & 0.39 & 0.09 & -0.15 & 0.12 \\ 
Heterogeneous & BC-SBW & -0.45 & -0.39 & 0.42 & 0.07 & -0.21 & 0.15 \\ 
Unadjusted & SBW & -0.50 & -0.61 & 0.56 & -0.18 & -0.56 & 0.42 \\ 
Unadjusted & H-SBW & -0.52 & -0.61 & 0.57 & -0.26 & -0.60 & 0.46 \\ 
Homogeneous & BC-HSBW & -0.53 & -0.62 & 0.58 & 0.05 & -0.09 & 0.07 \\ 
Heterogeneous & BC-HSBW & -0.53 & -0.72 & 0.63 & 0.03 & -0.19 & 0.14 \\ 
Unadjusted & BC-SBW & -0.82 & -0.93 & 0.88 & -0.55 & -0.84 & 0.71 \\ 
Unadjusted & OLS & -0.91 & -0.84 & 0.88 & -0.81 & -0.59 & 0.71 \\ 
Unadjusted & GLS & -0.87 & -0.91 & 0.89 & -0.78 & -0.44 & 0.63 \\ 
Unadjusted & BC-HSBW & -0.93 & -0.99 & 0.96 & -0.61 & -0.58 & 0.60 \\ 
Homogeneous & OLS & -1.75 & -1.21 & 1.50 & 1.12 & -1.08 & 1.10 \\ 
Homogeneous & GLS & -1.76 & -1.45 & 1.61 & 1.22 & -0.66 & 0.98 \\ 
Heterogeneous & GLS & -1.18 & -17.60 & 12.47 & 0.27 & -0.32 & 0.29 \\ 
Heterogeneous & OLS & -0.85 & -22.90 & 16.21 & 0.22 & -0.50 & 0.39 \\ 
 \hline
 \end{tabular}
\end{threeparttable}
\end{table}

\begin{table}[h!]
\centering
\begin{threeparttable}
\caption{Point estimates and confidence intervals: all estimators}
%\begin{longtable}{llrrrrrr}
\label{tab:confintmain}
\begin{tabular}{lllrlr}
\hline
 &  & \multicolumn{2}{c}{Primary dataset} & \multicolumn{2}{c}{Early expansion 
 excluded} \\
  \hline
Weight type & Adjustment & Estimate (95\% CI) & Difference & Estimate (95\% CI) & Difference \\ 
  \hline
H-SBW & Homogeneous & -2.33 (-3.54, -1.11) & 0.01 & -2.09 (-3.24, -0.94) & 0.19 \\ 
  H-SBW & Heterogeneous & -2.24 (-3.47, -1.00) & 0.10 & -2.06 (-3.36, -0.77) & 0.22 \\ 
  H-SBW & Unadjusted & -2.34 (-2.88, -1.79) & - & -2.28 (-2.87, -1.70) & - \\ 
  BC-HSBW & Homogeneous & -2.05 (-3.30, -0.80) & 0.17 & -1.94 (-3.27, -0.61) & 0.28 \\ 
  BC-HSBW & Heterogeneous & -1.98 (-3.21, -0.75) & 0.24 & -1.93 (-3.55, -0.32) & 0.29 \\ 
  BC-HSBW & Unadjusted & -2.22 (-2.91, -1.52) & - & -2.22 (-3.14, -1.31) & - \\ 
  SBW & Homogeneous & -2.35 (-3.76, -0.95) & 0.04 & -2.05 (-3.19, -0.91) & 0.16 \\ 
  SBW & Heterogeneous & -2.28 (-3.58, -0.98) & 0.11 & -2.03 (-3.35, -0.72) & 0.18 \\ 
  SBW & Unadjusted & -2.39 (-2.99, -1.79) & - & -2.21 (-2.75, -1.68) & - \\ 
  BC-SBW & Homogeneous & -2.07 (-3.14, -1.00) & 0.13 & -1.99 (-3.33, -0.66) & 0.23 \\ 
  BC-SBW & Heterogeneous & -2.00 (-3.07, -0.92) & 0.20 & -2.00 (-3.65, -0.34) & 0.23 \\ 
  BC-SBW & Unadjusted & -2.19 (-2.94, -1.45) & - & -2.23 (-3.12, -1.33) & - \\ 
   \hline
\end{tabular}
    \begin{tablenotes}
      \item[] Note: ``Difference'' column reflects difference between adjusted and unadjusted estimators
        \item[] Note: Confidence intervals are estimated using the leave-one-state-out jackknife and quantiles from t-distribution with $m_1 - 1$ degrees of freedom.
    \end{tablenotes}
\end{threeparttable}
\end{table}

\newpage

\newpage

\begin{landscape}

\begin{table}[H]\caption{Leave one state out jackknife point estimates: all estimators, primary dataset}\label{tab:loojackknifec1}
\centering
%\begin{tabular}
\begin{tabular}{lrrrr|rrrr|rrrr}
\hline
 & \multicolumn{4}{c}{Homogeneous} & \multicolumn{4}{c}{Heterogeneous} & \multicolumn{4}{c}{Unadjusted} \\ 
 \hline
 %\cmidrule(lr){2-5} \cmidrule(lr){6-9} \cmidrule(lr){10-13}
Left out state & BC-HSBW & BC-SBW & H-SBW & SBW & BC-HSBW & BC-SBW & H-SBW & SBW & BC-HSBW & BC-SBW & H-SBW & SBW \\ 
\hline
All & -2.05 & -2.07 & -2.33 & -2.35 & -1.98 & -2.00 & -2.24 & -2.28 & -2.22 & -2.19 & -2.34 & -2.39 \\ 
AR & -2.40 & -2.45 & -2.59 & -2.61 & -2.30 & -2.36 & -2.46 & -2.52 & -2.38 & -2.49 & -2.33 & -2.45 \\ 
AZ & -2.06 & -2.14 & -2.28 & -2.31 & -2.02 & -2.06 & -2.26 & -2.25 & -2.25 & -2.24 & -2.35 & -2.39 \\ 
CA & -1.94 & -1.93 & -2.17 & -2.08 & -1.96 & -1.90 & -2.12 & -2.04 & -2.13 & -2.14 & -2.24 & -2.19 \\ 
CO & -2.02 & -2.03 & -2.34 & -2.34 & -2.00 & -1.98 & -2.34 & -2.34 & -2.21 & -2.22 & -2.36 & -2.44 \\ 
CT & -2.04 & -2.07 & -2.27 & -2.28 & -1.98 & -2.00 & -2.20 & -2.22 & -2.23 & -2.25 & -2.34 & -2.39 \\ 
HI & -2.08 & -2.09 & -2.30 & -2.33 & -2.01 & -2.03 & -2.20 & -2.26 & -2.26 & -2.23 & -2.32 & -2.39 \\ 
IA & -2.01 & -2.10 & -2.29 & -2.34 & -1.93 & -2.04 & -2.18 & -2.26 & -2.19 & -2.23 & -2.31 & -2.38 \\ 
IL & -2.04 & -1.98 & -2.36 & -2.29 & -2.00 & -1.94 & -2.31 & -2.26 & -2.20 & -2.14 & -2.35 & -2.40 \\ 
KY & -1.83 & -1.85 & -2.03 & -1.99 & -1.79 & -1.82 & -1.99 & -1.97 & -2.15 & -2.18 & -2.25 & -2.32 \\ 
MD & -2.06 & -2.11 & -2.40 & -2.40 & -2.02 & -2.04 & -2.32 & -2.33 & -2.26 & -2.27 & -2.42 & -2.46 \\ 
MI & -2.03 & -2.02 & -2.31 & -2.29 & -1.96 & -1.88 & -2.23 & -2.16 & -2.29 & -2.24 & -2.42 & -2.39 \\ 
MN & -2.02 & -2.03 & -2.34 & -2.38 & -1.99 & -1.96 & -2.30 & -2.33 & -2.19 & -2.14 & -2.34 & -2.39 \\ 
ND & -1.93 & -2.01 & -2.26 & -2.29 & -1.85 & -1.93 & -2.17 & -2.21 & -2.05 & -2.15 & -2.24 & -2.34 \\ 
NJ & -2.04 & -2.04 & -2.32 & -2.33 & -2.01 & -2.01 & -2.28 & -2.30 & -2.28 & -2.25 & -2.41 & -2.45 \\ 
NM & -1.95 & -1.92 & -2.19 & -2.17 & -1.94 & -1.91 & -2.20 & -2.19 & -2.19 & -2.13 & -2.32 & -2.32 \\ 
NV & -2.06 & -2.06 & -2.36 & -2.37 & -1.99 & -1.99 & -2.27 & -2.31 & -2.23 & -2.19 & -2.36 & -2.40 \\ 
OH & -2.43 & -2.10 & -2.67 & -2.72 & -2.41 & -2.20 & -2.67 & -2.64 & -2.34 & -2.19 & -2.38 & -2.43 \\ 
OR & -2.08 & -2.08 & -2.35 & -2.37 & -2.05 & -2.05 & -2.33 & -2.36 & -2.22 & -2.19 & -2.33 & -2.39 \\ 
RI & -2.09 & -2.08 & -2.34 & -2.34 & -2.01 & -2.01 & -2.22 & -2.26 & -2.25 & -2.21 & -2.34 & -2.39 \\ 
WA & -2.02 & -2.06 & -2.26 & -2.32 & -1.99 & -2.03 & -2.24 & -2.31 & -2.19 & -2.16 & -2.26 & -2.34 \\ 
WV & -2.07 & -2.07 & -2.34 & -2.36 & -2.00 & -2.01 & -2.24 & -2.29 & -2.23 & -2.21 & -2.34 & -2.40 \\ 
 \hline
\end{tabular}
%\end{tabular}
\begin{tablenotes}
  \item Note: Row labeled ``All'' reflects the original estimate
\end{tablenotes}
\end{table}

\begin{table}\caption{Leave one state out jackknife point estimates: all estimators, early expansion excluded}\label{tab:loojackknifec2}
\begin{tabular}{lrrrr|rrrr|rrrr}
\hline
 & \multicolumn{4}{c}{Homogeneous} & \multicolumn{4}{c}{Heterogeneous} & \multicolumn{4}{c}{Unadjusted} \\ 
 \hline
% \cmidrule(lr){2-5} \cmidrule(lr){6-9} \cmidrule(lr){10-13}
Left out state & BC-HSBW & BC-SBW & H-SBW & SBW & BC-HSBW & BC-SBW & H-SBW & SBW & BC-HSBW & BC-SBW & H-SBW & SBW \\ 
\hline
All & -1.94 & -1.99 & -2.09 & -2.05 & -1.93 & -2.00 & -2.06 & -2.03 & -2.22 & -2.23 & -2.28 & -2.21 \\ 
AR & -1.82 & -1.85 & -1.92 & -1.88 & -1.71 & -1.73 & -1.77 & -1.75 & -2.39 & -2.44 & -2.27 & -2.16 \\ 
AZ & -1.73 & -1.74 & -1.93 & -1.88 & -1.64 & -1.66 & -1.85 & -1.81 & -2.18 & -2.15 & -2.21 & -2.17 \\ 
CO & -1.87 & -1.95 & -2.12 & -2.08 & -1.86 & -1.92 & -2.09 & -2.00 & -2.17 & -2.23 & -2.30 & -2.24 \\ 
HI & -2.09 & -2.09 & -2.05 & -2.03 & -2.08 & -2.09 & -2.04 & -1.95 & -2.36 & -2.30 & -2.25 & -2.20 \\ 
IA & -1.91 & -1.99 & -2.07 & -2.03 & -1.85 & -1.95 & -2.01 & -1.98 & -2.18 & -2.27 & -2.28 & -2.21 \\ 
IL & -1.75 & -1.68 & -2.01 & -1.86 & -1.62 & -1.56 & -1.94 & -1.79 & -2.06 & -2.03 & -2.27 & -2.16 \\ 
KY & -1.76 & -1.87 & -1.95 & -1.93 & -1.70 & -1.85 & -1.89 & -1.89 & -2.08 & -2.20 & -2.20 & -2.16 \\ 
MD & -1.98 & -1.96 & -2.24 & -2.09 & -1.85 & -1.87 & -2.07 & -1.96 & -2.31 & -2.36 & -2.41 & -2.35 \\ 
MI & -1.86 & -1.95 & -2.07 & -2.09 & -1.97 & -1.94 & -2.14 & -1.97 & -2.26 & -2.33 & -2.31 & -2.26 \\ 
ND & -1.95 & -1.98 & -2.10 & -2.06 & -1.86 & -1.94 & -1.95 & -1.97 & -2.09 & -2.22 & -2.17 & -2.19 \\ 
NM & -1.81 & -1.88 & -1.94 & -1.88 & -1.88 & -1.89 & -1.98 & -1.86 & -2.17 & -2.15 & -2.22 & -2.15 \\ 
NV & -2.01 & -2.08 & -2.21 & -2.20 & -1.92 & -2.00 & -2.10 & -2.01 & -2.27 & -2.36 & -2.38 & -2.32 \\ 
OH & -2.40 & -2.40 & -2.48 & -2.41 & -2.45 & -2.48 & -2.49 & -2.47 & -2.43 & -2.43 & -2.36 & -2.25 \\ 
OR & -1.97 & -2.07 & -2.07 & -2.08 & -2.00 & -2.05 & -2.08 & -1.97 & -2.16 & -2.22 & -2.27 & -2.21 \\ 
RI & -2.00 & -2.06 & -2.07 & -2.03 & -2.00 & -2.05 & -2.08 & -2.01 & -2.25 & -2.29 & -2.28 & -2.21 \\ 
WV & -1.95 & -1.99 & -2.11 & -2.04 & -1.91 & -1.98 & -2.07 & -2.03 & -2.24 & -2.24 & -2.29 & -2.22 \\ 
 \hline
\end{tabular}
\begin{tablenotes}
  \item Note: Row labeled ``All'' reflects the original estimate
\end{tablenotes}
\end{table}

\end{landscape}

\clearpage

\clearpage

\section{Simulation Study}\label{app:simstudy}

This final section presents simulation results evaluating the performance of our proposed estimators in an idealized setting where the outcome model and measurement error models are known. The first subsection outlines our simulation study. The second subsection presents selected results, including the bias, mean-square error, and coverage rates for our proposed estimators and variance estimates. The final subsection provides additional results analyzing the performance of H-SBW and our proposed variance estimator absent measurement error.

\subsection{Study design}

For our simulations we generate populations of $M_1 = 5000$ states, each with $p_s$ CPUMAs, so that we obtain a population of $N_1 = \sum_{s=1}^{M_1} p_s$ CPUMAs. We draw $p_s \stackrel{iid}\sim \lfloor Exp(0.1) + 10\rfloor$ so that the average number of regions per state is approximately 20. We then generate a 3-dimensional covariate vector $X_{sc}$:

\begin{align}\label{eqn:pheccovariates}
\mu_s \stackrel{iid}\sim MVN(0, \Sigma_B) \\
X_{sc} \mid \mu_s \stackrel{iid}\sim MVN(\mu_s, \Sigma_V) 
\end{align}

Define $\Sigma_X = \Sigma_V + \Sigma_B$. Let $\sigma^2_{x, j}$ be the j-th diagonal element of $\Sigma_X$. Across all simulations, we fix this number to be constant and equal to 2 (i.e. $\sigma^2_{x, j} = \sigma^2_x = 2$). We also fix the off-diagonal elements of both $\Sigma_V$ and $\Sigma_B$ to be equal and so that $Cor(X_j, X_k) = 0.25$. Finally, let $\rho_{x, j}$ denote the within-state correlation of the observations (i.e. $Cor(X_{j, sc}, X_{j, sd})$ for $c \ne d$). We set this value to be equal for all covariates, so that $\rho_{x, j} = \rho_x$, but vary this parameter across simulations.

We then generate outcomes according to the model:

\begin{align*}
Y_{sc} \mid X_{sc} \sim MVN(X_{1, sc} + X_{2, sc} + X_{3, sc}, \Omega)
\end{align*}
where $\Omega$ is a block-diagonal matrix representing homoskedastic and equicorrelated errors outlined below.

\begin{align}
    \label{eqn:phecvariance1}\Omega &= \begin{pmatrix}
    \Omega_1 & 0 & 0 & ... & 0 \\
    0 & \Omega_2 & 0 &  ...  & 0 \\
    & & ...  & & \\
    0 & 0 & 0 &, ..., & \Omega_{M_1} 
    \end{pmatrix} \\
    \label{eqn:phecvariance2}\Omega_s &= \begin{pmatrix}
    \sigma^2_{\epsilon} + \sigma^2_{\varepsilon} & \sigma^2_{\varepsilon} & \sigma^2_{\varepsilon} & ... & \sigma^2_{\varepsilon} \\
    \sigma^2_{\varepsilon} & \sigma^2_{\epsilon} + \sigma^2_{\varepsilon} & \sigma^2_{\varepsilon}& ... & \sigma^2_{\epsilon} \\
    & & ... & & \\
    \sigma^2_{\varepsilon} & \sigma^2_{\varepsilon} & \sigma^2_{\varepsilon} & ... & \sigma^2_{\epsilon} + \sigma^2_{\varepsilon}
    \end{pmatrix}
\end{align}
In other words $\sigma^2_{\varepsilon}$ represents the variance component from a state-level random effect and $\sigma^2_{\epsilon}$ represents a variance component from a CPUMA-level random effect, as described in the main paper.

We next generate our noisy outcome and covariate estimates $(J, W)$:

\begin{align*}
(J_{sc}, W_{sc}) \mid (Y_{sc}, X_{sc}) \stackrel{iid}\sim MVN((Y_{sc}, X_{sc}), \Sigma_{\nu, sc})
\end{align*}

\begin{align*}
    \Sigma_{\nu, sc} = \begin{pmatrix}
    \sigma^2_{\nu, sc} & 0 & 0 & 0 \\
    0 & \sigma^2_{\nu, sc} & 0 & 0 \\
    0 & 0 & \sigma^2_{\nu, sc} & 0 \\
    0 & 0 & 0 & \sigma^2_{\nu, sc}
    \end{pmatrix}
\end{align*}

We allow $\sigma^2_{\nu, sc}$ to either be constant or a function of the sample size of an underlying survey that generates the estimate. In the latter case we simulate these sample sizes $r_{sc} \stackrel{iid}\sim Unif(300, 2300)$. We calculate $\sigma_{v, sc}^2$ that satisfies the following equality: $\sigma^2_{v, sc} = \sigma^{2\star}_v/r_{sc}$. This reflects the model assumed in the ``heterogeneous adjustment.'' We also define $\sigma_v^2$ as the expected value of $\sigma^2_{v, sc}$:

\begin{align*}
    \sigma_{\nu}^2 =
     \sigma_\nu^{2\star}\mathbb{E}[1/r_{sc}]
\end{align*}
We then define parameter $\tau$:

\begin{align*}
    \tau = \sigma^2_x/(\sigma^2_x + \sigma^2_{\nu})
\end{align*}

We consider different values of $\tau$ throughout our simulations. This parameter reflects the extent of the measurement error, with $\tau = 1$ meaning that the covariates are measured without error, and the noise increasing as $\tau$ goes to zero.

Define $\rho^\star = \sigma^2_{\varepsilon}/(\sigma^2_{\epsilon} + \sigma^2_{\varepsilon} + \sigma^2_{\nu})$. $\rho^\star$ represents the within-state correlation of the outcome model errors given the true covariates $X$, including the measurement errors in the outcome. We fix this to be 0.25 for our primary simulation results. We caveat that $\rho^\star$ only represents an optimal value in a setting without measurement error. This is due to the contribution of additional terms in the variance of the estimator from the noisy covariate measurements.  

We then consider all 18 combinations of the following parameters:

\begin{itemize}
    \item $r_{sc} \stackrel{iid}\sim Unif(300, 2300); r_{sc} = 1$ 
    \item $\rho_x \in \{0, 0.25, 0.5\}$
    \item $\tau \in \{0.85, 0.9, 0.95\}$
\end{itemize}

For each parameterization we take 1000 random samples of size $m_1 = 25$ states (with $n_1$ total CPUMAs that average around 450). For each sample we estimate a series of H-SBW weights that set $\delta = 0$ and the target $\upsilon_0 = (1, 1, 1)$. Let $\rho$ denote the assumed $\rho^\star$ in the H-SBW objective. We generate weights for all combinations of input datasets $Z$ and correlation-parameters $\rho$:

\begin{itemize}
    \item $Z \in \{W$, $X$, $\hat{X}^{hom}$, $\hat{X}^{het}$, $\hat{X}^{cor}$\}
    \item $\rho \in \{0, 0.25, 0.5\}$
\end{itemize}

We estimate the variance for each estimator using the leave-one-state-out jackknife, described in Section~\ref{sec:methods}. 

Note: for $\hat{\kappa}$ we use the empirical covariance matrix of $W$, the estimated means $\bar{W}$, and $\hat{\Sigma}_{\nu, sc}$, where we draw $\hat{\Sigma}_{\nu, sc}$ from $\Sigma_{\nu, sc} + \mathcal{N}(0, 0.001*450*I_d)$. In other words, when averaged together we assume that $\hat{\Sigma}_{\nu}$ have a fairly precise estimate of $\Sigma_{\nu}$.

\subsection{Selected results}\label{ssec:resultsI}

\begin{figure}[H]
\begin{center}
    \caption{Simulation study: estimator bias}\label{fig:simbias}
    \label{fig:loveplotc1}
    \includegraphics[scale=0.5]{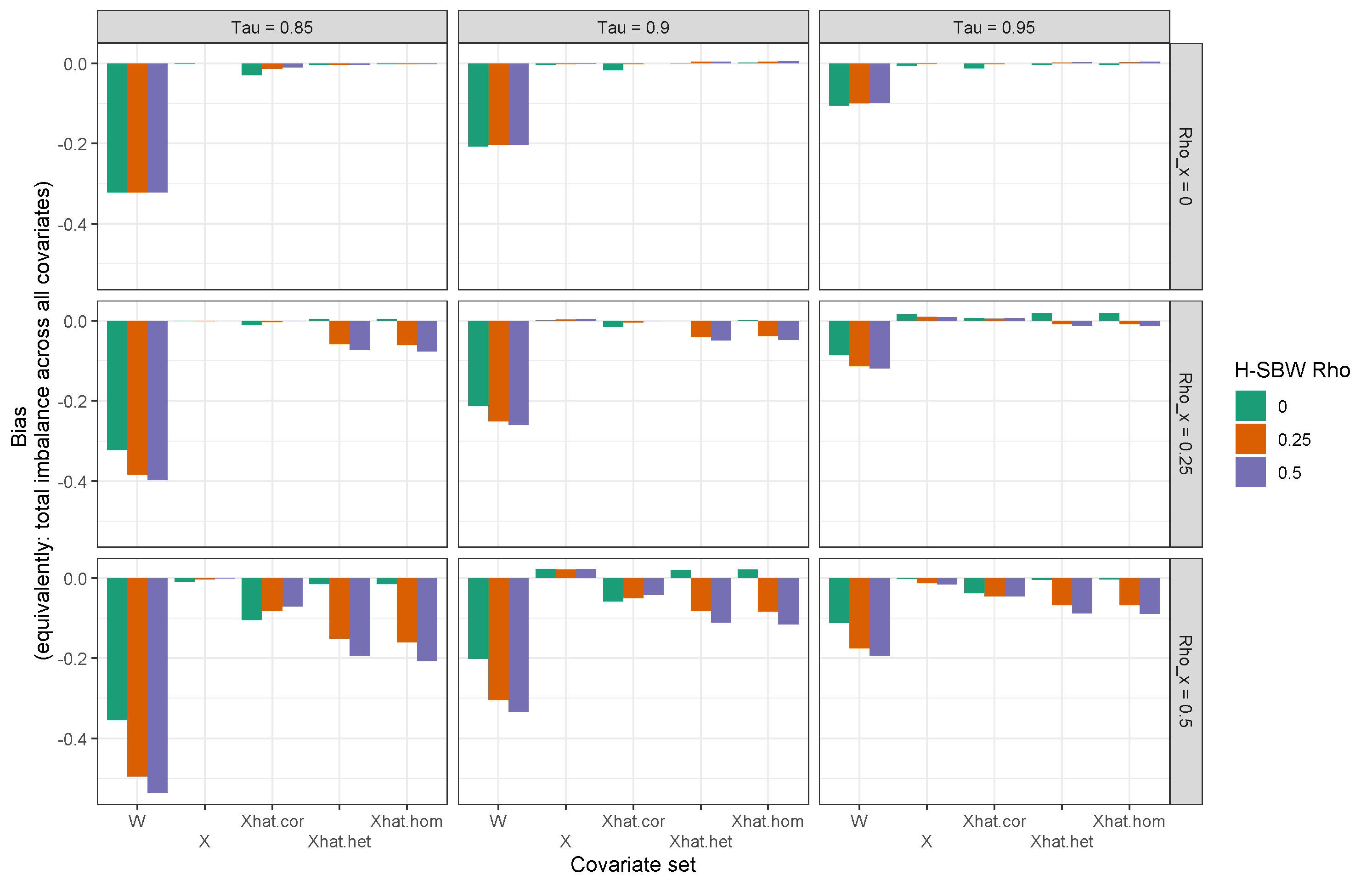}
    \subcaption{Averaged across 1000 simulations for each specification}
\end{center}
\end{figure}

We present the results where the ``heterogeneous adjustment'' model is correct. The results for the homoskedastic measurement errors are quite similar and we observe in the text where they differ. Figure~\ref{fig:simbias} displays the bias associated with each estimator. From left to right the panels reflect different values of $\tau$: the left-most panels have the most measurement error while the right-most panels have the least. From top to bottom the panels reflect different values of $\rho_x$: the top-most panel has no correlation structure among the covariates, while the bottom-most panels are more highly correlated within state. Within each panel we organize the results by the input covariate set: $W$ represents the estimators generated without any covariate adjustment; $X$ reflects estimators generated on the true covariates; ``Xhat-het'' represents the heterogeneous adjustment ($\hat{X}^{het}$), ``Xhat-hom'' represents the homogeneous adjustment ($\hat{X}^{hom}$), and ``Xhat-cor'' represents the correlated adjustment ($\hat{X}^{cor}$), described in Appendix~\ref{app:adjustmentdetails}. The estimators are colored by the assumed value of $\rho$ in the H-SBW objective, where $\rho = 0$ is equivalent to SBW. Across all of the present simulations, we set $\rho^\star$ equal to 0.25.

We highlight two general results. First, if we know the true values of $X$, all of our estimators are unbiased. Second, balancing on $W$ results in bias. This bias increases as $\tau$ decreases, and, when the covariates are dependent, increases with $\rho$. These results are consistent with our propositions in Appendices~\ref{app:AsecI} and ~\ref{app:AsecII}.

When attempting to mitigate this bias by using some estimate of $\mathbb{E}[X \mid W]$, balancing on $\hat{X}^{hom}$, $\hat{X}^{het}$, or $\hat{X}^{cor}$ results in approximately unbiased estimates for all values of $\rho$ when the covariates uncorrelated. When $X$ are correlated across observations, setting $\rho > 0$ results in biased estimates for $\hat{X}^{het}$ or $\hat{X}^{hom}$, where the bias increases with $\rho$. The SBW estimators remain approximately unbiased, even when the data are correlated. We speculate in Remark~\ref{remark:sbwspeculation} why this may be the case. In settings where $\rho_x = 0.25$ the estimators using that $\hat{X}^{cor}$ remain approximately unbiased. However, when $\rho_x = 0.5$ even these estimators appear to have some bias. In further investigations (results not displayed but available on request), this bias decreases with the number of states. Regardless of the adjustment set, all biases on the adjusted datasets are smaller in absolute magnitude than the corresponding biases associated with estimators that naively balance $W$. 

All of these simulations had heterogeneous measurement error. When examining the results when the errors are homoskedastic we find that the estimators that balance on $\tilde{X}^{het}$ have a small bias even when $\tau = 0$ or $\rho = 0$. Assuming this model is correct when it is not appears to have some cost. This may help explain the slightly worse performance we found when using the heterogeneous adjustment in our validation study in Section~\ref{sec:results}.

\begin{figure}[H]
\begin{center}
    \caption{Simulation study: estimator variance}\label{fig:simvar}
    \label{fig:loveplotc1}
    \includegraphics[scale=0.5]{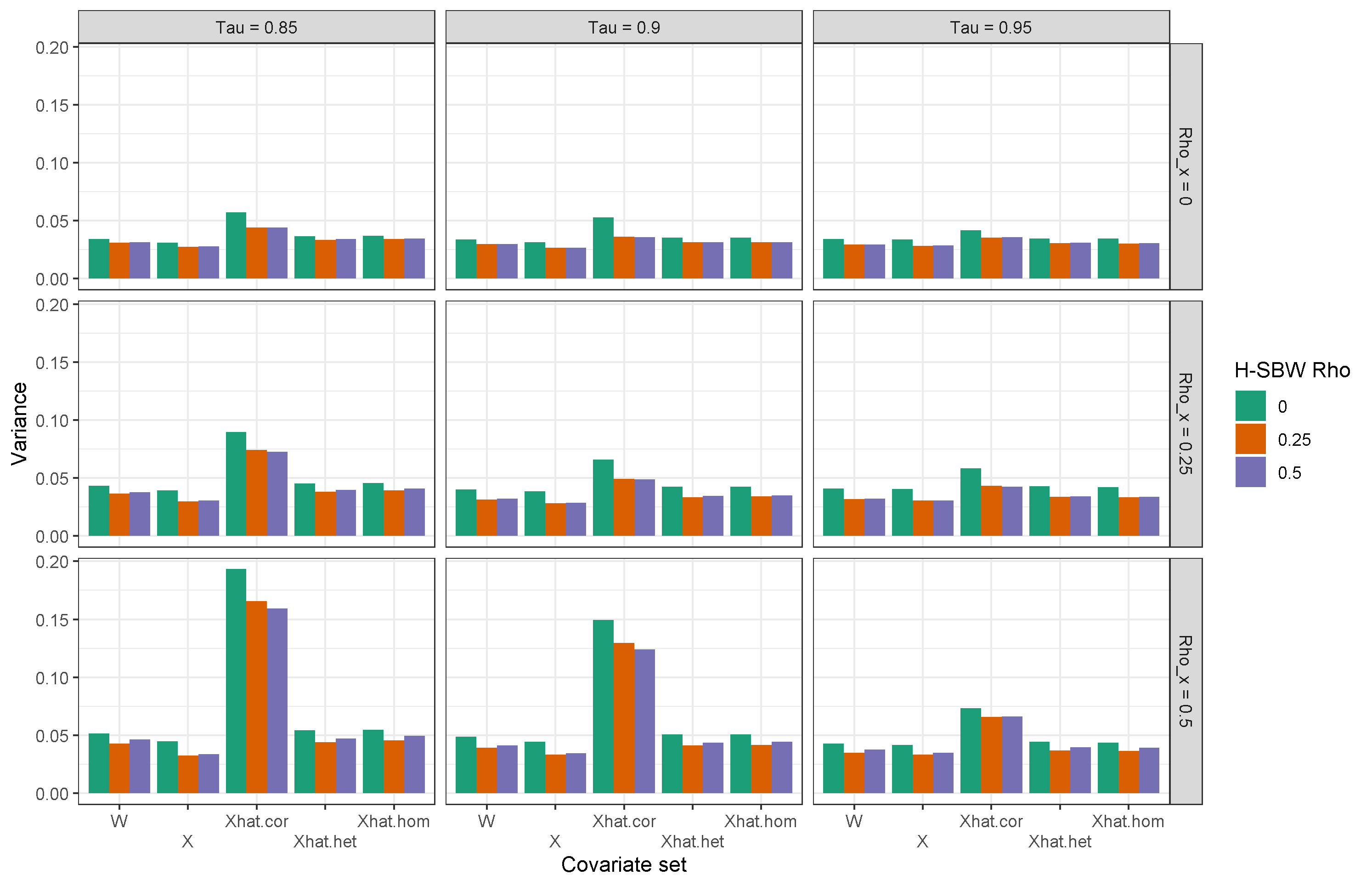}
    \subcaption{Averaged across 1000 simulations for each specification}
\end{center}
\end{figure}

Figure~\ref{fig:simvar} displays the variance associated with these same estimators. Throughout we obtain modest reductions in variance as we increase $\rho$. Choosing either $\rho = 0.25$ or $\rho = 0.5$ gives similar reductions relative to choosing $\rho = 0$ (again equivalent to SBW). These variance reductions may increase if we increased $\rho^\star$, which we fix in this simulation study to be 0.25. For example, Section~\ref{appssec:simstudyresults2}, Figure~\ref{fig:hsbwvarx}, considers a wider range of parameterizations of $(\rho, \rho^\star)$ in a context without measurement error, and shows that larger variance reductions are possible when $\rho^\star$ is higher. Admittedly, these settings may be unlikely in practice.\footnote{In the setting without measurement error $\rho^\star$ represents the optimal value.}

These results also show that balancing on $\hat{X}^{cor}$ can lead to a much more variable estimate, particularly when the covariates have larger within-state correlations. Only in the setting with uncorrelated covariates and little measurement error does the variability induced by this adjustment appears to have less cost.

\begin{figure}[H]
\begin{center}
    \caption{Simulation study: estimator mean-square-error}\label{fig:simmse}
    \label{fig:loveplotc1}
    \includegraphics[scale=0.5]{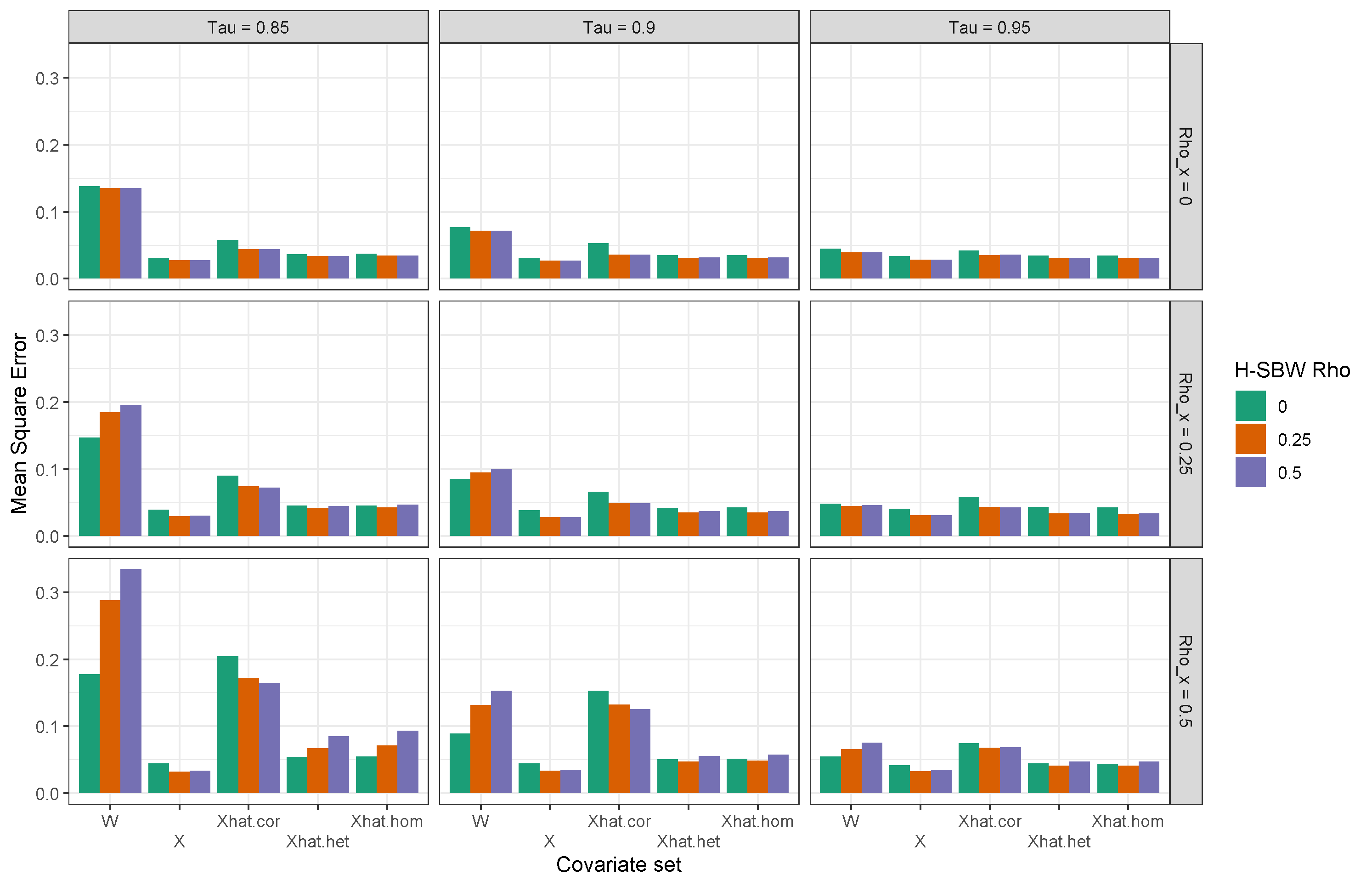}
    \subcaption{Averaged across 1000 simulations for each specification}
\end{center}
\end{figure}

Figure~\ref{fig:simmse} displays the MSE of these estimators. The estimators generated on $\hat{X}^{cor}$ have larger MSE, largely driven by the increase in variability we observed previously. Despite the increase in bias for H-SBW with $\hat{X}^{hom}$ or $\hat{X}^{het}$ we still often find modest MSE reductions when using H-SBW relative to SBW. This appears more likely as the magnitude of the measurement error decreases and/or the within-state correlation decreases (i.e. moving towards the top-right of the plot). Whether an MSE improvement is possible also depends on $\rho^\star$: if we were to set $\rho^\star = 0$, we would expect the MSE of these estimators to increase for all estimators as $\rho$ increases, even when we observe $X$ (see also Section~\ref{appssec:simstudyresults2}). The results are similar when considering homoskedastic measurement errors, though the space where we see MSE improvements for H-SBW relative to SBW on the uncorrelated adjustments appears to increase slightly (results available on request). 

\begin{figure}[H]
\begin{center}
    \caption{Simulation study: jackknife coverage rates}\label{fig:simcoverage1}
    \includegraphics[scale=0.5]{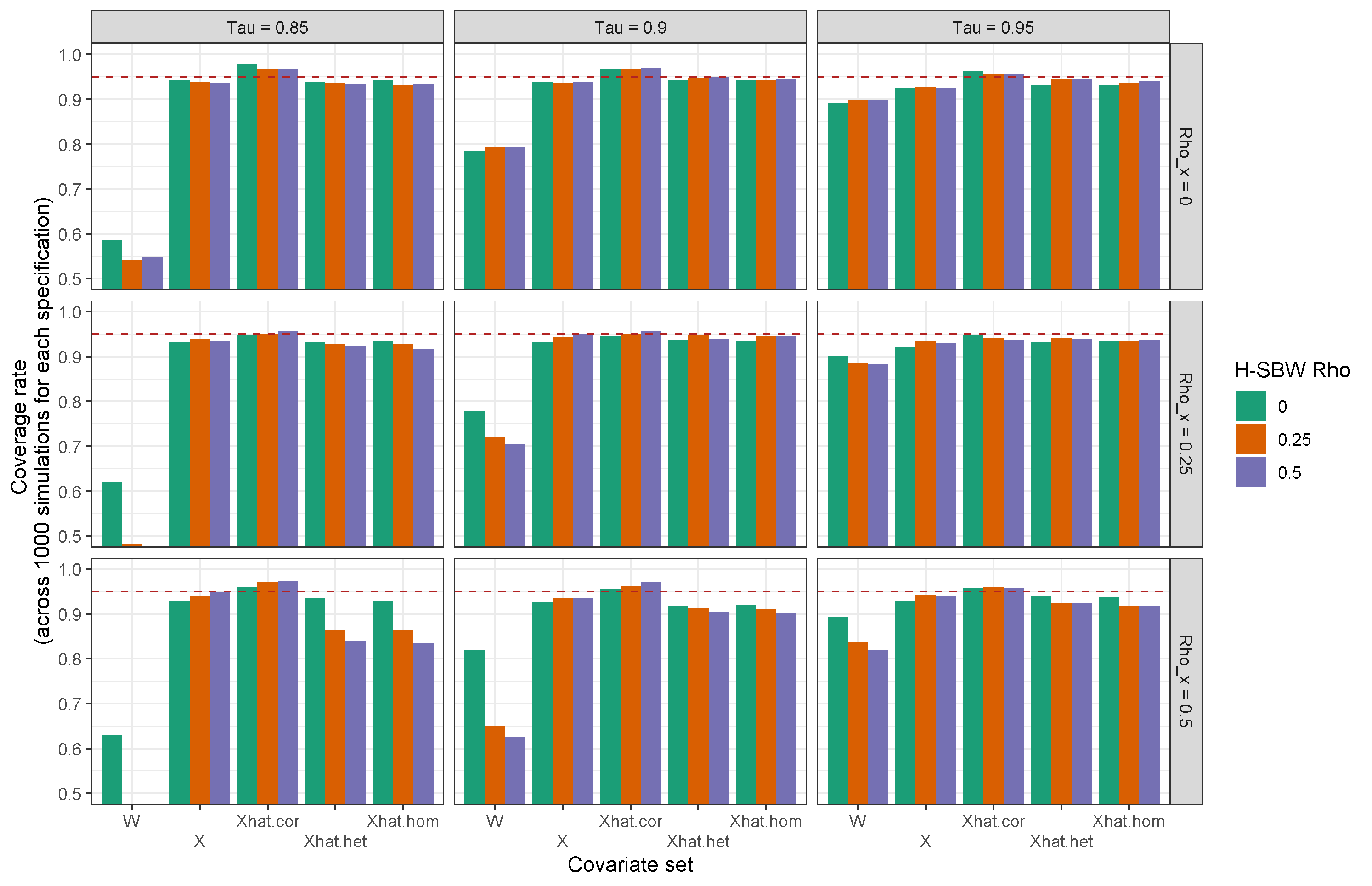}
    \subcaption{Averaged across 1000 simulations for each specification}
\end{center}
\end{figure}

\begin{figure}[H]
\begin{center}
    \caption{Confidence interval length}\label{fig:simcoverage2}
    \includegraphics[scale=0.5]{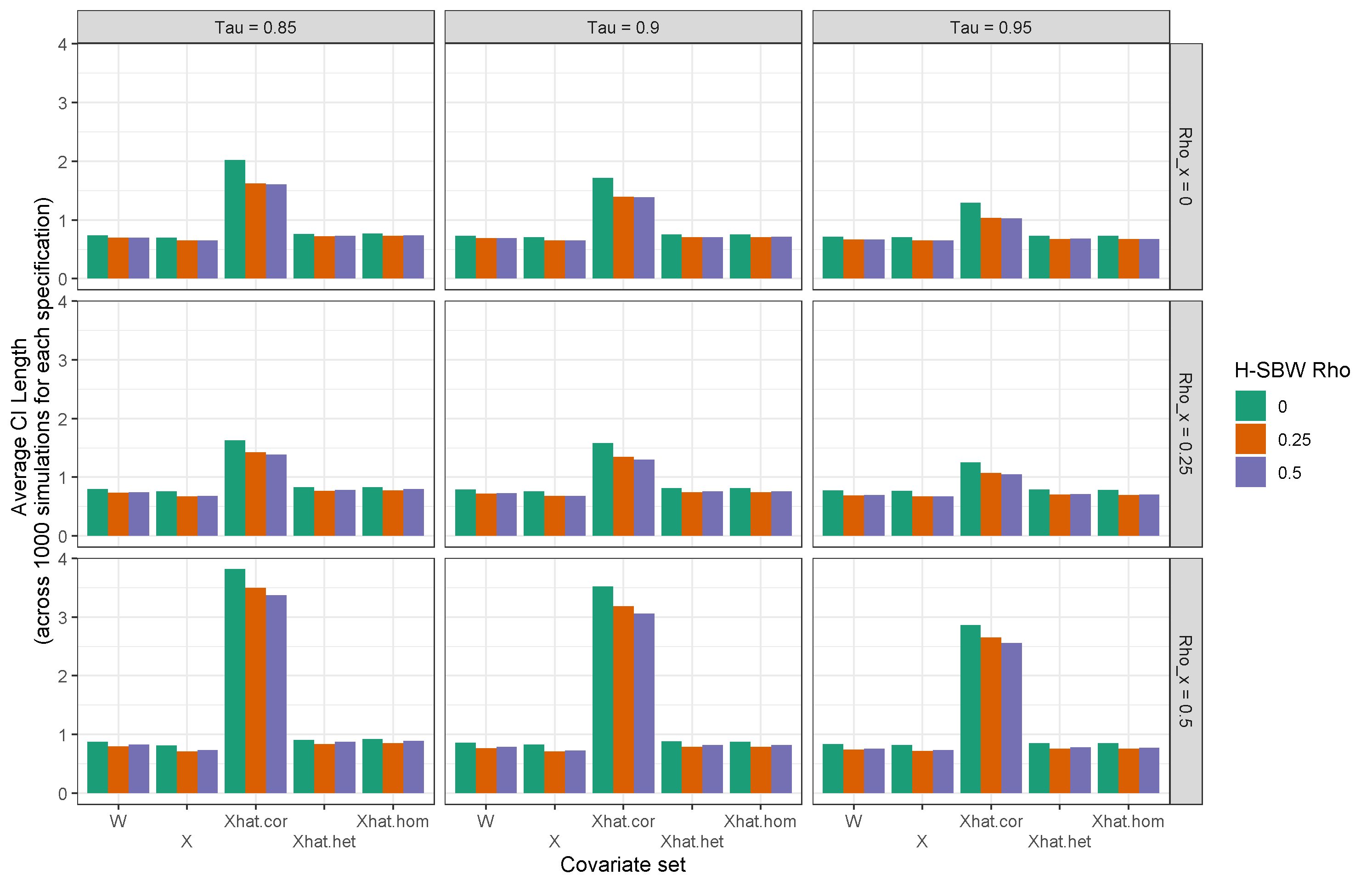}
    \subcaption{Averaged across 1000 simulations for each specification}
\end{center}
\end{figure}

We next evaluate the performance of the leave-one-state-out jackknife procedure and evaluate confidence interval coverage (for 95\% confidence intervals) and length, and display these results in Figures~\ref{fig:simcoverage1} and ~\ref{fig:simcoverage2}. We use the standard normal quantiles to generate confidence intervals throughout.

Figure~\ref{fig:simcoverage1} shows that using SBW ($\rho = 0$) we obtain approximately nominal coverage rates across all specifications that use $X$ or some version of $\hat{X}$. However we fail to ever achieve nominal coverage rates when balancing on $W$. Even when the amount of measurement error is small ($\tau = 0.95$) we only obtain at best slightly below 90 percent coverage rates. Confidence interval coverage appears to deteriorate as we increase $\rho_x$, even when balancing on the true covariates $X$.\footnote{This is consistent with results in \cite{cameron2008bootstrap}.} Unsurprisingly the rates worsen for estimators generated on $\hat{X}^{het}$ and $\hat{X}^{hom}$ in the settings where we found the highest bias. We also obtain slightly less than desired rates for estimators using $X$. This may be due to the fact that we do not use any degrees-of-freedom adjustment for our confidence intervals (we examine this further in Section~\ref{appssec:simstudyresults2}). On the other hand, coverage rates are accurate or conservative for estimators generated on $\hat{X}^{cor}$. The positive bias of the jackknife variance estimates appears to drive this result.\footnote{We almost always find a positive or negligible bias for the unscaled jackknife variance estimates, consistent with Proposition~\ref{prop:jackknife}.}

Figure~\ref{fig:simcoverage2} compares the mean confidence interval lengths using the leave-one-state-out jackknife. The H-SBW estimator is associated with slightly more precise estimates, reflecting that the estimators have decreased variability under our assumed correlation structures. We again see that even when we choose $\rho$ sub-optimally we obtain more precise inferences than when using SBW. This suggests benefits to using H-SBW even when our estimate of $\rho$ is inaccurate. As the previous results would suggest, the confidence intervals using $\hat{X}^{cor}$ are quite large. The results are similar when considering homoskedastic measurement errors. 

We emphasize five takeaways from this simulation study. First, the variance improvements of SBW relative to H-SBW are likely modest. That said, the extent to which we can improve on SBW also depend on $\rho^\star$, which we fix in this study, and these improvements can be greater as $\rho^\star$ increases, which we explore more in Section~\ref{appssec:simstudyresults2} below.\footnote{They also depend on how $p_s$ varies across states. We draw $p_s$ from only one distribution throughout our simulations.} Second, H-SBW can increase the bias of our proposed estimators in the context of measurement error and dependent data. However, the bias is generally small relative to the bias of balancing on the noisy covariate measurements $W$, and MSE improvements using H-SBW are still possible relative to SBW with the simple covariate adjustments ($\hat{X}^{hom}$ or $\hat{X}^{het}$). Third, despite the improved theoretic properties, accounting for the correlation in the data when using H-SBW (i.e. balancing on $\hat{X}^{cor}$ and the data are measured with error) may not be worth the cost in variance given the sample size of states that we consider here. Fourth, we find no evidence that the ``heterogeneous adjustment'' improves our subsequent estimates even when it reflects the true model. This adjustment may even induce bias when the model is incorrect.\footnote{This finding may in part reflect the distribution of sample sizes we generated, which we took to be uniform. Perhaps with a different distribution these results would differ.} Finally, confidence interval coverage when using the leave-one-state-out jackknife variance estimates give close to nominal coverage rates for our unbiased estimators -- even when using the standard normal quantiles in a setting with 25 states. 

This simulation study assumes throughout that we know the true data generating model for the outcome and that are data are gaussian. Because in practice we would not have such knowledge, this study complements our validation study in Section~\ref{sec:results}, which has more direct bearing on understanding how these estimators might perform in an applied setting.

\subsection{Additional results: H-SBW without measurement error}\label{appssec:simstudyresults2}

We consider additional simulations for the setting where $X$ is known and the outcome model follows homoskedastic but possibly correlated errors. For these results we only vary $\rho^\star \in \{0, 0.25, 0.5, 0.75, 0.99\}$, and fix $\rho_x = 0.25$ throughout. 

\begin{figure}[H]
\begin{center}
    \caption{Variance of H-SBW estimator for known covariates}\label{fig:hsbwvarx}
    \includegraphics[scale=0.5]{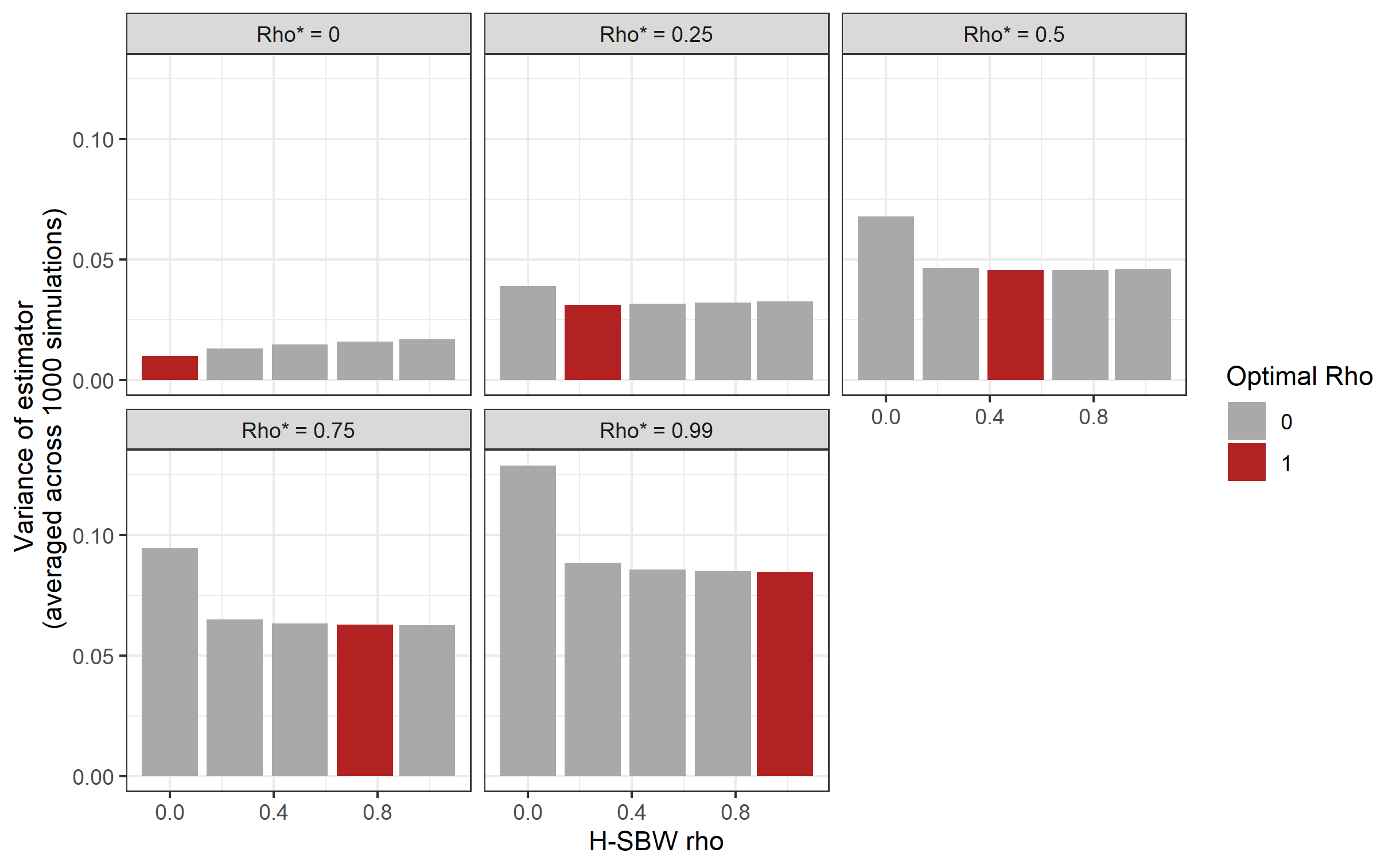}
    \subcaption{Averaged across 1000 simulations for each specification}
\end{center}
\end{figure}

Figure~\ref{fig:hsbwvarx} displays the empirical variance of the H-SBW estimators averaged over 1000 simulations. Each panel reflects different values of $\rho^\star$, while the x-axis throughout displays the assumed value of $\rho$ in the H-SBW objective. The red bars indicate when $\rho$ is optimally selected, $\rho = \rho^\star$. This selection corresponds with the lowest variance estimator as we expect. Consistent with our previous results, when $\rho^\star > 0$ any assumed $\rho$ improves the variance of the resulting estimator relative to SBW. This requires in general that our assumed correlation structure of the error terms is correct.

\begin{figure}[H]
\begin{center}
    \caption{H-SBW coverage for known covariates}\label{fig:hsbwcoveragex}
    \includegraphics[scale=0.5]{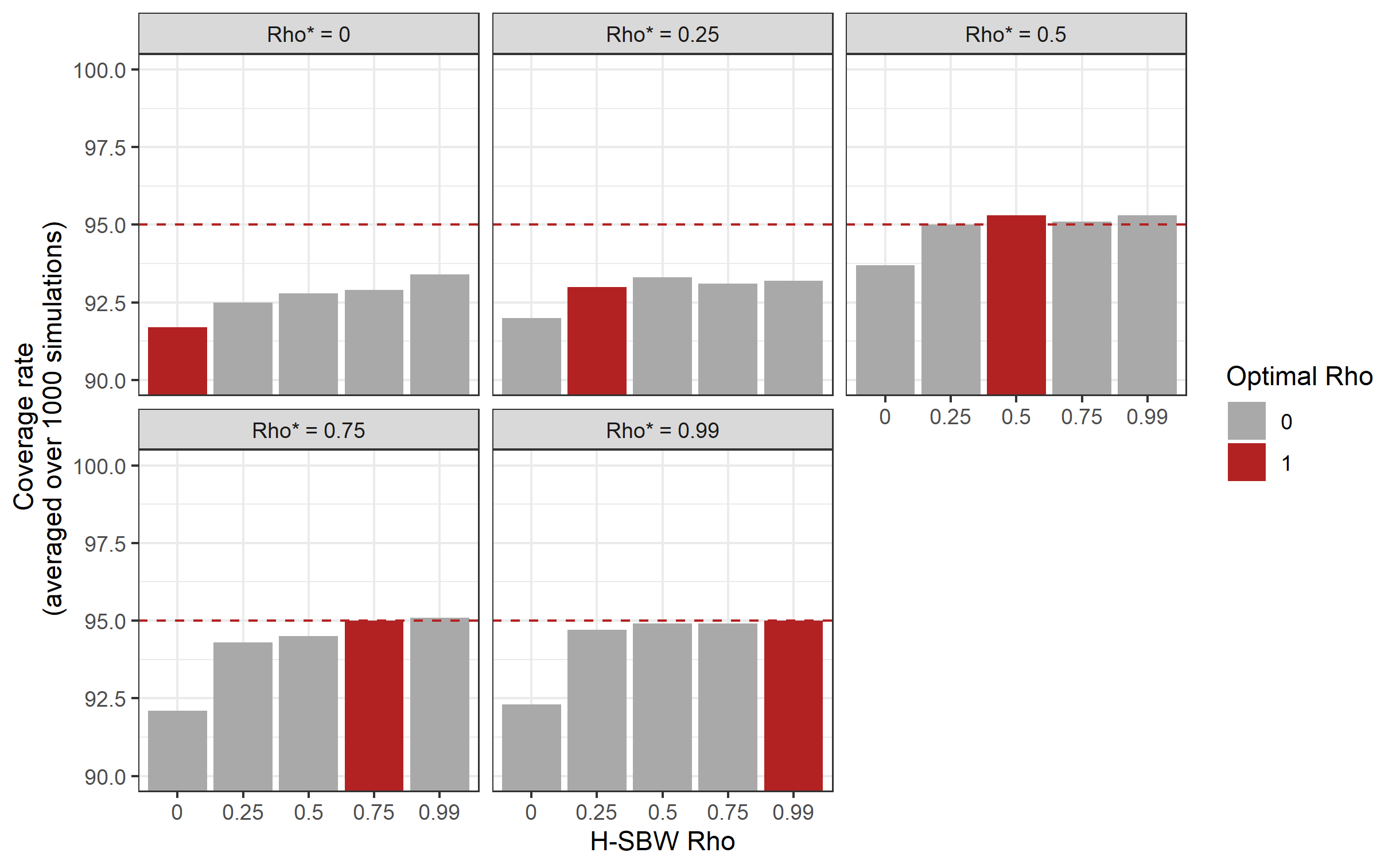}
    \subcaption{Averaged across 1000 simulations for each specification}
\end{center}
\end{figure}

We conclude by examining the confidence interval coverage for the leave-one-state-out-jackknife. Figure~\ref{fig:hsbwcoveragex} displays the results. Consistent with our previous findings we obtain slightly less than nominal coverage rates, particularly when $\rho^\star$ is small. This slight undercoverage is likely due in part to our use of standard normal quantiles to generate confidence intervals.\footnote{Consistent with Proposition~\ref{prop:jackknife} and the previous results we again find that the unscaled variance estimates are all either positively or negligibly biased.} Interestingly, coverage appears to improve when setting $\rho > 0$ even when $\rho^\star = 0$. We speculate that by more evenly dispersing the weights across states, H-SBW may be increasing the effective degrees of freedom of the estimator (see, e.g., \cite{cameron2015practitioner}), illustrating another potential benefit of this approach. Analyzing the theoretic properties of these variances estimators in this setting is beyond the scope of the present study but would provide interesting future work.

\end{document}